\documentclass[twocolumn,secnumarabic,amssymb, nobibnotes, aps,prx,showpacs]{revtex4}
\usepackage{amsmath,amsthm,amssymb}
\usepackage[dvips]{graphicx,color}
\usepackage{color}

\begin{document}

\title{\textcolor{black}{Statistical properties of fluctuations of time series representing appearances of words in nationwide blog data and their applications: An example of modelling fluctuation scalings of nonstationary time series}} 
\title{\textcolor{black}{Statistical properties of fluctuations of time series representing appearances of words in nationwide blog data and their applications: An example of modelling fluctuation scalings of nonstationary time series}}
\author{Hayafumi Watanabe$^{1,2}$}\email[E-mail: ]{hayafumi.watanabe@gmail.com}
\author{Yukie Sano$^3$}
\author{Hideki Takayasu$^4$}
\author{Misako Takayasu$^5$}
\affiliation{$^1$Hottolink,Inc., 6 Yonbancho Chiyoda-ku, Tokyo 102-0081, Japan}
\affiliation{$^2$Risk Analysis Research Center, The Institute of Statistical Mathematics, 10-3 Midori-cho, Tachikawa, Tokyo 190-8562, Japan}
\affiliation{$^3$Faculty of Engineering, Information and Systems, University of Tsukuba, Tennodai, Tsukuba, Ibaraki 305-8573 Japan}
\affiliation{$^4$Sony Computer Science Laboratories, 3-14-13 Higashi-Gotanda, Shinagawa-ku, Tokyo 141-0022, Japan}
\affiliation{$^5$Institute of Innovative Research, Tokyo Institute of Technology, 4259 Nagatsuta-cho, Midori-ku, Yokohama 226-8502, Japan}
\begin{abstract}
To elucidate the non-trivial empirical statistical properties of fluctuations of a typical non-steady time series representing the appearance of words in blogs, we investigated approximately \textcolor{black}{three billion Japanese blog articles} over a period of six years and analyse some corresponding mathematical models.
First, we introduce a solvable non-steady extension of the random diffusion model, which can be deduced by modelling the behaviour of heterogeneous random bloggers.
Next, we deduce theoretical expressions for both the temporal and ensemble fluctuation scalings of this model, and demonstrate that these expressions can reproduce all empirical scalings over eight orders of magnitude. 
Furthermore, we show that the model can reproduce other statistical properties of time series representing the appearance of words in blogs, such as functional forms of the probability density and correlations in the total number of blogs.
As an application, we quantify the abnormality of special nationwide events by measuring the fluctuation scalings of 1771 basic adjectives.
\par \par \par
\end{abstract}

\maketitle
\section{Introduction} 
In order to understand human behaviour with high accuracy, the use of data from social media is rapidly spreading in both practical applications (such as marketing, television shows, politics, and finance) and basic sciences (such as sociology, physics, psychology, and information science) \cite{preis2012quantifying, ugander2011anatomy, ceron2014every, ginsberg2009detecting, sakaki2010earthquake, grajales2014social, yu2012survey}.
In such analyses of social media data, one of the most important basic objects is the time series representing the appearance of considered keywords. That is, a sequence of daily counts of the appearances of a considered word within a huge social media data set. This quantity is mostly used to measure temporal changes in social concerns related to the considered word. \par
%
%
%
Our research focuses on the ``fluctuation'' (i.e., occurrence of random noise) in the time series. We aim to describe this fluctuation precisely, whereas the majority of previous research has focused on ``trends'' in the time series (i.e., nonrandom parts of the time series) for practical reasons.
The reasons why we focus on fluctuation are as follows: \textcolor{black}{ 
(i) The information regarding noise is important for extracting essential information from the data in precise observations. 
For example, this can be used to eliminate noise, detect anomalies, etc. }
(ii) The fluctuation of a time series of social media data obeys a statistical law known as ``fluctuation scaling'', which can be observed in various complex systems relating to both natural and human phenomena \cite{xu2015taylor, eisler2008fluctuation,onnela2010spontaneous, sato2010fluctuation, gerlach2014scaling, sano2010macroscopic, 10.1371/journal.pone.0109004}. Thus, it is also important to understand the properties of fluctuation in social media data in the context of general complex systems science or physical sciences. \par

\textcolor{black}{Fluctuation scaling (FS), which is also known as ``Taylor's law'' \cite{taylor1961aggregation} in ecology, is a power law relation between the system size (e.g., a mean) and the magnitude of fluctuation (e.g., a standard deviation). }
FS is observed in various complex systems, such as random work on a complex network \cite{PhysRevLett.100.208701}, internet traffic \cite{argollo2004separating}, river flows \cite{argollo2004separating}, animal populations \cite{xu2015taylor}, insect numbers \cite{xu2015taylor, eisler2008fluctuation}, cell numbers \cite{eisler2008fluctuation}, foreign exchange markets \cite{sato2010fluctuation}, the download numbers of Facebook applications \cite{onnela2010spontaneous}, word counts of Wikipedia \cite{gerlach2014scaling}, academic papers \cite{gerlach2014scaling}, old books \cite{gerlach2014scaling}, crimes \cite{10.1371/journal.pone.0109004}, and Japanese blogs \cite{sano2010macroscopic}. \par 
Note that physicists have studied linguistic phenomena using concepts of complex systems \cite{link1} such as competitive dynamics \cite{abrams2003linguistics}, statistical laws \cite{altmann2015statistical}, and complex networks \cite{cong2014approaching}. Our study can also be positioned within this context, that is, we study properties of the time series of word counts in nationwide blogs (a linguistic phenomenon) using FS, which is one of the concepts of complex science or statistical physics.
By this viewpoint, we can analyse fluctuations very accurately.
\par
 A certain type of FS can be explained by the random diffusion (RD) model \cite{PhysRevLett.100.208701}.
The RD model, which is described by a Poisson process with a random time-variable Poisson parameter, has been introduced as a mean field approximation for a random walk on a complex network. 
It can be demonstrated that the fluctuation of this model obeys FS, with an exponent of $0.5$ for a small system size ( i.e., a small mean) or $1.0$ for a large system size (i.e., a large mean). 
Because this model is based only on a Poisson process,  it is not only applicable to random walks on complex networks, but also to a wide variety of phenomena related to random processes. For instance, this model can reproduce a type of FS concerning the appearance of words in Japanese blogs \cite{sano2009, PhysRevE.87.012805}. 
However, owing to the assumption of stationarity for the RD model, this steady model cannot be applied to describe unsteady properties, as observed in real time series regarding the appearance of words in blogs.
 \par
There exists a pioneering study regarding the relations between FS and unsteady time series. In Ref. \cite{argollo2004separating}, Argollo de Menezes et al. introduced a method of separating ``internal fluctuations'' corresponding to individual factors and ``external fluctuations'' corresponding to unsteady shared factors.
Moreover, they showed that there are two types of FS for internal fluctuations with exponents of 0.5 or 1.0, by applying this method to empirical data regarding internet routers (0.5), a microchip (0.5), the World Wide Web (1.0), and the highway system (1.0).
However, a theoretical basis for these FSs has not been clarified. \par
 In our study, in order to validate the model we explore not only the fluctuation scalings (a scaling between the mean and variance), but also the functional forms of probability distributions. 
 Although the vast majority of previous theoretical and empirical studies have focused only on scalings \cite{giometto2015sample, cohen2013taylor}, there have been a few previous studies that investigated the relations between fluctuation scalings and the distributions. A. Fronczak et al. described a relationship with the canonical distribution that is deduced from the second law of thermodynamics \cite{fronczak2010origins}. S. Wayne et al. demonstrated a relationship with Tweedy distributions that was introduced by Tweedy in 1984 in order to explain fluctuation scalings, and is related to scale invariance of the family of probability distributions \cite{kendal2011taylor, kendal2004taylor}. Joel E Cohen examined a relationship with random sampling of a skewed distribution, such as the log normal distribution \cite{cohen2015random}. 
\par
In this paper, we first introduce a simple nonsteady extention of the RD model to describe nonsteady time series that obey FS, such as word appearance in blogs. 
Second, we derive three types of mathematical expressions for FSs in this model: the raw time series of word appearances, the time series scaled by the total number of blogs, \textcolor{black}{and ensemble scalings at fixed times}. In addition, we 
demonstrate that these expressions reproduce the empirical scalings over eight orders of magnitude, by using five billion Japanese blog articles from 2007. Furthermore, we show that the model can also reproduce other statistical properties, such as the shapes of probability density functions. 
Third, we apply our model to the quantification of the abnormalities of special nationwide events, and the temporal dependence of an abnormality regarding a particular word.
Finally, we conclude with a discussion. \par
\section{Data set}
\begin{figure*}
\includegraphics[width=12cm]{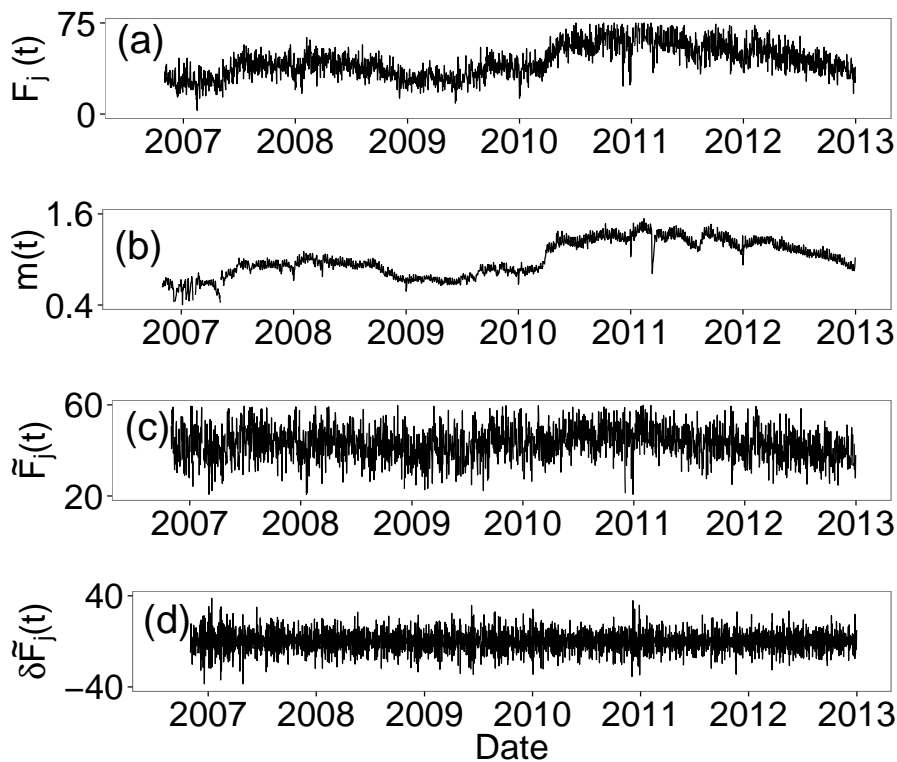}
\caption{
(a) An example of daily time series of raw word appearances for the ``yowai (weak)'', $F_j(t)$. 
(b) The daily time series of the normalised total number of blogs, $m(t)$ (see Appendix H).  
(c) The daily time series of word appearances scaled by the normalised total number of blogs, for ``yowai'',  $\tilde{F}_j(t)=F_j(t)/m(t)$ \textcolor{black}{(defined by Eq. \ref{def_tilde_f}.)} 
(d) The differential time series of word appearances scaled by the normalised total number of blogs, for ``yowai'', \textcolor{black}{$\delta \tilde{F}_i(t) \equiv F(t)/m(t)-F(t-1)/m(t-1)$ (defined by Eq. \ref{def_dtF}). }
From these figures, we can confirm that the time-variation of raw word appearances $F_j(t)$ shown in the panel (a) is almost the same as that of the total number of blogs $m(t)$ shown in the panel (b). }
\label{tseries}
\end{figure*}

In our data analysis, we analyse a time series representing the frequencies with which words appear in Japanese blogs per day. 
In order to obtain this time series, we employed a large database of Japanese blogs ("Kuchikomi@kakaricho"), which is provided by Hottolink Inc. This database contains three billion articles from Japanese blogs, covering 90 percent of Japanese blogs since November 1st 2006. 
Fig. \ref{tseries} shows a example of the time series.

\section{Model}
It was reported in Ref. \cite{PhysRevLett.100.208701} that the FS of a blog has two scaling regions, the exponents of which are 0.5 for a small mean region (Poisson region) and 1.0 for a large mean region (non-Poisson region), 
and this FS can be explained by the (steady) RD model. The RD model represents a mixture of Poisson models, and was originally introduced from a mean field average approximation of the random walk on a complex network, in order to understand the FS of transport on a complex network \cite{PhysRevLett.100.208701}.
However, because the original \textcolor{black}{(steady)} RD model represents a steady probabilistic process, it is unable to describe non-steady effects on the FS, such as changes in the number of bloggers. 
Thus, for a theoretical analysis we introduce and analyse a simple nonstationary extension of the RD model (extended RD model) to describe nonstationary time series. 
 \par
\textcolor{black}{
\subsection{Steady RD model}
}
\textcolor{black}{
The original (steady) RD model described in \cite{sano2010macroscopic}, \cite{PhysRevLett.100.208701}, which is a Poisson process consisting of a stochastic process whose Poisson parameter (the mean value) varies randomly, is defined as follows for $t=1,2,3, \cdots T$:}
\textcolor{black}{
\begin{eqnarray}
F_j(t) &\sim& Poi(c_j \cdot \Lambda(t)) \label{r0} \\
\Lambda(t) &\sim& Unif(1,\Delta_u) \label{r1}
\end{eqnarray}
where $X \sim Poi(A)$ is defined by a random variable $X$ that obeys the Poisson distribution and has Poisson parameter $A$, $X \sim Unif(A,B)$ is defined such that $X$ obeys a uniform distribution with support $[A-B,A+B]$, and $0 \leq \Delta_u \leq 1$ . 
This equation indicates that the random variable $F_j(t)$, which represents the $j$-th observable at time $t$, is sampled from a Poisson distribution with a Poisson parameter $c_j \cdot \Lambda(t)$. Here,  
$c_j \geq 0$ is a scale factor of the Poisson parameter of the model, and $\Lambda(t)$, which is related to the total number of blogs, is a random factor that obeys the independent uniform distribution defined by Eq. \ref{r1}.  
In the case of a time series of blogs, the observable $F_j(t)$ corresponds to the frequency with which the $j$-th word occurs on the $t$-th day, $c_j$ corresponds to the temporal mean of the frequency $c_j=\sum_{t=1}^{T}F_j(t)/T$ $(T \to \infty)$, and $\Lambda$ is related to the total number of blogs (scaled by its temporal mean).}
\par
\textcolor{black}{
\subsection{Extended RD model}
}
\textcolor{black}{
We extend the (steady) RD model to precisely describe non-stationary effects (i.e., time-variances in the usages of words and the total number of blogs) as follows: 
(i)The scale parameter is modified from a constant $c_j$ to a time-varying parameter $c_j(t)$. 
(ii)The distribution of the random part $\Lambda$, which is related to the total number of blogs, is modified from a steady uniform distribution to an arbitrary distribution with time varying mean $m(t)$ and standard deviation $\Delta_m(t)$.
 Then, the extended RD model, which is a nonstationary Poisson process consisting of a stochastic process whose Poisson parameter (the mean value) varies randomly, is defined as follows for $t=1,2,3,\cdots,T$:}
\begin{eqnarray}
F_j(t) &\sim& Poi(c_j(t) \cdot \Lambda_j(t)) \quad (j=1,2,\cdots,W(t)). \label{macro0} \\
\Lambda_j(t) &\sim& Distribution \nonumber \\
&& s. t.  \nonumber \\
&& <\Lambda_j(t)>=m(t), \nonumber \\ 
&& <(\Lambda_j(t)-<\Lambda_j(t)>)^2>=\Delta_m(t)^2.  \label{macro0abc}
\end{eqnarray}
\textcolor{black}{
\textcolor{black}{The first equation indicates that the random variable $F_j(t)$ is sampled from a Poisson distribution whose Poisson parameter takes a value $c_j(t) \cdot \Lambda_j(t)$.} Furthermore, $c_j(t) \geq 0$ is a scale factor of the Poisson parameter of the model, $\Lambda_j(t)$ is a random factor, and \textcolor{black}{$W(t)$ is the total number of types of words at time $t$.} In the case of a time series of blogs, as in the original steady RD model, the observable $F_j(t)$ corresponds to the frequency with which the $j$-th word occurs on the $t$-th day, and larger values of $c_j(t)$ indicate that the j-th word appears more frequently at time $t$.}
 \par
 
%
\textcolor{black}{$\Lambda_j(t)$ is a non-negative random variable with mean $m(t) \geq 0$ and standard deviation $\Delta_m(t) \geq 0$. }  
Here, $m(t)$ is a shared time-variation factor for the whole system, and 
we assume that $(\sum_{t=1}^{T}m(t)/T=1)$, for normalization.
In the case of a time series of blogs, $m(t)$ closely corresponds to the normalised number of blogs. \par
For particular settings, this corresponds to the following known models: 
\textcolor{black}{
(i) In the case that $c_j(t)$ is a constant $c_j(t)=c_0$ and $\Lambda_j(t)$ is equal to 1, namely, the probability density function of $\Lambda_j(t)$ is the delta function $P_{\Lambda_j(t)}(x)=\delta(x-1)$, this represents the steady Poisson process with the parameter $c_0$. 
(ii) In the case that the probability density function of $\Lambda_j(t)$ is the delta function $P_{\Lambda_j(t)}(x)=\delta(x-m(t))$, this represents the non-steady Poisson process with the parameter $c_j(t) \cdot m(t)$.
}
(iii) In the case that $\Lambda_j(t) \sim Unif(1, \Delta/3)$ \textcolor{black}{and $c_j(t)=c_0$},
 this represents the original (steady) RD model given by \cite{PhysRevLett.100.208701, sano2010macroscopic}. 
Therefore, the proposed model (i.e., the extended RD model) represents an extension of Poisson processes and the steady RD model.  \par
For convenience of analysis, we assume that $c_j(t)$ can be decomposed into 
a scale component $\check{c}_j$, which corresponds to the temporal mean of the count of the j-th word in an observation period, and a time variance component $r_j(t)$, such that
\begin{equation}
c_j(t)=\check{c}_j \cdot r_j(t). \label{cj}
\end{equation}
Here, we also assume \textcolor{black}{for normalization} that $(\sum_{t=1}^{T}r_j(t)/T=1)$.
In addition, we assume for simplicity that $\Delta_m(t)$ is decomposed as
\begin{equation}
\Delta_m(t)=m(t)^{\beta_{m}} \Delta_0(t), \label{Deltam}
\end{equation}
where $\beta_m$ is a real constant and $\Delta_0(t)>0$ is a part which does not depend on $m(t)$.
\textcolor{black}{Note that this assumption constitutes a simplification of the correlation between the mean $m(t)$ of $\Lambda_j(t)$ and the corresponding standard deviation $\Delta_m(t)$. 
 For instance, with the condition that $\beta_{m}=0$ the standard deviation $\Delta_m(t)$ is immutable, regardless of the mean $m(t)$, 
  and with the condition that $\beta_{m}=1$ the standard deviation $\Delta_m(t)$ is proportional to the mean $m(t)$. 
}

  \begin{table*}
 \begin{tabular}{cl cc}
 \hline
 & Parameter & Meaning & Estimation \\
 \hline
  \multicolumn{4}{l}{$F \sim Poi(c_j(t) \cdot \Lambda_j(t))$} \\
  (i)& \multicolumn{3}{l}{ $c_j(t)=\check{c}_j \dot r_j(t)$} \\
  &\textbullet$\check{c}_j$ & scale factor of the $j$-th word & $\sum^{T}_{t=1}F_j(t)/T$ \\
  &\textbullet$r_j(t)$ & time-variant factor of the $j$-th word  &unable to estimate accurately$^{*1}$ \\
   (ii)  & \multicolumn{3}{l}{$\Lambda_j(t) \sim \{$distribution with mean $m(t)$ and standard deviation $\Delta_m(t)\}$, $\Delta_m(t)=\Delta_0(t) \cdot m(t)^{\beta_m}$} \\
  &\textbullet $m(t)$ & mean of $\Lambda_j(t)$ (scaled number of blogs) & Appendix H\\ 
  &\textbullet$\Delta_0(t)$ & scale parameter of the standard deviation of $\Lambda_j(t)$ &  0.021  \\
  &\textbullet$\beta_m$ & relation parameter between the standard deviation and the mean &  1.0$^{*2}$ \\ 
   \hline
  \multicolumn{4}{l}{${}^{*1}$ Instead of a direct estimation, we use lower or upper bounds on $\{r_j(t)\}$ for a rigorous analysis. } \\
  \multicolumn{4}{l}{${}^{*1}$ Exceptionally, when we calculate the density function, we estimate $c_j(t)$ roughly using the moving average (Eq. \ref{moving_average}, Eq. \ref{moving_median}). } \\
  \multicolumn{4}{l}{${}^{*2}$ For $ 0.5 \leq \beta_{m} \leq 1.5$, the model does not contradict empirical data. We use $\beta_m=1.0$ for simplicity in our empirical analysis.} \\
 \end{tabular}
 \caption{Summary of the model and parameters}
 \label{table_parameter}
 \end{table*}%
 
   \begin{table*}
 \begin{tabular}{clcc}
 \hline
 & $\sigma^2$ & $a_1:$ Coefficients of  $\mu$ &  $a_2$: Coefficients of  $\mu^2$  \\
 \hline \hline
   (i)& Temporal scaling &  \multicolumn{2}{r}{($\mu=E[F_j]=E[\check{F_j}]=\check{c}_j$)} \\
    &  $V[F_j]$ &1 & $V[r_j]+(V[r_j]+1)\cdot (V[m]+(1+V[m]) \cdot E[\Delta_0^2])$ \\
    &  -Lower bound &1&  $V[m]+(1+V[m]) \cdot E[\Delta_0^2] $  \\
    &  $V[\check{F_j}]$ &$E[\frac{1}{m}]$& $V[r_j]+ (V[r_j]+1)\cdot E[\Delta_0^2]$   \\
    &  -Lower bound &$E[\frac{1}{m}]$&  $E[\Delta_0^2]$   \\
    &  $V[\delta \check{F_j}]$ & $2 \cdot E[\frac{1}{m}]$ & $V[\delta r_j]+ 2 \cdot (V[r_j]+1) \cdot  E[\Delta_0^2]$ \\
    &  -Lower bound & $2 \cdot E[\frac{1}{m}]$ & $ 2 \cdot E[\Delta_0^2]$   \\
  \hline
  (ii)&Ensemble scaling& \multicolumn{2}{r}{($\mu=E_c[F(t)]=c \cdot m(t)$)} \\
    &$V_c[F]$ &1& $V_c[r(t)]+\Delta_0(t)^2 \cdot (1+V_c[r(t)])$ \\	
    &-Lower bound &1& $\Delta_0(t)$  \\
    && \multicolumn{2}{r}{($\mu=\overline{E_c[F(t)]}=c \cdot \overline{m(t)}$)} \\
    &$V_c[\delta{F_j}]$ & $2$ & $  V_c[\delta (r(t) \cdot m(t))]/\overline{m(t)^2}+2 \cdot \Delta_0(t)^2 \cdot (1+V_c[r(t)])^{*}$ \\
    &-Lower bound & $2$ & $2 \cdot \Delta_0^2(t)$ \\
   \hline
   \multicolumn{3}{l}{*We assume that $\Delta_0(t) \approx \Delta_0(t-1)$, $V^{\zeta}_c[r(t)] \approx V^{\zeta}_c[r(t-1)]$ for simplicity. }
 \end{tabular}
 \caption{Summary of the coefficients of the scalings of the extended RD model and corresponding lower bounds on $\{r_j(t)\}$ for the conditions that $\beta_m=1$, $\sigma^2 \approx a_1 \cdot \mu+a_2 \cdot \mu^2$. See Appendix B. }
 \label{results}
 \end{table*}%

\par
The extended RD model we introduce is determined by five parameters, $\check{c}_j$, $r_j(t)$, $m(t)$, $\Delta_0(t)$, and $\beta_{m}$.
Hence, in our study we investigate the precise dependence of FSs and their accompanying phenomena on these five parameters.
When comparing the model with empirical data, $\check{c}_j$ is estimated by the temporal average of counts of the j-th word $\check{c}_j(t)=\sum_{t=1}^T F_j(t)/T$; $m(t)$ is estimated by ensemble median of counts of the words at the time t, as described in Appendix H;, $\Delta_0(t)=0.021$ (We assume that $\Delta_0(t)$ is constant for simplicity of empirical analysis).  
\textcolor{black}{For $ 0.5 \leq \beta_{m} \leq 1.5$, the model does not contradict empirical data, and is not easy to differentiate between different values for the parameter $\beta_m$. Thus, we set $\beta_m=1.0$ in this paper for simplicity. A summary of the parameters of the model is presented in table \ref{table_parameter}.} \par
\textcolor{black}{
Note that $r_j(t)$ cannot accurately be estimated using data. Therefore, in order to analyse the data rigorously, we do not perform a direct estimation of this parameter. Instead of a direct estimation, we compare the theory and data by considering lower or upper bounds on $\{r_j(t)\}$. Exceptionally, when we calculate density functions we roughly estimate $r_j(t)$ 
using the moving average (see Eq. \ref{moving_average} and Eq. \ref{moving_median}).  }
\par
%
\subsection{Derivation of the extended RD model from a micro model of blogger behaviour}
The extended RD model (macro model) described above can be deduced from a model that describes simplified heterogeneous blogger behaviour. 
The details of the derivation are provided in Appendix A. In this section, we only present the results. \par
\subsubsection{The random blogger model (the micro model)}
We consider the following model for a system consisting of $N(t)$ bloggers and $W(t)$ words. 
The bloggers perform the following behaviour from time $t=1$ to $t=T$: 
\begin{enumerate}
\item The i-th blogger writes his or her blog randomly with probability $p^{(i)}(t)$ ($0 \leq p^{(i)}(t) \leq 1$) $(i=1,2,3,\cdots,N(t))$. 
Here, by $S(t)$ we denote the set of bloggers who write his or her blog at the time t.
\item The $i$-th blogger who writes a blog in step 1 writes the j-th word $f^{(i)}_j(t)$ times.  $f^{(i)}_j(t)$ is sampled from a Poisson distribution with Poisson parameter  $\lambda^{(i)}_j(t)$ $(j=1,2,3,\cdots,W(t))$. 
\item The total count of the $j$-th word of the system calculated by  $F_j(t)=\sum_{i \in S(t)}f^{(i)}_j(t)$, and the total number of bloggers who write blogs in the system is $M(t)=\sum_{i \in S(t)}1$.
\end{enumerate} \par
\subsubsection{The extended RD model deduced from the blogger model}
Here, we consider the probability distribution of $F_j(t)$. 
Under the conditions that $M(t)$ can be observed and $N(t)>>1$, the distribution of $F_j(t)$ can be approximated as follows: 
\begin{equation}
F_j(t) \sim Poi(c_j(t) \cdot \Lambda_j(t)).
\end{equation}
\textcolor{black}{
Here, the scale factor $c_j(t)$ is given by}
\begin{equation}
c_j(t) =  \mu^{(\lambda)}_j(t) \cdot \sum_{t=1}^{T}M(t),
\end{equation}
\textcolor{black}{and the mean $m(t)$ and the variance $\Delta_m(t)$ of $\Lambda_j(t)$ are given by }
\begin{equation}
m(t) = \frac{M(t)}{\sum_{t=1}^{T}M(t)}, 
\end{equation}
\begin{equation}
\Delta_m(t)=m(t)^{\beta_m} \cdot \Delta_0(t),  
\end{equation}
\begin{equation}
\Delta_0(t) = \sqrt{(1-\frac{M(t)-1}{N(t)-1})} \cdot \frac{1}{\sum_{t=1}^{T}M(t)} \cdot \sigma^{(\lambda)}(t) \label{Delta}, 
\end{equation}
\begin{equation}
\beta_m = 0.5,
\end{equation}
%
\textcolor{black}{where $\mu_j^{(\lambda)}(t) \approx \sum^{N(t)}_{i=1}\lambda_j^{(i)}(t)/N(t)$, 
$\sigma^{(\lambda)}(t)^2 \cdot \mu_j^{(\lambda)}(t)^2 \approx \sum^{N(t)}_{i=1}\{\lambda_j^{(i)}(t)-\mu_j^{(\lambda)}(t)\}^2/N(t)$, and the specific form of the distribution function of $\Lambda_j(t)$ is determined by the parameters $\{p^{(i)}(t)\}$ and $\{\lambda_j^{(i)}(t)\}$. } \par
\textcolor{black}{These equations indicate that we can connect the macro parameters of the extended RD model, given \textcolor{black}{in Eqs. \ref{macro0}, \ref{cj}, and \ref{Deltam}}, with the statistics for the parameters for micro bloggers, $\mu_j^{(\lambda)}(t)$ and $\sigma^{(\lambda)}(t)^2$．}
\par
Under the condition that $\lambda^{(i)}_j(t)=\lambda^{(0)}_j(t) \quad (i=1,2,\cdots,N(t))$, we can obtain that $\Delta_0(t)=0$ from Eq. \ref{Delta}. 
Thus, 
the model represents Poisson processes in the case of homogeneous bloggers. 
In other words, the condition that the model exhibits the particular properties of the RD model
is that the bloggers are heterogeneous.
\par
Note that the fact that $\beta_m=0.5$ can be deduced from the blogger model. 
\textcolor{black}{However, as will be mentioned in the following section, 
\textcolor{black}{empirical observations are not contradicted in the range $0.5 \leq  \beta_{m} \leq 1.5 $.}
Thus, we will need more precise observations in the future in order to verify that $\beta_{m}=0.5$. That is, to verify the validity of the blogger model. 
}
\par
\par
\par
\section{Properties of the model}
In this section, we investigate the statistical properties of the extended RD model, and compare them with the corresponding properties of blog data.
Note that \textcolor{black}{as mentioned above, the model does not contradict empirical data for $ 0.5 \leq \beta_{m} \leq 1.5$}, and is not easy to differentiate regarding the parameter $\beta_m$.
 For simplicity, we present only the case of $\beta_{m}=1.0$ in this section. Discussions concerning general $\beta_m$ are given 
in Appendix B. (The results of this section can be obtained by substituting  $\beta_m=1$ into the results of Appendix B.). \par
\textcolor{black}{Table \ref{results} presents a summary of the fluctuation scalings employed in this section.} 
  
 \par
\subsection{Temporal fluctuation scaling}
First, we discuss the temporal fluctuation scaling (TFS).
The TFS of variable the $A(t)$ ($t=1,2,\cdots T$) 
is defined by the scaling between the temporal mean $E[A]$ and the temporal variance $V[A]$, 
\begin{equation}
V[A] \propto E[A]^{\alpha}. \label{TFS_of_A}
\end{equation}
Here, the temporal mean $E[A]$ and temporal variance $V[A]$ are defined by 
\begin{equation}
E[A] \equiv \sum^{T}_{t=1}A(t)/T,
\end{equation}
\begin{equation}
V[A] \equiv \sum^{T}_{t=1}(A(t)-E[A])^2/T.
\end{equation}
 \par
 Note that the above definition of the TFS in Eq. \ref{TFS_of_A} is expressed in terms of the variance, although the standard deviation is usually used in observations.
 Under this condition, the TFS expressed by the standard deviation can be written as $V[A]^{1/2} \propto E[A]^{\alpha/2}$. 
\textcolor{black}{ In addition, we assume in this section that $T>>1$ for simplicity, and that the following approximations are rigorous in the limit $T \to \infty$.}
 \par
%
\subsubsection{TFS of raw data} 
Here, we investigate the TFS of the time series of the raw counts of word appearances $F_j(t)$ (see Fig. \ref{tseries}(a)), 
which is determined by the steady RD model described in Ref. \cite{sano2009}. 
From Eq. \ref{V_F_ex} in Appendix B, we can obtain the temporal mean 
\begin{eqnarray}
E[F_j(t)]
\approx \check{c_j}
\label{E_F}
\end{eqnarray}
and the temporal variance 
\begin{eqnarray}
&&V[F_j] \nonumber \\ 
&\approx&\check{c}_j+ \nonumber \\
&&\check{c}_j^2 \{ V[r_j]+(V[r_j]+1)\cdot (V[m]+(1+V[m]) \cdot E[\Delta_0^2]) \}. \nonumber \\
\label{V_F}
\end{eqnarray} 
\textcolor{black}{
By inserting  Eq. \ref{E_F} into Eq. \ref{V_F}, we obtain the relationship between the variance and the mean of $F_j(t)$ given by 
\begin{eqnarray}
&&V[F_j]  \approx E[F_j]+E[F_j]^2 \nonumber \\
&\cdot& \{ V[r_j]+(V[r_j]+1)\cdot (V[m]+(1+V[m]) \cdot E[\Delta_0^2]) \}. \nonumber \\
\label{V_FE}
\end{eqnarray}
}
%
\textcolor{black}{
In the case that $r_j(t)=1$ ($t=1,2,\cdots,T$), which indicates the $j$-th word is steady,
\textcolor{black}{from Eq. \ref{V_F}} we can also obtain the following more simple expression: 
\begin{eqnarray}
V[F_j]& \approx &\check{c}_j+ 
\check{c}_j^2 \{(V[m]+(1+V[m]) \cdot E[\Delta_0^2]) \}, 
\label{V_F2}
\end{eqnarray}
\textcolor{black}{which can be written as a function of the mean as 
\begin{eqnarray}
V[F_j]& \approx &E[F_j]+ 
E[F_j]^2 \{(V[m]+(1+V[m]) \cdot E[\Delta_0^2]) \}. \nonumber \\
\label{V_F2E}
\end{eqnarray}
}
In addition, because $V[r_j] \geq 0$ this equation also gives a lower bound on $V[F_j]$ over $\{r_j(t)\}$. From this result, we can deduce the following scaling relations: 
\begin{equation}
V[F_j] \approx
\begin{cases}
\check{c}_j & \text{($\check{c}_j << 1$)} \\
\check{c}_j^2 \{(V[m]+(1+V[m]) \cdot E[\Delta_0^2])  & \text{($\check{c}_j >> 1$)}, 
\end{cases}
\end{equation}
and the corresponding scaling as a function of the mean is written as
\begin{equation}
V[F_j] \approx
\begin{cases}
E[F_j] & \text{($E[F_j] << 1$)} \\
E[F_j]^2 \{(V[m]+(1+V[m]) \cdot E[\Delta_0^2])  & \text{($E[F_j] >> 1$)}, \label{scaling_EF} 
\end{cases}
\end{equation}
where the conditions of the scaling with a single exponent are only that $V[m]=0$ and $E[\Delta_0^2]=0$. 
That is, the time series is a steady Poisson process (i.e., $m(t)=1$ and $r_j(t)=1$.). }
\par
\begin{figure}
\centering
\begin{minipage}{0.5\hsize}
\includegraphics[width=6.2cm]{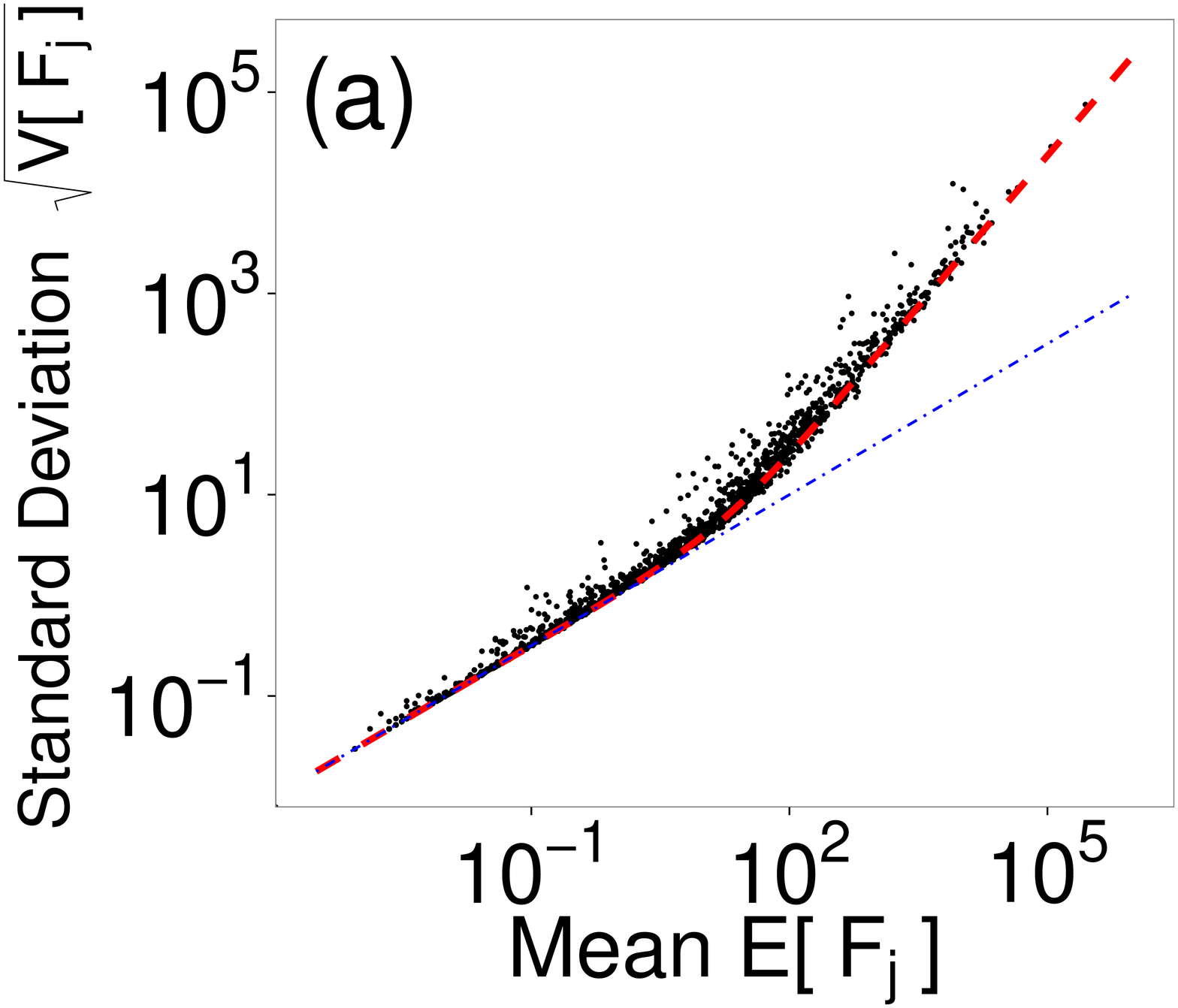}
\end{minipage}
\begin{minipage}{0.5\hsize}
\centering
\includegraphics[width=6.2cm]{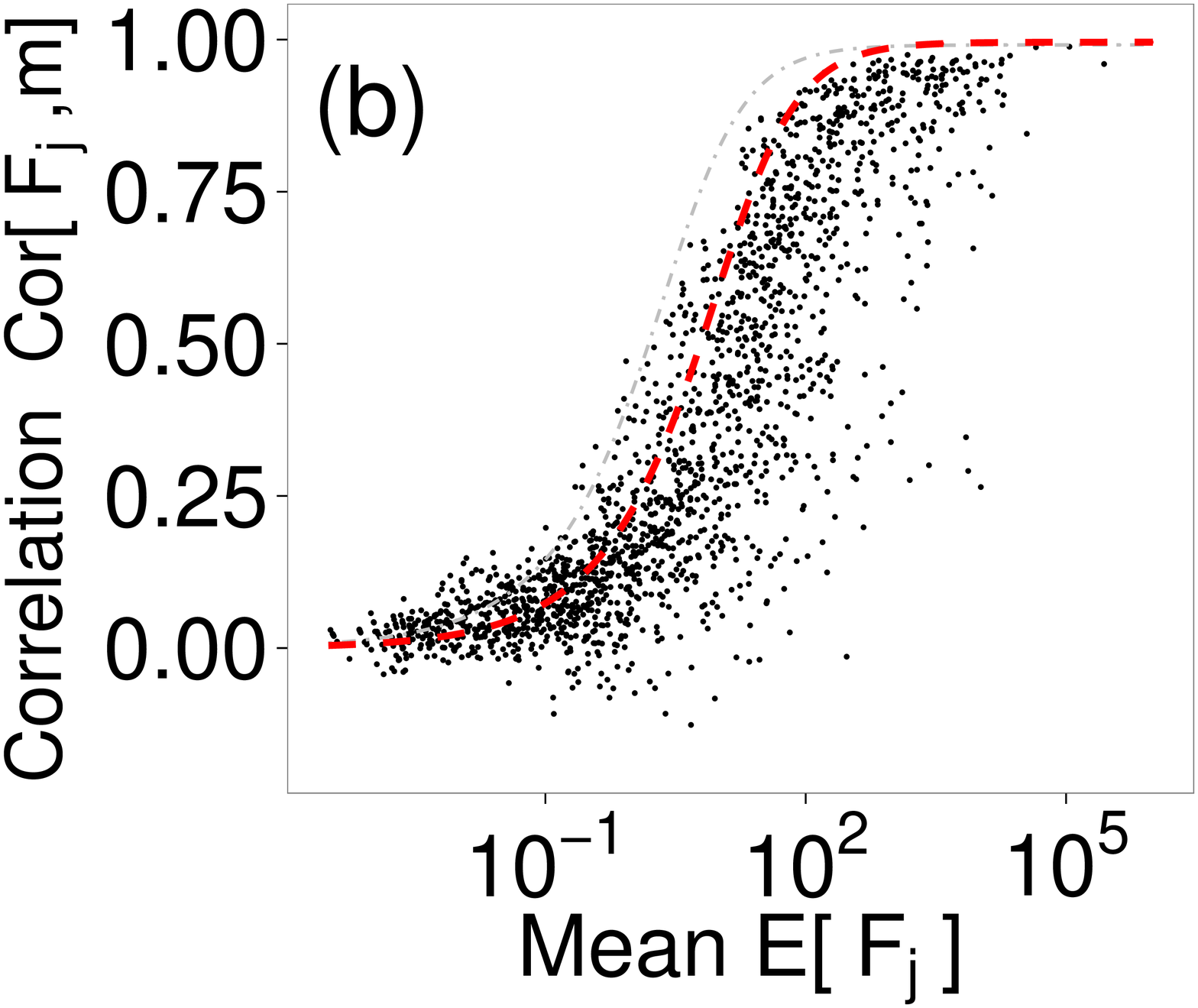}
\end{minipage}
\caption{(a) TFS of raw time series, $V[F_j]$.  The data shown are empirical results of 1771 adjectives (black triangles),
 corresponding to the theoretical curve given by \textcolor{black}{Eq. \ref{V_F2} or Eq. \ref{V_F2E}} (red dashed line). The blue dash-dotted line indicates $y=x^{0.5}$, which corresponds to the Poisson distribution. 
(b) The correlations of raw time series between individual time series $Cor[F_j,m]$ and the scaled total number of blogs $m(t)$ (i.e., the shared Poisson parameter).  
The data shown are empirical results of 1771 adjectives (black plus triangles),
the theoretical curve given by Eq. \ref{cor1} (red dashed line), and the theoretical upper bound, which is considered for the finite number of observables given in Eq. \ref{cor_up} (grey dashed-dotted line). From these panels, we can confirm that the empirical data is in good agreement with the theoretical curve.}
\label{fig_mean_sd}
\end{figure}

\textcolor{black}{
From Fig. \ref{fig_mean_sd} (a), we can confirm that the theoretical curve given by Eq. \ref{V_F2E} is in agreement with the lower bound of empirical data. 
} 
\textcolor{black}{
Here, because $r_j(t)$ cannot accurately be estimated using data, we consider the lower bound for comparing the theory with empirical data. 
Similarly, in later sections we will also employ lower or upper bounds.}
\par
\textcolor{black}{
Note that the steady RD model given by Eq. \ref{r0} explains the temporal FS of raw data $F_j(t)$ by approximating the unsteady time series of blogs using a steady time series (i.e.,  $\Lambda_j(t) \sim Unif(1,V[m]/3)$ and $c_j(t)=\check{c_j}$). 
 However, the model cannot consistently explain certain properties under this condition, which will be discussed later.   
}

\par 
\subsubsection{TFS of the data scaled by the total number of blogs} 
\begin{figure}
\begin{minipage}{0.5\hsize}
\centering
\includegraphics[width=6.2cm]{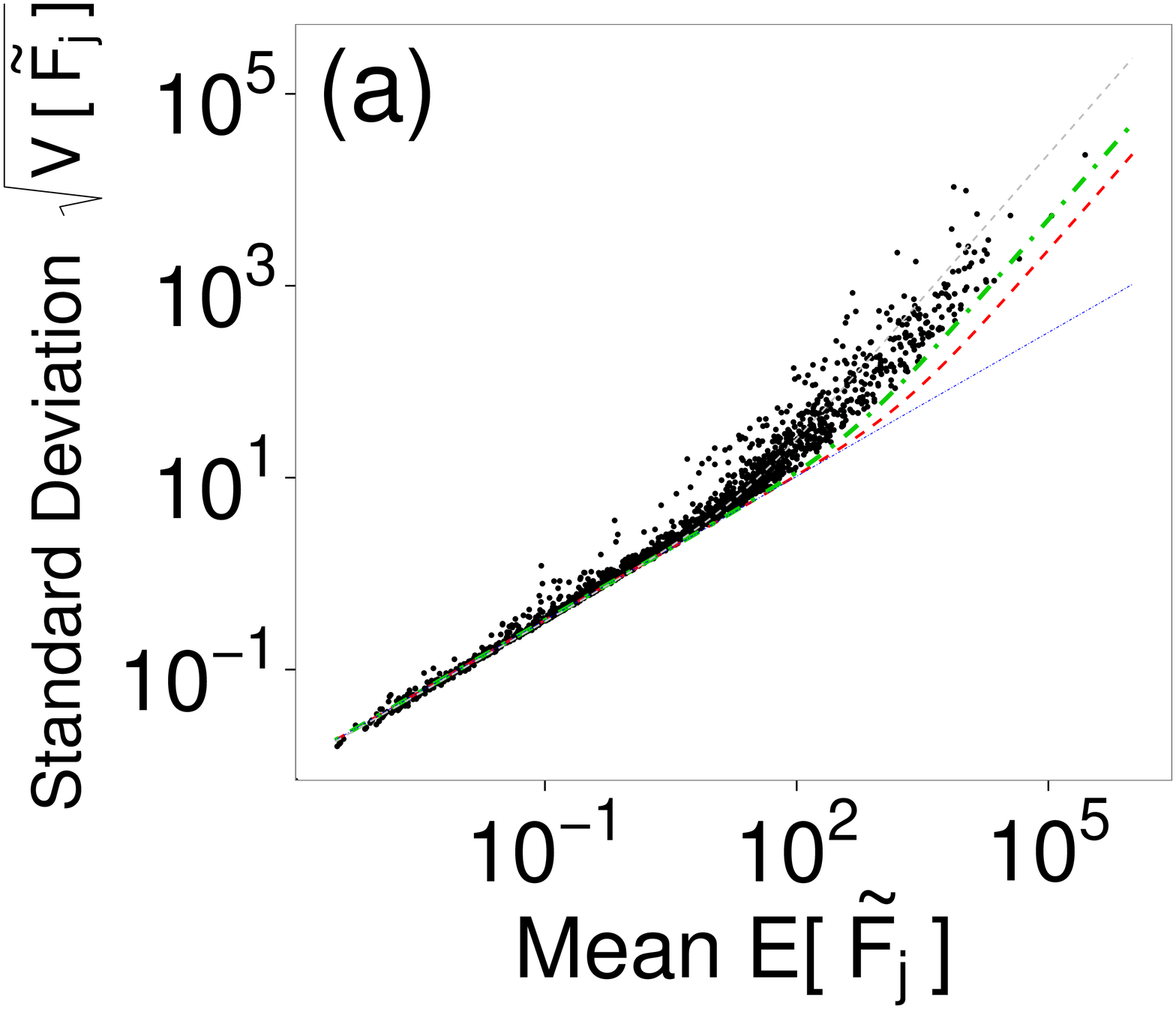}
\end{minipage}
\begin{minipage}{0.5\hsize}
\centering
\includegraphics[width=6.2cm]{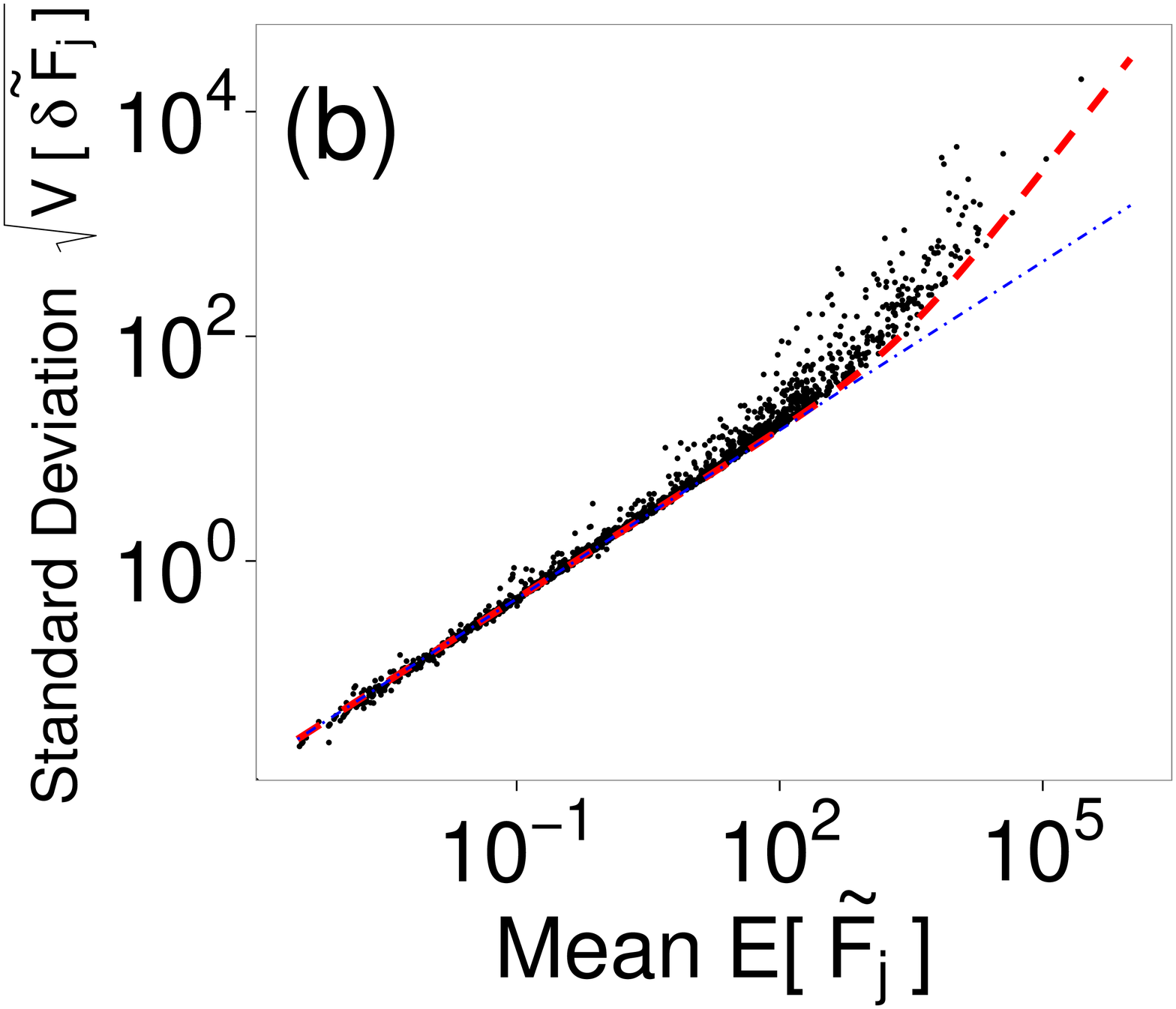}
\end{minipage}
\caption{(a) TFS of normalised time series, $V[\tilde{F}_j(t)]^{0.5}$ (\textcolor{black}{$\tilde{F}_j(t)$ is defined by Eq. \ref{def_tilde_f}}).  
 The data shown are empirical results of 1771 adjectives (black triangles), and the corresponding results of theoretical curves given by \textcolor{black}{Eq. \ref{V_tilde_F2} or Eq. \ref{V_tilde_FH2} (red dashed line)}.
 The green dashed-dotted line indicates the correction of the theoretical lower bound, which is considered in the nonstationary $V[r_j]$ given by Eq. \ref{cor_V_tilde_F2E}. 
 In addition, the thin blue dash-dotted line indicates $\propto x^{0.5}$, and thin grey dashed line is the theoretical curve of the raw time series given by \textcolor{black}{Eq. \ref{V_F2E}}.   
(b) \textcolor{black}{The TFS of the differential of the time series of word appearances $V[\delta \tilde{F}_j]$ ($\delta \tilde{F}_j(t)$ is defined by Eq. \ref{def_dtF}).}
The black triangles indicate the actual data, the red dashed line indicates the theoretical lower bound given by \textcolor{black}{Eq. \ref{v_delta_tilda_f2} or Eq. \ref{v_delta_tilda_f2E}}, and the blue dash-dotted line indicates $\propto x^{0.5}$. We can confirm that the theoretical curve is in accordance with the empirical data from the panel (b).}
\label{fig_mean_sd_t}
\end{figure}

Next, we discuss the time series of word appearances scaled by the total numbers of blogs 
\textcolor{black}{
\begin{equation}
\tilde{F}_j(t)=F_j(t)/m(t). \label{def_tilde_f}
\end{equation}
}
In practice, this value is used to estimate $c_j(t)=\check{c}_j\cdot r_j(t)$, which corresponds to the original time deviation of the $j$-th word separated from the effects of deviations in the total number of blogs $m(t)$ (see Figs. \ref{tseries}(b) and (c)).   
From Eq. \ref{V_tilde_F_ex} in Appendix B, we can obtain that
\begin{eqnarray}
E[\tilde{F}_j]
\approx \check{c_j} \label{E_tilde_F}
\end{eqnarray}
\begin{eqnarray}
V[\tilde{F}_j]
\approx E[\frac{1}{m}]\check{c}_j+
\check{c}_j^2& \{V[r_j]+ (V[r_j]+1)\cdot E[\Delta_0^2] )\}. \nonumber  \\
\label{V_tilde_F}
\end{eqnarray}
Thus, by combining Eq. \ref{E_tilde_F} and Eq. \ref{V_tilde_F} we obtain the relationship between the variance and mean of $\tilde{F}_j(t)$ given by
\begin{eqnarray}
V[\tilde{F}_j]
&\approx& E[\frac{1}{m}] \cdot E[\tilde{F}_j]+E[\tilde{F}_j]^2 \nonumber \\
&\cdot& \{V[r_j]+ (V[r_j]+1)\cdot E[\Delta_0^2] )\}. \nonumber  \\
\label{V_tilde_FE}
\end{eqnarray}
In the case that $r_j(t)=1$ or $V[r_j]=0$, which means that the $j$-th word is steady, we can obtain the simple expression  
\begin{eqnarray}
V[\tilde{F}_j]
\approx E[\frac{1}{m}]\check{c}_j+\check{c}_j^2 \cdot E[\Delta_0^2],   
\label{V_tilde_F2}
\end{eqnarray}
\textcolor{black}{
and the corresponding variance can be written as a function of the mean as 
\begin{eqnarray}
V[\tilde{F}_j]
\approx E[\frac{1}{m}]E[\tilde{F}_j]+E[\tilde{F}_j]^2 \cdot E[\Delta_0^2],   
\label{V_tilde_FH2}
\end{eqnarray}
}
where because $V[r_j] \geq 0$, this equation also gives a lower bound on $V[F_j]$.
\par
 Fig. \ref{fig_mean_sd_t} (a) shows a comparison between 
 Eq. \ref{V_tilde_FH2} 
 and the corresponding data. 
From this figure, we can confirm that the lower bound of the data disagrees slightly with 
 Eq. \ref{V_tilde_FH2} 
 as given by the red dashed curve. 
The reason for this disagreement is that $V[r_j]$ cannot be neglected for all $j$ (by comparison $E[\Delta_0^2]$). 
In fact, the corrected lower bound with respect to $V[r_j]$ (green dashed-dotted line) is given by 
\textcolor{black}{ 
\begin{equation}
V[\tilde{F}_j]\geq E[\tilde{F}_j] \cdot E[\frac{1}{m}]+E[\tilde{F}_j]^2 \cdot \{V_{min}+ E[\Delta_0^2] \}, 
\label{cor_V_tilde_F2E}
\end{equation}
which is confirmed to be in agreement with the actual data.}

Here, we assume that $V_{min}=min_j\{V[r_j]\}$ is roughly estimated by $min_{\{j \in \check{c}_j \geq 100 \}}\{V[F_j/(\check{c}_j \cdot m )]\}$. 
Note that we limit the words by $\check{c}_j \geq 100$ in order to focus on the words that are dominant in the term of $\check{c}_j^2$ in Eq. \ref{V_tilde_F}. \par
\par
Next, we investigate the temporal FS of the time difference of a scaled time series of word appearances (see Fig. \ref{tseries}(d)), 
\textcolor{black}{
\begin{eqnarray}
\delta \tilde{F}_j(t) &\equiv& \tilde{F}_j(t)-\tilde{F}_j(t-1) \nonumber \\
&=&F_j(t)/m(t)-F_j(t-1)/m(t-1). \label{def_dtF}
\end{eqnarray}
}
By taking the difference, we can reduce the nonstationary effects $r_j(t)$ on temporal FSs. 
From Eq. \ref{delta_V_mf_a} in Appendix B, the variance is given as follows: 
\begin{eqnarray}
&&V[\delta \tilde{F}_j] \approx 2 \cdot \check{c}_j \cdot E[\frac{1}{m}] \nonumber \\ 
&+&\check{c}_j^2 \cdot \{V[\delta r_j] 
+ 2 \cdot (V[r_j]+1) \cdot  E[\Delta_0^2] \}.
\end{eqnarray}
%
\textcolor{black}{Using the facts that $V[r_j] \geq 0$ and $V[\delta r_j] \geq 0$,  we can obtain the lower bound} 
\textcolor{black}{
\begin{eqnarray}
V[\delta \tilde{F}_j] &\geq& 2 \cdot \check{c}_j \cdot E[\frac{1}{m}]+\check{c}_j^2 \cdot \{ 2 \cdot  E[\Delta_0^2] \}, \nonumber \\ 
\label{v_delta_tilda_f2}
\end{eqnarray}
}
\textcolor{black}{
and by using Eq. \ref{E_tilde_F}, we can obtain the corresponding lower bound as a function of the mean as
\begin{eqnarray}
V[\delta \tilde{F}_j]&\geq& 2 \cdot E[\tilde{F}_j] \cdot E[\frac{1}{m}]+E[\tilde{F}_j]^2 \cdot \{ 2 \cdot  E[\Delta_0^2] \},  \nonumber \\ \label{v_delta_tilda_f2E}.
\end{eqnarray}
}
The condition of scaling with a single exponent is that $E[\Delta_0^2]=0$. That is, the Poisson parameter does not fluctuate. 
In terms of the blogger model, this condition also corresponds to the case that bloggers are homogeneous.
\par
From Fig. \ref{fig_mean_sd_t} (b), we can confirm that the theoretical curve given by Eq. \ref{v_delta_tilda_f2} or \textcolor{black}{Eq. \ref{v_delta_tilda_f2E}} is in agreement with the corresponding lower bound of the empirical data. 
\subsection{Ensemble fluctuation scaling}
\begin{figure*}
\begin{minipage}[c]{0.48\hsize}
\includegraphics[width=6.3cm]{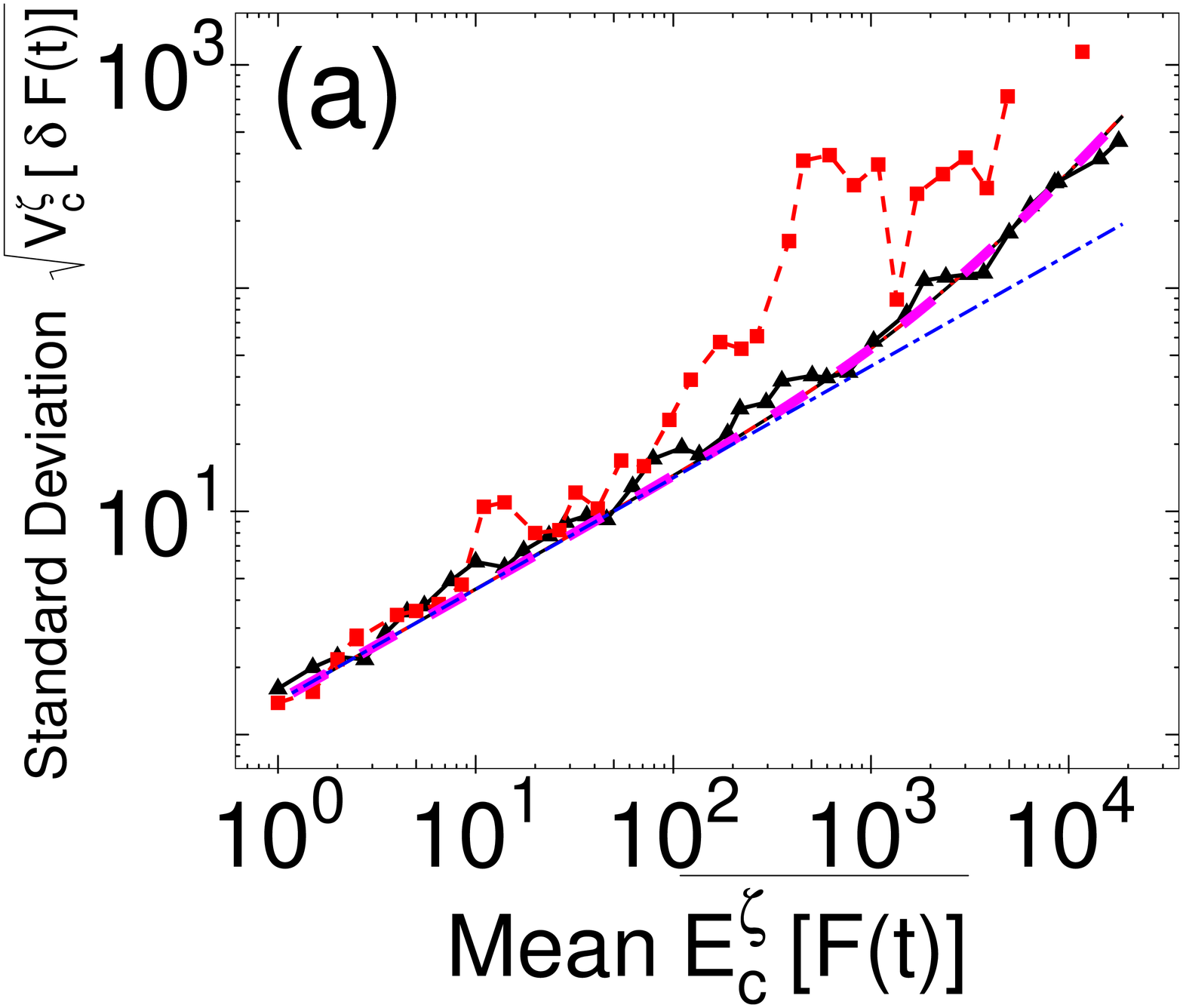}
\end{minipage}
\begin{minipage}[c]{0.48\hsize}
\includegraphics[width=6cm,angle=270]{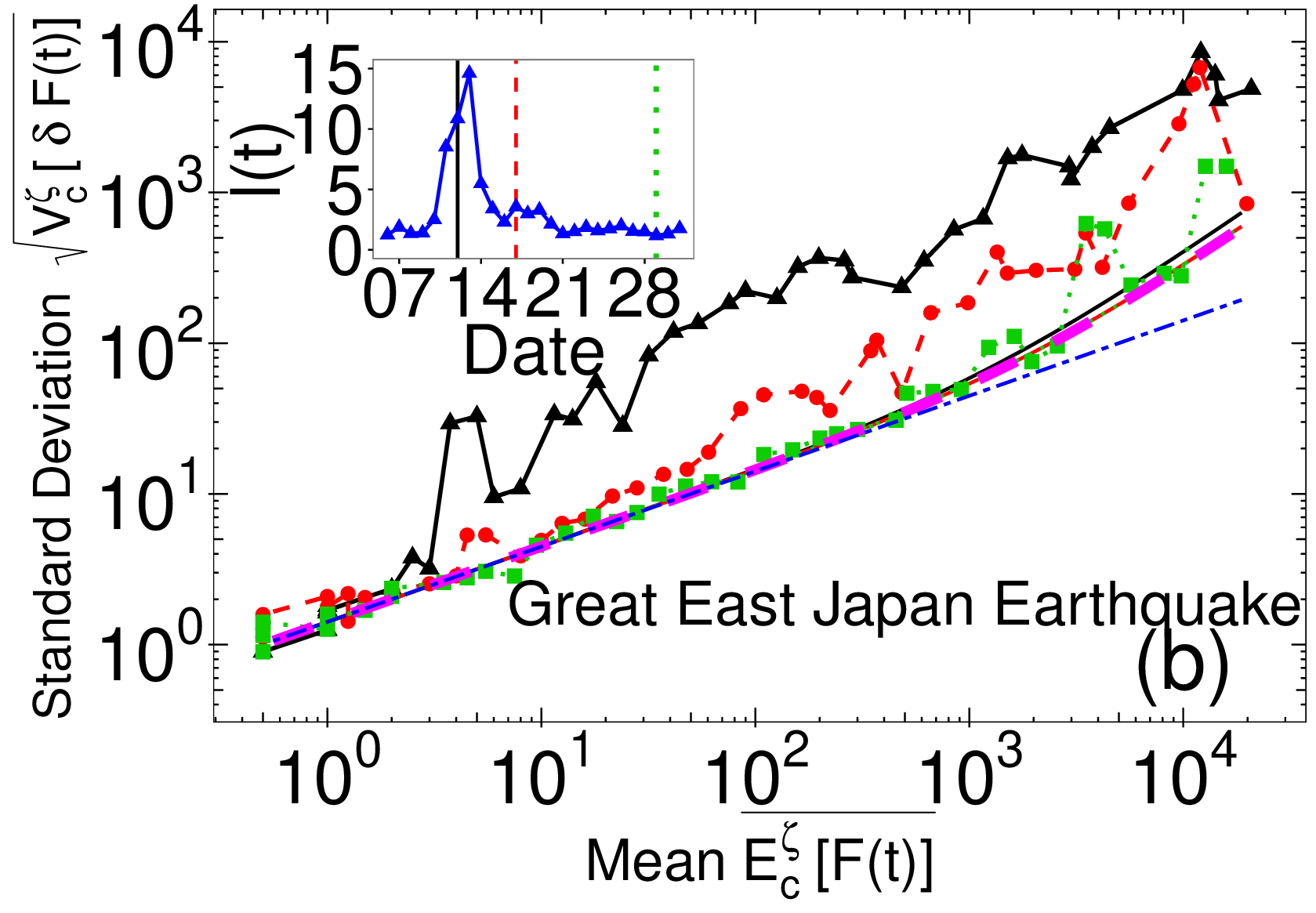}
\end{minipage}
\begin{minipage}[c]{0.49\hsize}
\includegraphics[width=6cm]{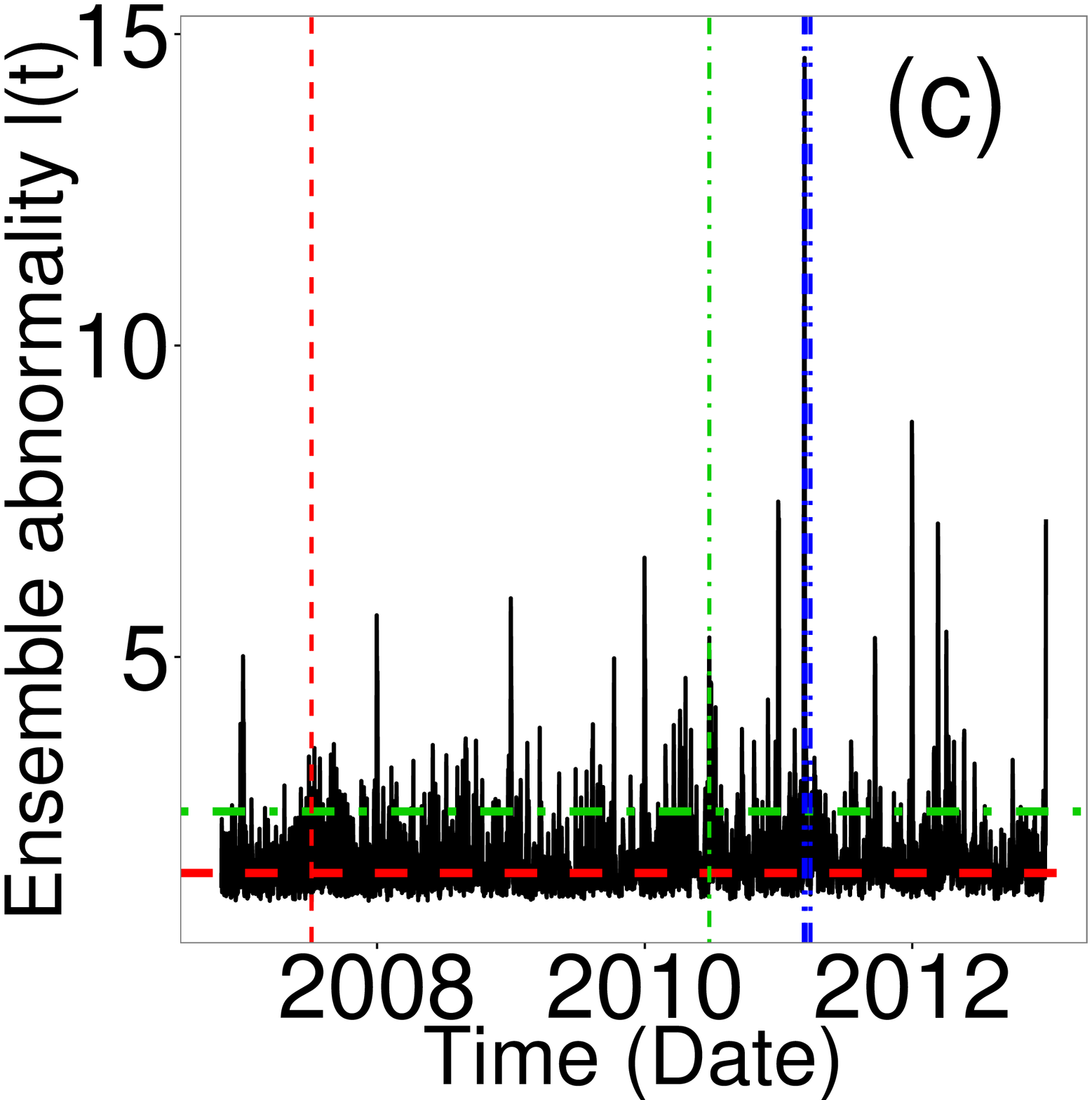}
\end{minipage}
\begin{minipage}[c]{0.49\hsize}
\includegraphics[width=6cm]{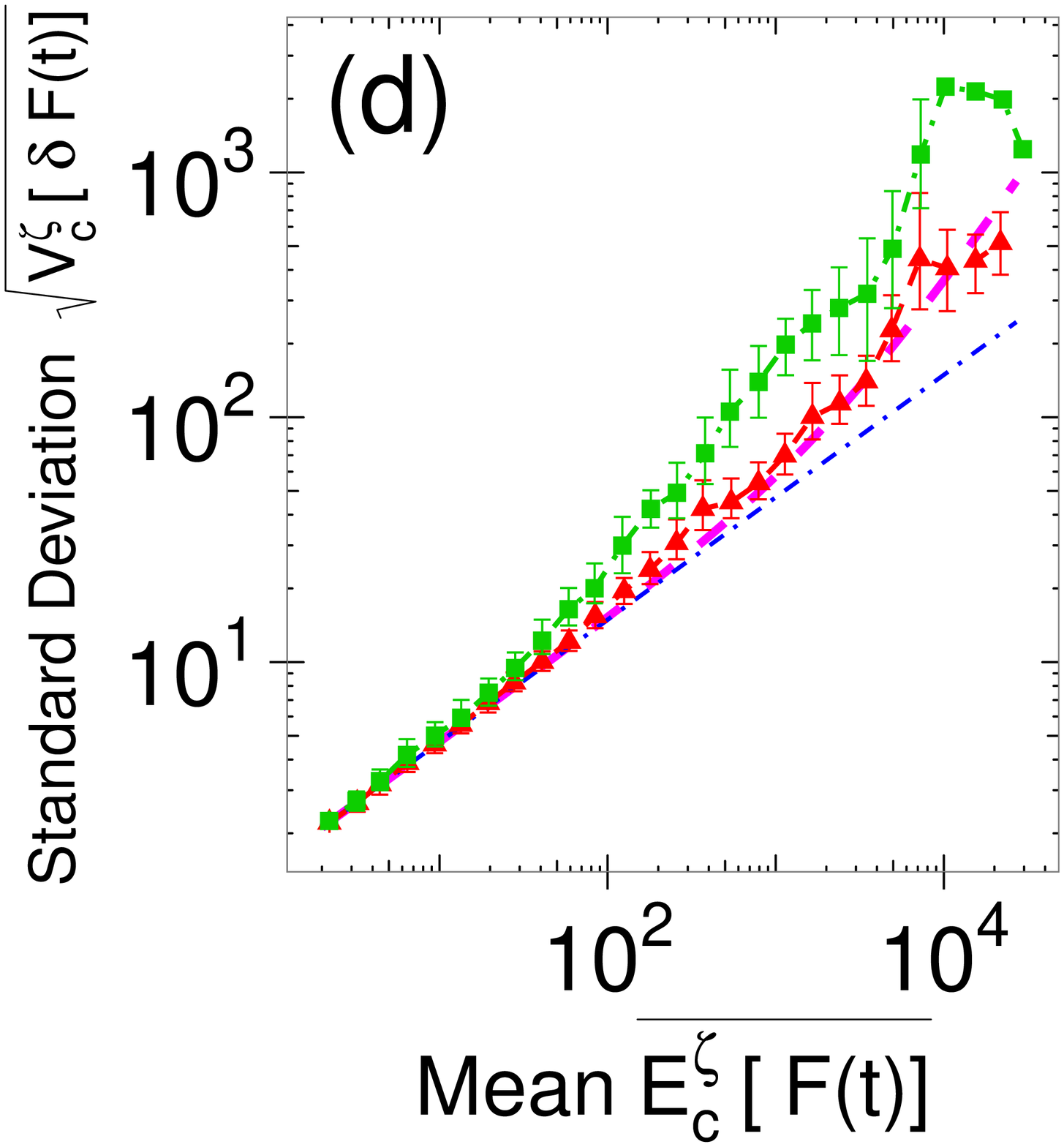}
\end{minipage}
\caption{
The box ensemble scaling of the differential of the time series of word appearances 
$E^{\zeta}_c[\delta F(t)]$, where $\delta F_j(t) \equiv F_j(t)-F_j(t-1)$ \textcolor{black}{(defined by Eq. \ref{def_dif_f})}. 
(a) The box EFSs: 06.07.2007 (black triangles) for a typical date; 26.06.2010 (red circles) in the case of a special nationwide event, the FIFA World Cup. The corresponding theoretical lower bounds given by \textcolor{black}{Eq. \ref{EFS_zeta_eq} or Eq. \ref{EFS_zeta_eqE}}: the black solid line for 06.07.2007, the red dashed line for 26.06.2010.
The thick purple dashed line indicates the theoretical curve in the case of the box size $\zeta_j=0$, as given by \textcolor{black}{Eq. \ref{EFS_c_eq} or Eq. \ref{EFS_c_eqE}} and the blue thin dash-dotted line is $y=x^{0.5}$.
(b) The corresponding figure for the great east Japan earthquake on 03.11.2011 (black triangles), 16.03.2011 (red circles), and 29.03.2011 (green circles); the inserted figure shows the deviations from theory, with $l(t)$ as defined by Eq. \ref{median_lm}, where vertical lines correspond to the dates of the main figure.   
  (c) \textcolor{black}{The time series of deviations from the theoretical lower bound $l(t)$ defined by Eq. \ref{median_lm} (i.e., the ensemble abnormality $l(t)$)}, is shown by the black line. 
  The vertical lines indicate the dates shown in panels (a) and (b): 06.07.2007 (the red dashed line), 26.06.2010 (the green dash-dotted line), 03.11.2011,  
  16.03.2011, and 29.03.2011 (the blue dash-double-dotted line). The horizontal lines are the 50th percentile of the set $\{l(t)\}$ (the red dashed line) and the 90th percentile (the green dash-dotted line).  (d) The relation between the deviation $l$ and the EFSs.  The median of the EFS regarding the dates on which $l$ is in the lower 50th percentile $y_{<50;50}(\mu)$, defined by Eq. \ref{medi_50}, where we use dates under the 50th percentile as shown by the horizontal red dashed line in panel (c).  The error bars, corresponding to the 25th percentile and the 75th percentile, $y_{<50;25}(\mu)$ and $y_{<50;75}(\mu)$.  The green lines indicate the corresponding statistics of the upper 90th percentile, $y_{>90;50}(\mu)$, for which we use the dates above the 90th percentile shown by the horizontal green dash-dotted line in the panel (c). The thick purple dashed line indicates the theoretical curve given by Eq. \ref{EFS_c_eq}, and the blue thin dash-dotted line indicates $y=x^{0.5}$.
}
\label{EFS}
\end{figure*}
Next, we consider ensemble fluctuation scaling (EFS). 
The EFS is defined by a power-law scaling between the ensemble mean and the ensemble variance at a fixed time, $V_c[A(t)] \propto E_c[A(t)]^{\alpha}$.
Here, the ensemble mean and the ensemble variance of the value of $A_j(t)$ at time $t$ are defined as follows:
\begin{equation}
E_c[A(t)]=\frac{\sum_{\{j:\check{c}_j=c \}}A_j(t)}{\sum_{\{j:\check{c}_j=c \}}1}, 
\end{equation}
\begin{equation}
V_c[A(t)]=E_c[\{A(t)-E_c[F(t)]\}^2],
\end{equation}
where we take the ensemble of words whose temporal means take the same particular value, $\check{c}_j=c$.
\par
Note that in the actual data analysis, we approximate the ensemble scalings by the following box ensemble scaling: 
\begin{equation}
E_c^{\zeta}[F(t)]=\frac{\sum_{\{j: c-\zeta \leq  \check{c}_j <  c+\zeta \}}F_j(t)}{\sum_{\{j:c-\zeta \leq  \check{c}_j <  c+\zeta \}}1}, 
\end{equation}
\begin{equation}
V_c^{\zeta}[F(t)]=E_c^{\zeta}[\{F(t)-E_c^{\zeta}[F(t)]\}^2],
\end{equation}
where the box ensemble scaling is in agreement with the ensemble scaling in the case that $\zeta=0$. 
In this paper, we assume that $\zeta=\zeta^{(0)} \cdot c$ with $\zeta^{(0)}=0.2$. \par
From Eq. \ref{EFS_zeta_a} in Appendix B, the box ensemble mean and the variance of $F_j(t)$ are given as: 
\begin{eqnarray}
E^{\zeta}_c[F(t)]
\approx c \cdot m(t) \label{en_E}
\end{eqnarray}
\begin{eqnarray}
&&V^{\zeta}_c[F(t)] \nonumber  \\
&\approx& c \cdot m(t) \nonumber \\
&+& c^2 \cdot m(t)^2 \cdot \{  V_c^{\zeta}[r(t)]+\Delta_0(t)^2 \cdot (1+V_c^{\zeta}[r(t)] \nonumber \\
&+&V_c^{\zeta}[\zeta](1+V_c^{\zeta}[r(t)]) \cdot (1+\Delta_0(t)^2) \}. \label{en_VV}
\end{eqnarray}
Here, we assume that $E_c[r(t)] \approx E_c^{\zeta_c}[r(t)] \approx 1$.
\textcolor{black}{Thus, by combining Eq. \ref{en_E} and Eq. \ref{en_VV} we obtain the relationship between the ensemble variance and ensemble mean of $\{{F}_j(t)\}$ given by 
\begin{eqnarray}
&&V^{\zeta}_c[F(t)] \nonumber  \\
&\approx& E^{\zeta}_c[F(t)] \nonumber \\
&+& E^{\zeta}_c[F(t)]^2 \cdot \{  V_c^{\zeta}[r(t)]+\Delta_0(t)^2 \cdot (1+V_c^{\zeta}[r(t)] \nonumber \\
&+&V_c^{\zeta}[\zeta](1+V_c^{\zeta}[r(t)]) \cdot (1+\Delta_0(t)^2) \}. \label{VcFE}
\end{eqnarray}
}
By using the fact that $V_c^{\zeta}[r(t)] \geq 0$, we can obtain the following theoretical lower bound over $\{r_j(t)\}$: 
\textcolor{black}{
\begin{eqnarray}
&& V^{\zeta}_c[F(t)] \geq  c \cdot m(t)   \nonumber \\
&+&   c^2  m(t)^2 \cdot \{ \Delta_0(t)^2
+V_c^{\zeta}[\check{\zeta}]\cdot(1+\Delta_0(t)^2) \},   \label{V_E_F2}  
\end{eqnarray} 
}
\textcolor{black}{and 
 as a function of the mean, 
\begin{eqnarray}
&& V^{\zeta}_c[F(t)] \geq  E^{\zeta}_c[F(t)]   \nonumber \\
&+&  E^{\zeta}_c[F(t)]^2 \cdot \{ \Delta_0(t)^2
+V_c^{\zeta}[\check{\zeta}]\cdot(1+\Delta_0(t)^2) \}. \nonumber \\
\label{V_zeta_CE}
\end{eqnarray}
}
In addition, in the case of the ensemble scaling, meaning $\zeta=0$, we can obtain the simpler relationship
\textcolor{black}{
\begin{eqnarray}
&& V_c[F(t)] \geq  c \cdot m(t)+ c^2 m(t)^2 \cdot \{ \Delta_0(t) \} 
\end{eqnarray}
}
\textcolor{black}{
and the corresponding expression written by the mean,   
\begin{eqnarray}
&& V_c[F(t)] \geq  E_c[F(t)]^2+ E_c[F(t)]^2 \cdot \{ \Delta_0(t) \}. \label{V_CFES} 
\end{eqnarray}
}
 \par
Here, $V_c^{\zeta}[r(t)]$ cannot be neglected in the observations, in the same manner as the temporal scaling $\tilde{F}(t)$.
Therefore, we also consider an ensemble scaling of the differential of appearances of a word, for comparing data.\par
\par
From Eq. \ref{EFS_zeta_delta_F} in Appendix B, the ensemble scalings of the differential of the time series of word appearances,
\textcolor{black}{
\begin{equation}
\delta F_j(t) \equiv F_j(t)-F_j(t-1), \label{def_dif_f}
\end{equation}
}
is given by 
\begin{eqnarray}
&&V^{\zeta}_c[\delta F(t)]  \nonumber \\
&\approx& 2 \cdot c \cdot (\overline{m(t)})  \nonumber \\
&+&c^2 \cdot \{V_c^{\zeta}[\check{\zeta}] \cdot V_c^{\zeta}[\delta (r_i(t) m(t))] \nonumber \\
&+&E_c^{\zeta}[\check{\zeta}]^2 \cdot V_c^{\zeta}[\delta (r(t) m(t))]+V_c^{\zeta}[\check{\zeta}] \cdot E_c^{\zeta}[\delta (r(t) m(t))]^2 \nonumber \\
&+&(1+V_c^{\zeta}[\check{\zeta}]) \cdot (V_c^{\zeta}[\delta (r(t) \cdot m(t))] \nonumber \\
&+& 2 \cdot \check{\Delta}_0^2 \cdot (\overline{m(t)^2} \nonumber \\
&+&\overline{m(t)^2}  \cdot V_c^{\zeta}[r(t)] )\},  \label{raw_dEFS}
\end{eqnarray}
where $\overline{A(t)}$ is defined by $\overline{A(t)} \equiv (A(t-1)+A(t))/2$ and we assume that $\Delta_0(t)$ is constant with $\Delta_0(t)=\check{\Delta_0}$ and $V^{\zeta}_c[r(t)] \approx V^{\zeta}_c[r(t-1)]$.  
By using the facts that $V_c^{\zeta}[\delta (r(t) m(t))] \geq 0$, $V_c^{\zeta}[r(t)] \geq 0$, and $E_c^{\zeta}[\delta (r(t) m(t))]^2 \geq \{\delta m(t)\}^2$, we can obtain the lower bound 
\textcolor{black}{
\begin{eqnarray}
&&V^{\zeta}_c[\delta F(t)] \nonumber \\
&\geq& 2 \cdot c \cdot (\overline{m(t)})  \nonumber \\
&&+c^2 \cdot \{V_c^{\zeta}[\check{\zeta}] \cdot \{\delta m(t)\}^2 +
2 \cdot \overline{m(t)^2} \cdot (1+V_c^{\zeta}[\check{\zeta}])\cdot \check{\Delta}_0^2 \},  \nonumber \\
\label{EFS_zeta_eq}
\end{eqnarray}
}
\textcolor{black}{
and the corresponding expression as a function of the mean is written as 
\begin{eqnarray}
&&V^{\zeta}_c[\delta F(t)] \nonumber \\
&\geq& 2 \overline{ E^{\zeta}_c[F(t)]}+\overline{ E^{\zeta}_c[F(t)]}^2 \nonumber \\
&\cdot& \{V_c^{\zeta}[\check{\zeta}] \cdot \{\delta m(t)\}^2/(\overline{m(t)^2} ) +
2 \cdot (1+V_c^{\zeta}[\check{\zeta}])\cdot \check{\Delta}_0^2 \}.  \nonumber \\
\label{EFS_zeta_eqE}
\end{eqnarray}
}
In addition, in the case of ensemble scaling, meaning $\zeta_j=0$ $(j=0,1,\cdots,W)$, we can obtain the relationship
\textcolor{black}{
\begin{eqnarray}
&&V_c[\delta F(t)] \geq 2 \{ c \cdot \overline{m(t)}+c^2 \overline{m(t)^2} \cdot  \check{\Delta}_0^2 \},   \label{EFS_c_eq}  
\end{eqnarray}
}
and as a function of the mean this is written as  
\begin{equation}
V_c[\delta F(t)] \geq 2 \{\overline{ E_c[F(t)]}+\overline{ E_c[F(t)]}^2 \cdot  \check{\Delta}_0^2 \}.  \label{EFS_c_eqE} 
\end{equation}
\par
%
Fig. \ref{EFS} (a) presents an example of a comparison between the theoretical lower bound given by \textcolor{black}{Eq. \ref{EFS_zeta_eq} or Eq. \ref{EFS_zeta_eqE}} and empirical data. Here, when we calculated the theoretical curve we assumed that $V^{\zeta}_c[\check{\zeta}]=1/12 \cdot (2 \cdot \zeta_0)^2$, by applying the approximation that $\check{\zeta}_j$ is distributed uniformly in the domain $[-\zeta_0,\zeta_0]$. 
From this figure, we can see that the empirical observation for $t=06.07.2007$, which is shown in the black triangles, is in accordance with the theoretical lower bound given by Eq. \ref{EFS_zeta_eq}, shown by the thin black solid line (this thin black solid line is almost completely overlapping with the thick purple dashed line). 
\textcolor{black}{This result implies that the uses of adjective words at the considered date are almost \textcolor{black}{the same as those} of the prior day, namely $|\delta r_j(t)| \approx 0$, because \textcolor{black}{it is required} for Eq. \ref{EFS_zeta_eq} to achieve equality that $V_c^{\zeta}[\delta (r(t) m(t))]=0$.}
 By contrast, the empirical observation for $t=26.06.2010$, which is denoted by the red squares, is not in accordance with the theoretical curve given by Eq. \ref{EFS_zeta_eq} or Eq. \ref{EFS_zeta_eqE}, as shown by the red dashed line (this line is also almost completely overlapping with the purple dashed line). 
This disagreement is caused by the FIFA World Cup, which constitutes a special nationwide event. 
\textcolor{black}{A special nationwide event \textcolor{black}{brings} about changes in extraordinary uses of words, including adjectives in blogs. That is, $|\delta r_j(t)|>>0$ for various words.
 Thus, it is not satisfied that $V_c[\delta(r(t) m(t))]=0$, which is a requirement for Eq. \ref{EFS_zeta_eq} to achieve equality.}
Therefore, the empirical observations differ from the theoretical lower bound.  \par
Fig. \ref{EFS} (b) presents an example of another significant nationwide event, the great east Japan earthquake (12.03.2011, 16.03.2011, and 29.03.2011).
From this figure, we can observe the separation of empirical curves from theoretical curves, and the relaxation towards the theoretical lower bound. The figure in Fig. \ref{EFS} (c), which shows the time dependency of the divergence of the empirical observations from the theoretical curves $l(t)$ (the details of this estimator of the deviation will be discussed later) clearly indicates the relaxation towards the theoretical curve directly. 
 In general, other nationwide events such as New Year, Christmas, typhoons, and solar eclipses also deviate largely from the theoretical lower bound given by Eq. \ref{EFS_zeta_eq} (see Appendix E Table \ref{abnormal_table}). \par
 Furthermore, the purple thick dash-dotted lines in Figs. \ref{EFS}(a) and (b) indicate the theoretical ensemble scaling (i.e., the box ensemble whose box size is zero, $\zeta_j=0$) given by \textcolor{black}{Eq. \ref{EFS_c_eq} or Eq. \ref{EFS_c_eqE}}, and this curve is also coincident with the actual data for dates with no special events occurring, $t=06.07.2007$ and $29.03.2011$.
This result implies that the effect of the box size, $\zeta^{(0)}=0.2$, is almost negligible in these cases.
Note that the effect of the box size becomes significant in the case that the total number of blogs varies greatly, because  $\{\delta m(t)\}^2$ in Eq. \ref{EFS_zeta_eq} is then not negligible.
\textcolor{black}{Examples of significant cases are the date of the great east Japan earthquake, on which the total number of blogs decreased by 10 percent from the previous term, and the date of a technical problem with our web crawler, on which the crawler could only collect around 30 percent of blogs compared with ordinary days. }
\par
 \textcolor{black}{We have seen typical cases of empirical EFS. Next, we investigate these in general terms.} 
To precisely measure the deviation of empirical observations from the theoretical lower bound given by Eq. \ref{EFS_zeta_eq},
we introduce the median of the logarithmic square of the deviations,
\begin{eqnarray}
&&L(t)= \nonumber \\ 
&&Median_{\{i: \check{c}_i>=300\}}[(\log(V_{\check{c}_i}^{\zeta_i}[F(t)])- \log(\check{c}_i+\gamma_{th}(t) \cdot \check{c}_i^2)^2], \nonumber \\ \label{median_lm_b}
\end{eqnarray}
and its exponential expression 
\begin{eqnarray}
l(t)=\exp(L(t)), \label{median_lm}
\end{eqnarray}
where $Median_{\{A\}}[B_i]$ is defined by the median of $B_i$ over a set $A$． 
\textcolor{black}{We call this deviation the ``ensemble abnormality $l(t)$''.}
We summarise the top 50 dates with \textcolor{black}{the large ensemble abnormality $l(t)$} in the table \ref{abnormal_table} in Appendix E,
and Fig. \ref{EFS} (c) indicates the time dependence of \textcolor{black}{this function}.
From this figure and table, we can confirm that the \textcolor{black}{ensemble abnormality} $l(t)$ (or $L(t)$) is time varying, and dates with large values (deviations) correspond to large nationwide events.
\textcolor{black}{Note that the reasons for using this quantity are as follows:
(i) We used the logarithm in order to avoid the problem that only words with a large mean $c_j$ are dominant.
(ii) We used the median to deal with outliers.} \par
In addition, we check the relation between \textcolor{black}{the ensemble abnormality $l(t)$} and the EFSs.  
The red triangles in Fig. \ref{EFS} (d) indicate the median of the EFS regarding the dates on which \textcolor{black}{$l(t)$} takes values in the lower 50th percentile. That is, roughly speaking, the statistics for ordinary dates (we calculate the EFS of the set of dates for which $l(t)$, as shown by the black line in Fig \ref{EFS}(c), is under the 50th percentile, as shown by the horizontal red dashed line), defined by
\begin{equation}
y_{<50}(\mu)=Median_{\{t| l(t) \leq q_l(0.5) \}}[s(\mu;t)], \label{medi_50}
\end{equation}
where by $s(\mu;t)$ we denote the empirical EFS at the time $t$; that is, the ensemble standard deviation as a function of the mean $\mu$ at the time $t$. Furthermore,
$q_{l}(x)$ denotes the temporal $100 \cdot x$ percentile of the set $\{\l(t)\}$. 
\textcolor{black}{Here, we also plot the error bars corresponding the 25th percentile and the 75th percentile, defined by $y_{(<50;25)}(\mu)=Percentile25_{\{t| l(t) \leq q_l(0.5) \}}[s(\mu;t)]$ 
and $y_{(<50;75)}(\mu)=Percentile75_{\{t| l(t) \leq q_l(0.5) \}}[s(\mu;t)]$ respectively, where $Percentile25_{\{A\}}[B_i]$ is defined by the 25th percentile of $B_i$ over a set $A$, and $Percentile75_{\{A\}}[B_i]$ is the corresponding value for the 75th percentile.}
From this figure, we can confirm that the empirical curve is in agreement with the theoretical lower bound shown in the purple dash-dotted line given by Eq. \ref{EFS_zeta_eq}. 
This result implies that the EFS for normal dates can be accurately described by Eq. \ref{EFS_c_eq}. 
In contrast, the green line in Fig. \ref{EFS} (d) shows the corresponding statistics for the set upper 90th percentile. That is, the statistics for dates with special events. This set consists of dates for which $l(t)$, as shown by the black line in Fig \ref{EFS} (c), is above the 90th percentile, as shown by the horizontal green dash-dotted line, defined by $y_{>90}(\mu)=Median_{\{t| l(t) \geq q_l(0.9) \}}[s(\mu;t)]$. 
 \textcolor{black}{
 From this observation, we can confirm that the empirical data for the set of dates with special events differs from the theoretical lower bound. Here, this result corresponds to the specific examples indicated by the red dashed line in Figs. \ref{EFS} (a) and \ref{EFS} (b).
 } \par

\section{Additional properties}

\begin{figure}
\begin{tabular}{cc}
\begin{minipage}[t]{0.5\hsize}
\includegraphics[width=4cm]{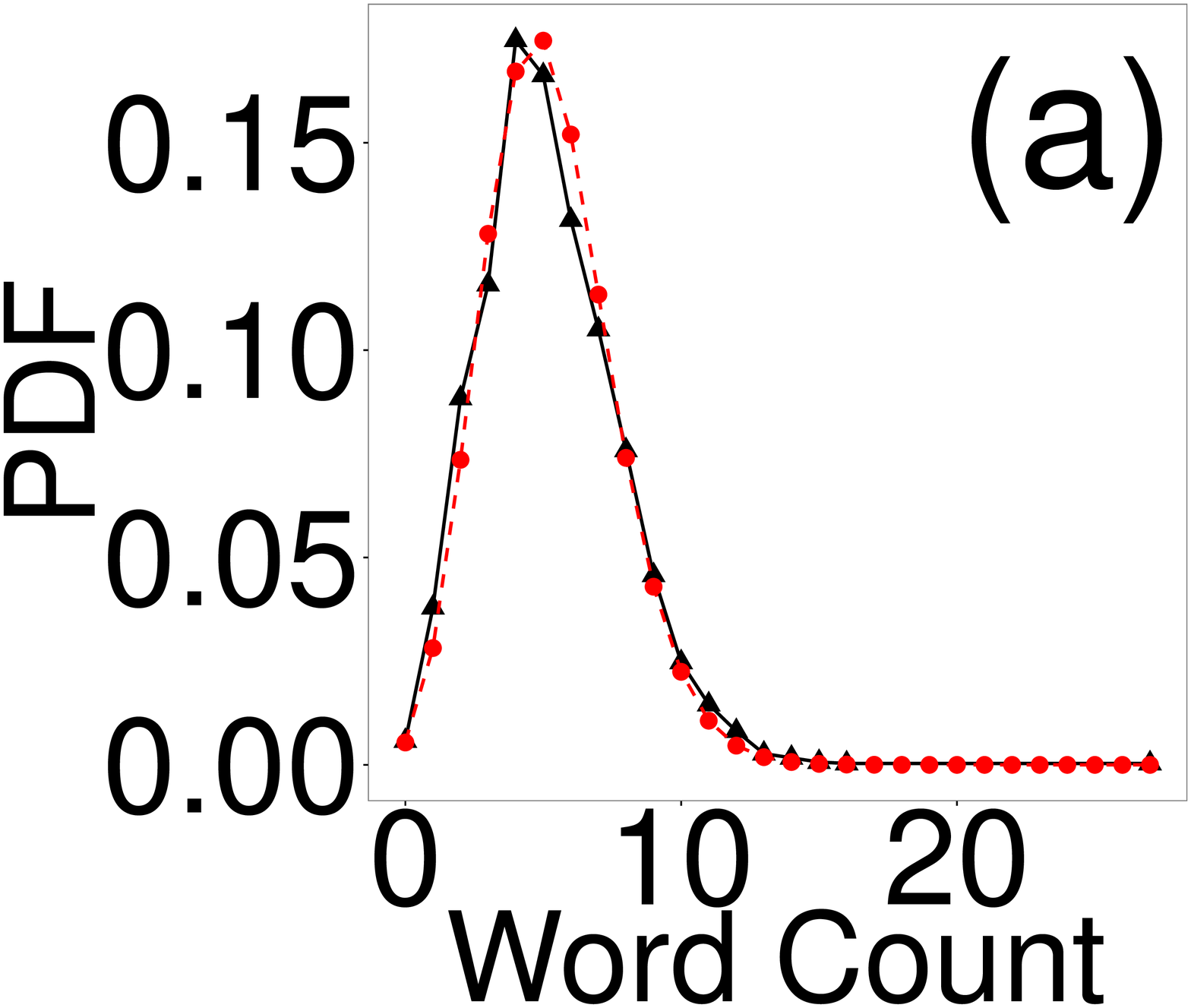}
\end{minipage}
\begin{minipage}[t]{0.5\hsize}
\includegraphics[width=4cm]{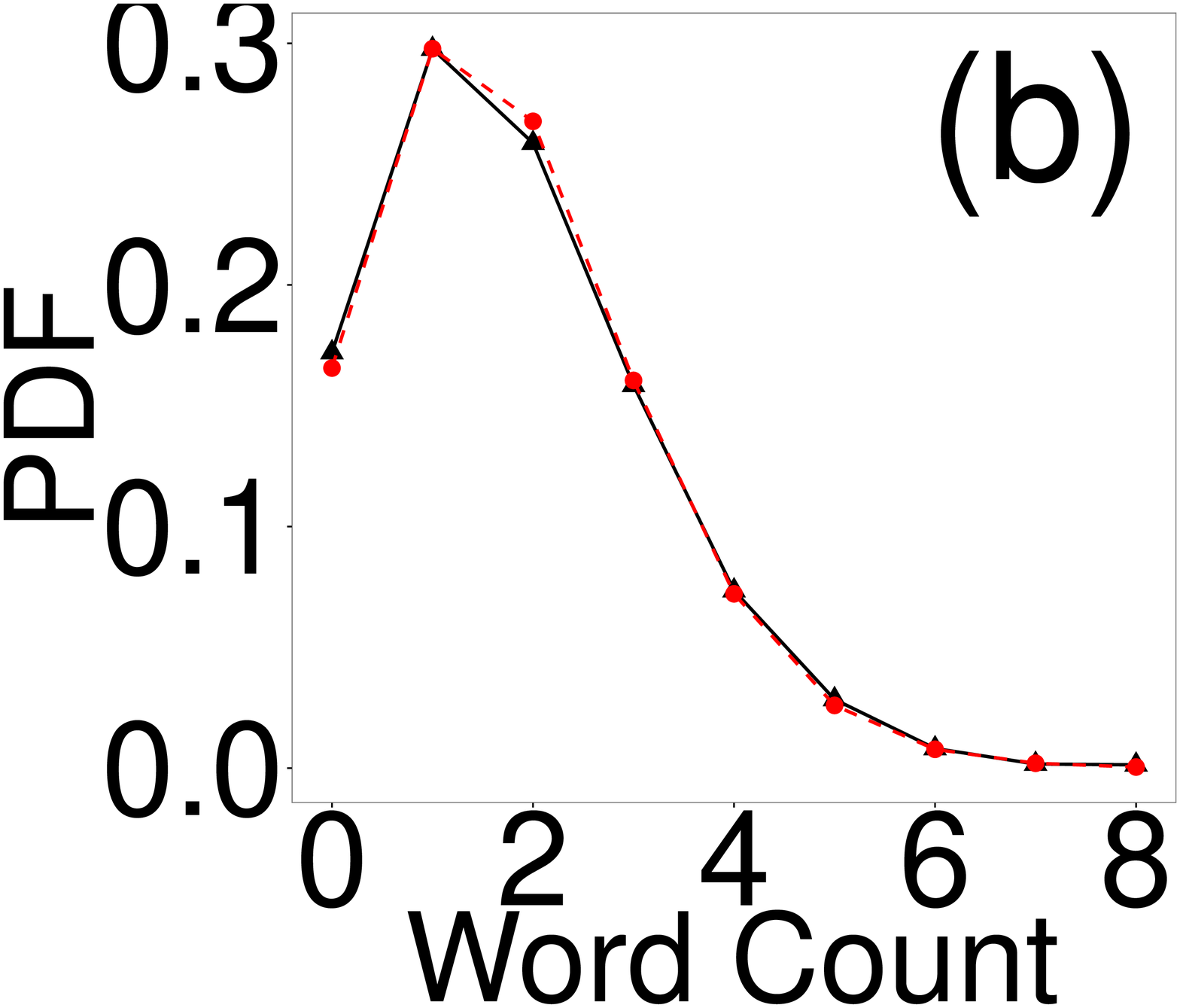}
\end{minipage}
\\
\begin{minipage}[c]{0.5\hsize}
\includegraphics[width=4cm]{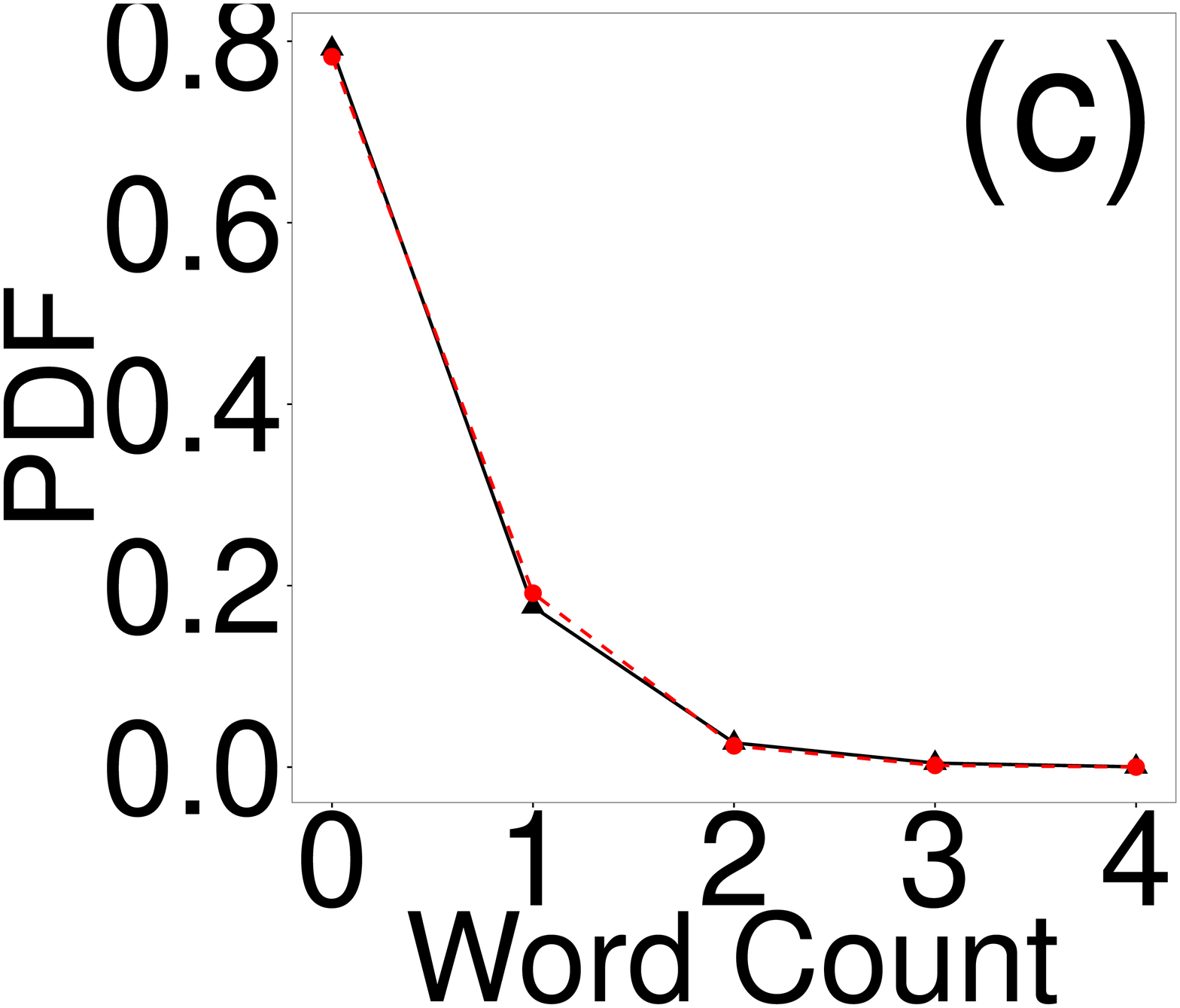}
\end{minipage}
\begin{minipage}[c]{0.5\hsize}
\includegraphics[width=4cm]{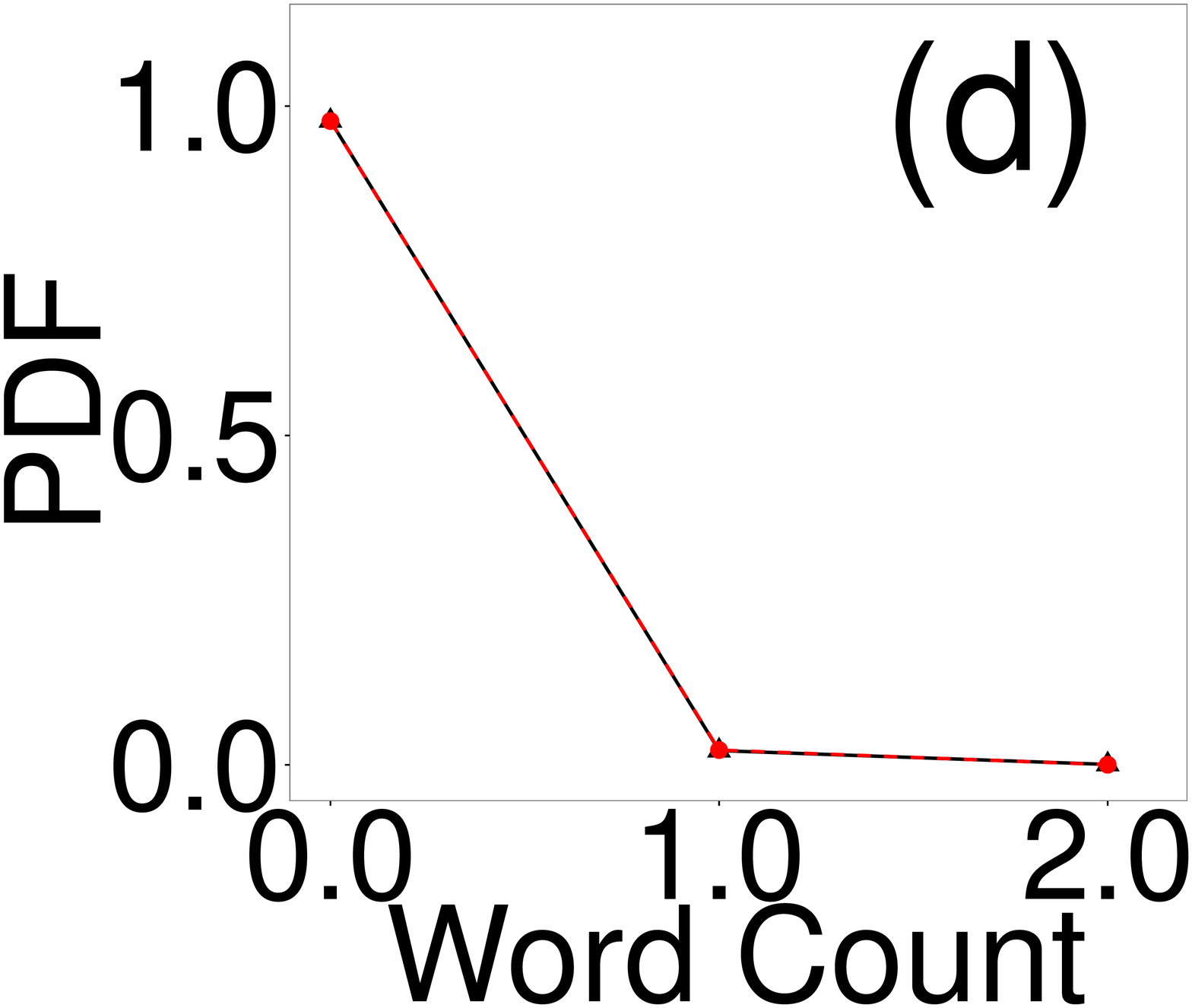}
\end{minipage}
\\
\multicolumn{1}{c}{
\centering
\begin{minipage}{1\hsize}
\includegraphics[width=6cm,angle=270]{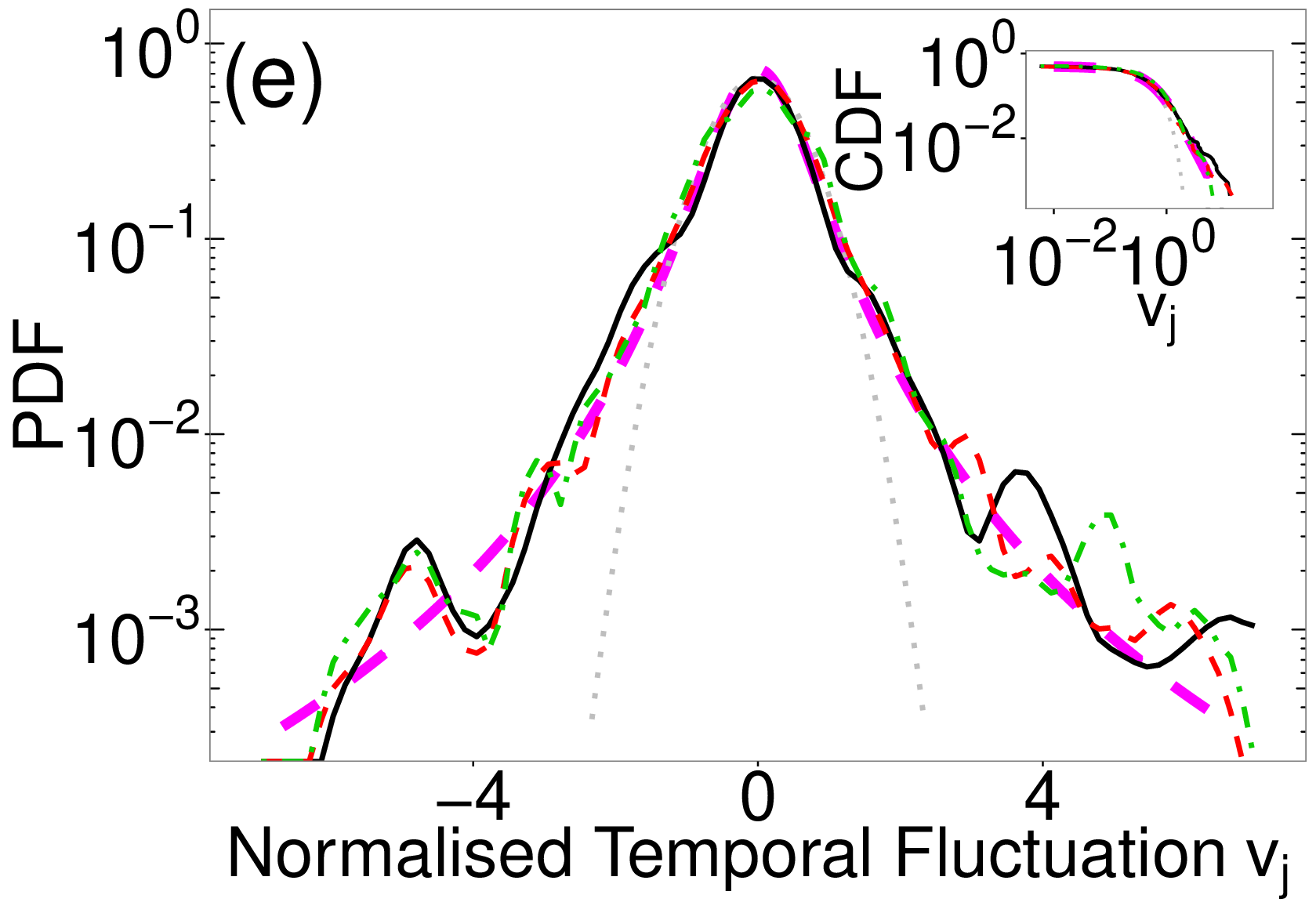}
\end{minipage}
}
\end{tabular}
\caption{
The temporal probability density function of word counts for fixed words. (a)-(d) The probability density function of word counts $F_j(t)$ for words whose mean word counts $\check{c}_j$ are small: \textcolor{black}{(a) ``jyamakusai'', which means ``bother'' ($\check{c}_j=5.67$) (b) ``komuzukashii'', which means ``difficult'' ($\check{c}_j=1.94$) (c) ``monoui'', which means ``languid'' ($\check{c}_j(t)=0.24$) 
(d) ``yuyusii'', which means ``momentous'', ($\check{c}_j=0.016$).  The black triangles indicate the empirical data and red circles indicate the Poisson distribution whose Poisson parameter corresponds to $\check{c}_j$. We can confirm that the empirical distribution obeys the Poisson distribution.}
\\
 \textcolor{black}{(e) The probability density functions of the differential of the scaled time series of word counts $v_j(t)=\delta({F_j(t)/m(t)})/\sigma_{v}(c_j(t),m(t),\Delta_0(t))$, in the case that the means of the word counts $\check{c}_j$ are very large: ``ii'', which means ``good'' ($\check{c}_j=110080$) is indicated by the black solid line; ``kurai'', which means ``dark'' ($\check{c}_j=45090$) is indicated by the red dashed line; and ``ooi'', which means ``many'' ($\check{c}_j=22535$) is indicated by the green dashed-dotted line.
 The peach coloured thick dashed line indicates the scaled t-distribution whose degree of freedom takes a value of 2.64, 
and the grey thin dotted line is the normal distribution that is in agreement with the central part of the empirical distributions, namely the normal distribution whose mean is 0 and standard deviation is 0.6. }
Here, $\sigma_{v}(c_j(t),m(t),\Delta_0(t))=\sqrt{2 \cdot \overline{c_j(t) /m(t)}+ 2 \cdot 
\overline{ c_j(t)^2  \cdot \Delta_0(t)^2}}$. The inserted figure displays the corresponding cumulative distribution functions in a log-log scale. 
 We can confirm that all distributions depict almost the same curve, as predicted by our theory, and also confirm that this common distribution can be approximated by a t-distribution. 
 }
\label{pdf_fig}
\end{figure}

\begin{figure*}
\begin{tabular}{cccc}
\begin{minipage}[c]{0.25\hsize}
\includegraphics[width=4cm]{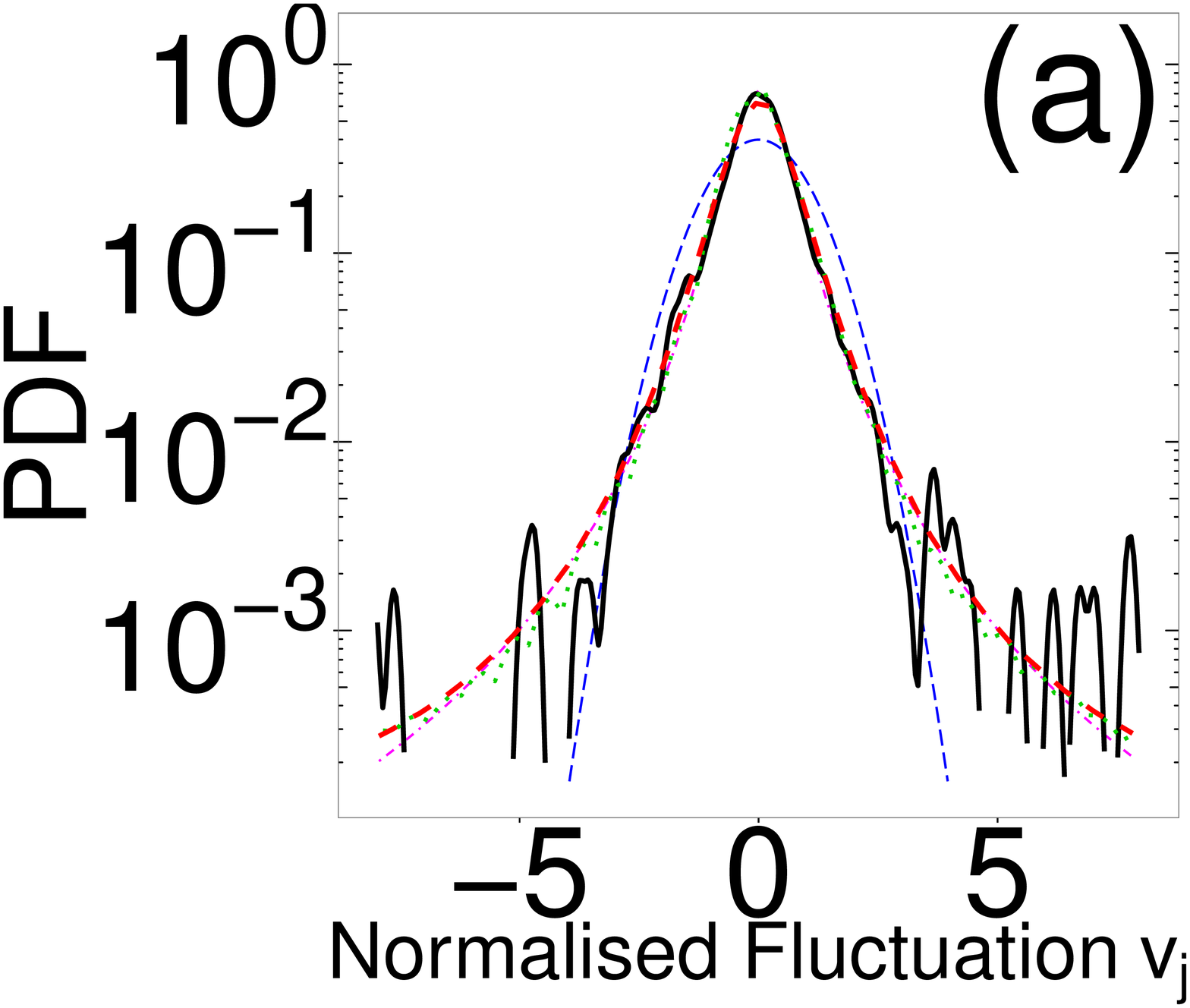}
\end{minipage}
\begin{minipage}[c]{0.25\hsize}
\includegraphics[width=4cm]{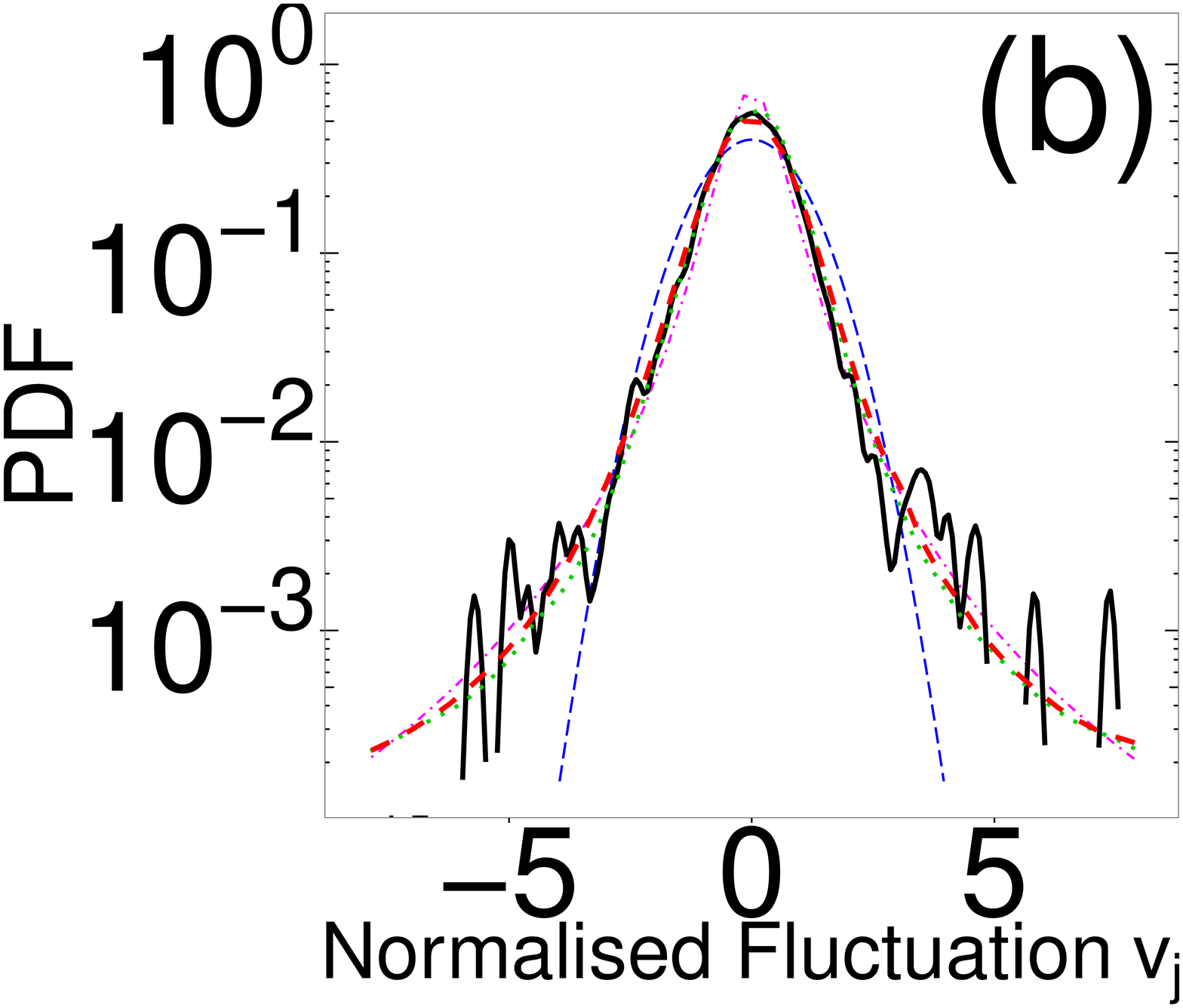}
\end{minipage}
\begin{minipage}[c]{0.25\hsize}
\centering
\includegraphics[width=4cm]{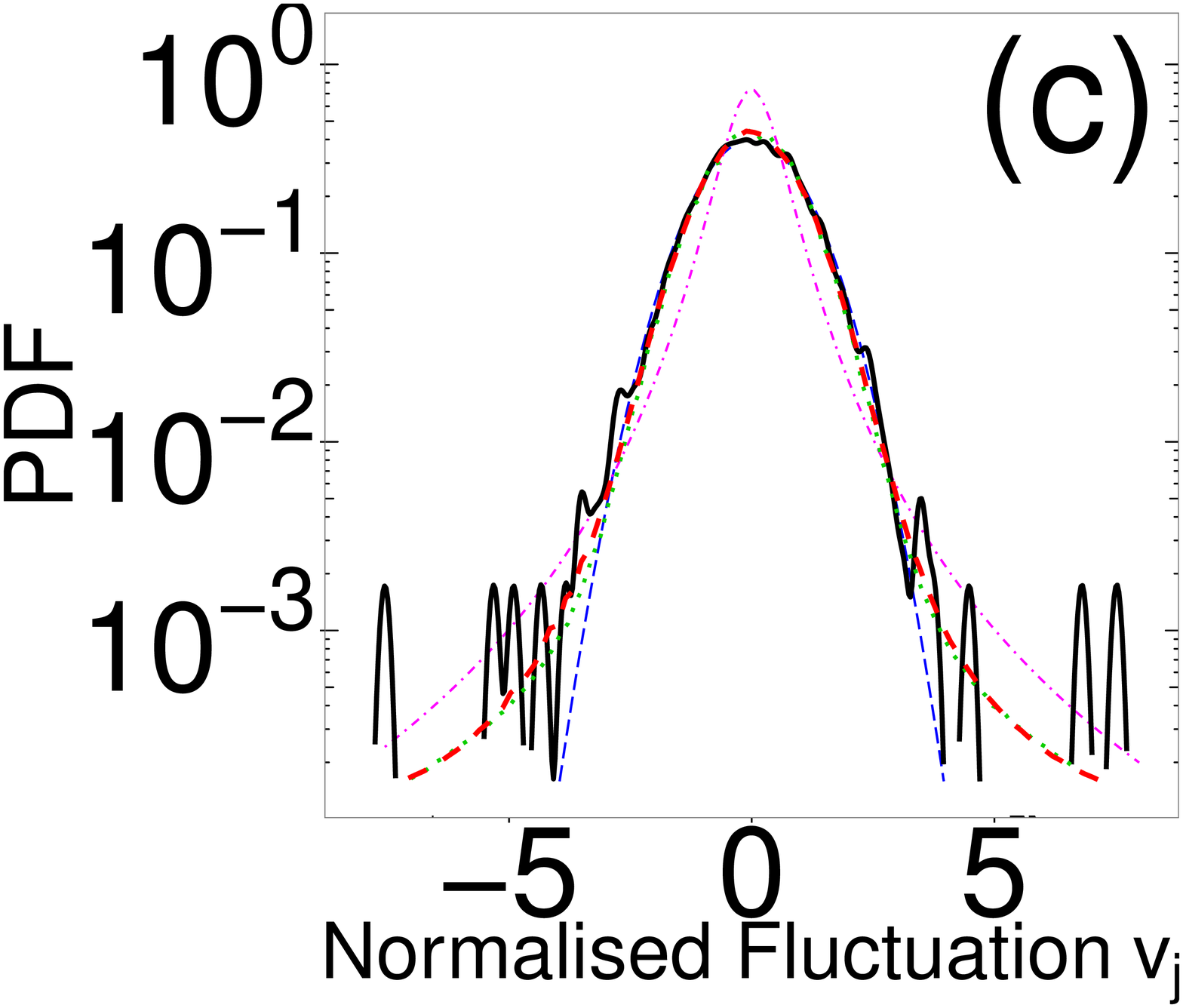}
\end{minipage}
\begin{minipage}[c]{0.25\hsize}
\includegraphics[width=4cm]{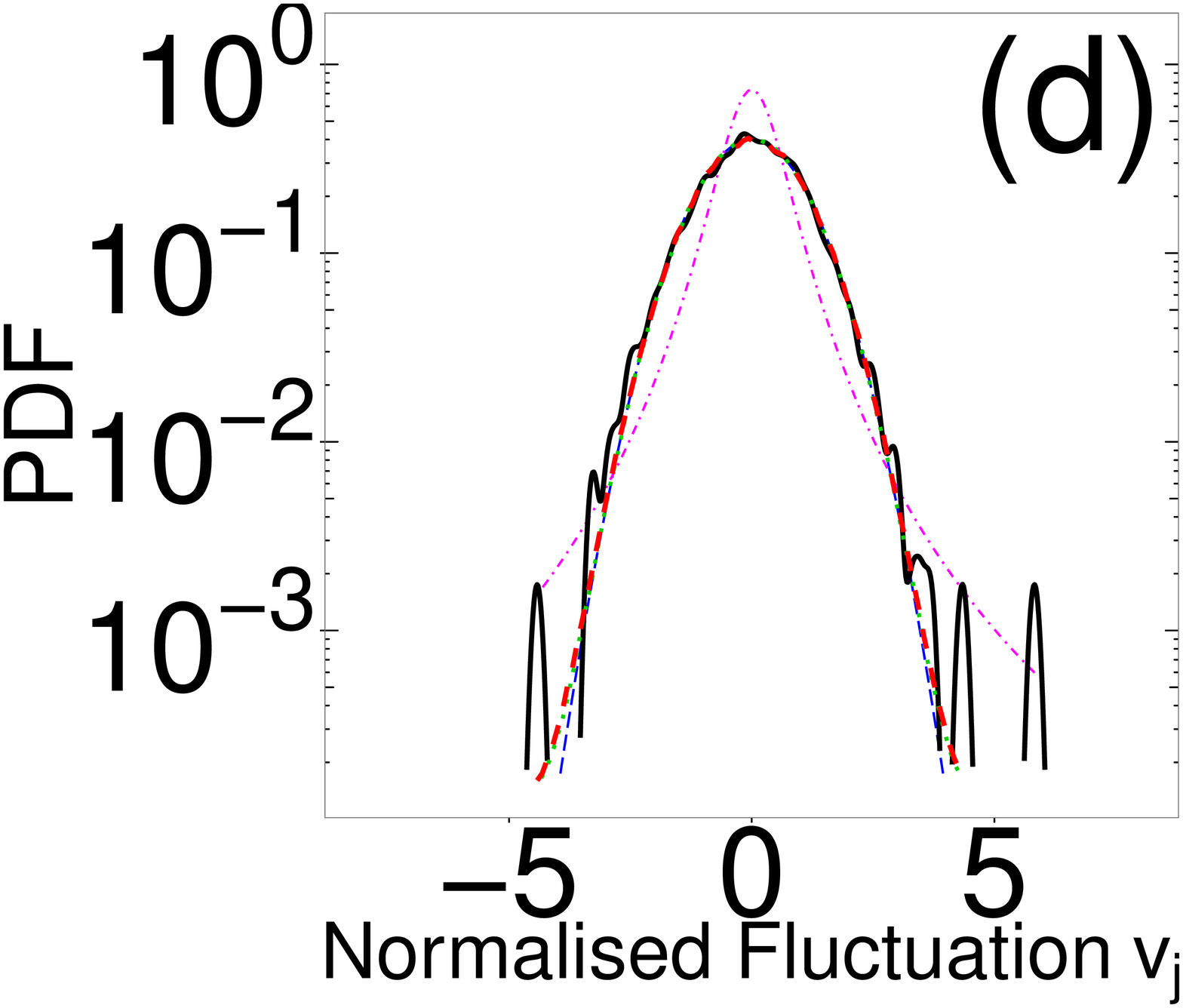}
\end{minipage}
\\
\begin{minipage}[c]{0.25\hsize}
\includegraphics[width=4cm]{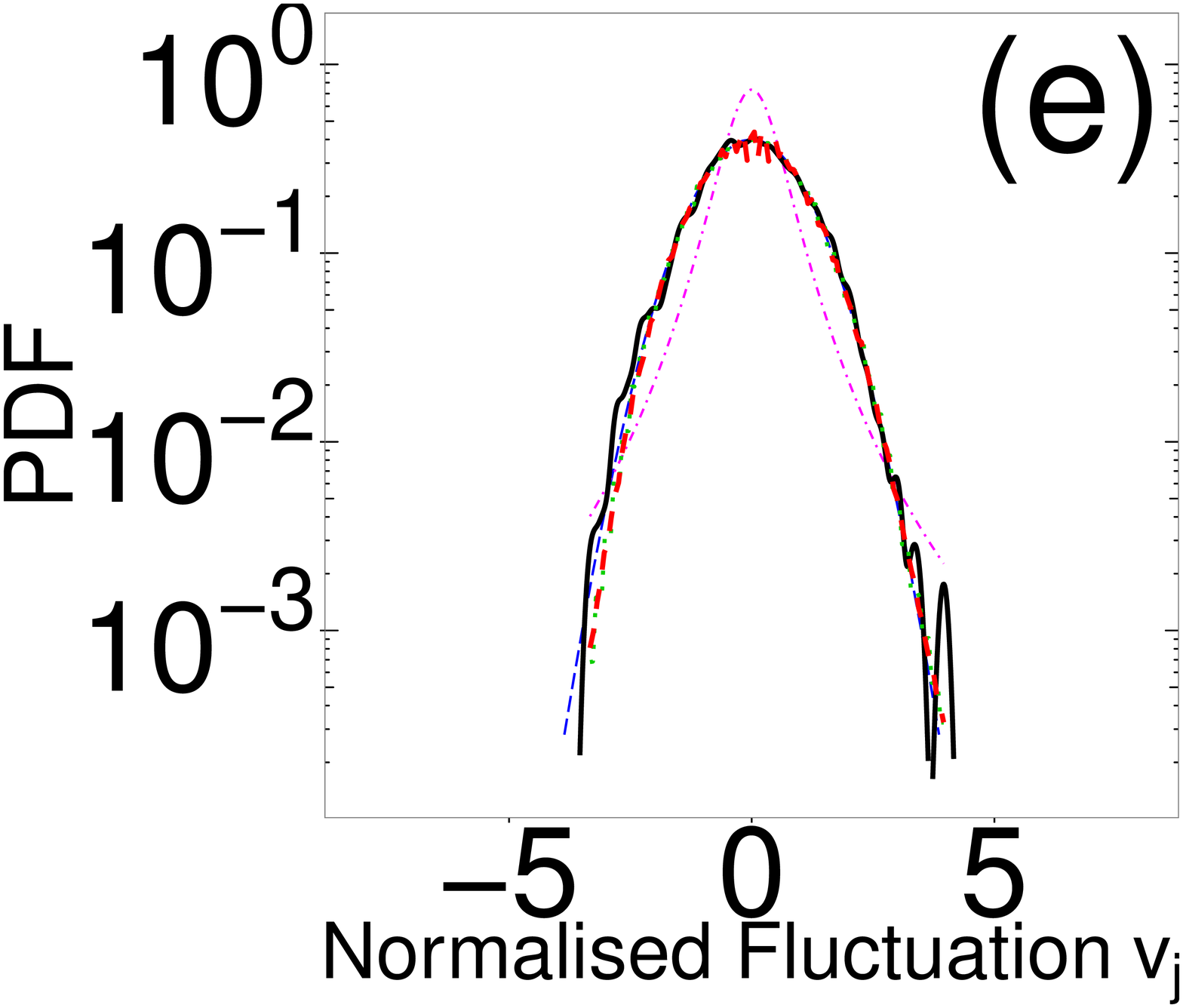}
\end{minipage}
\begin{minipage}[c]{0.25\hsize}
\includegraphics[width=4cm]{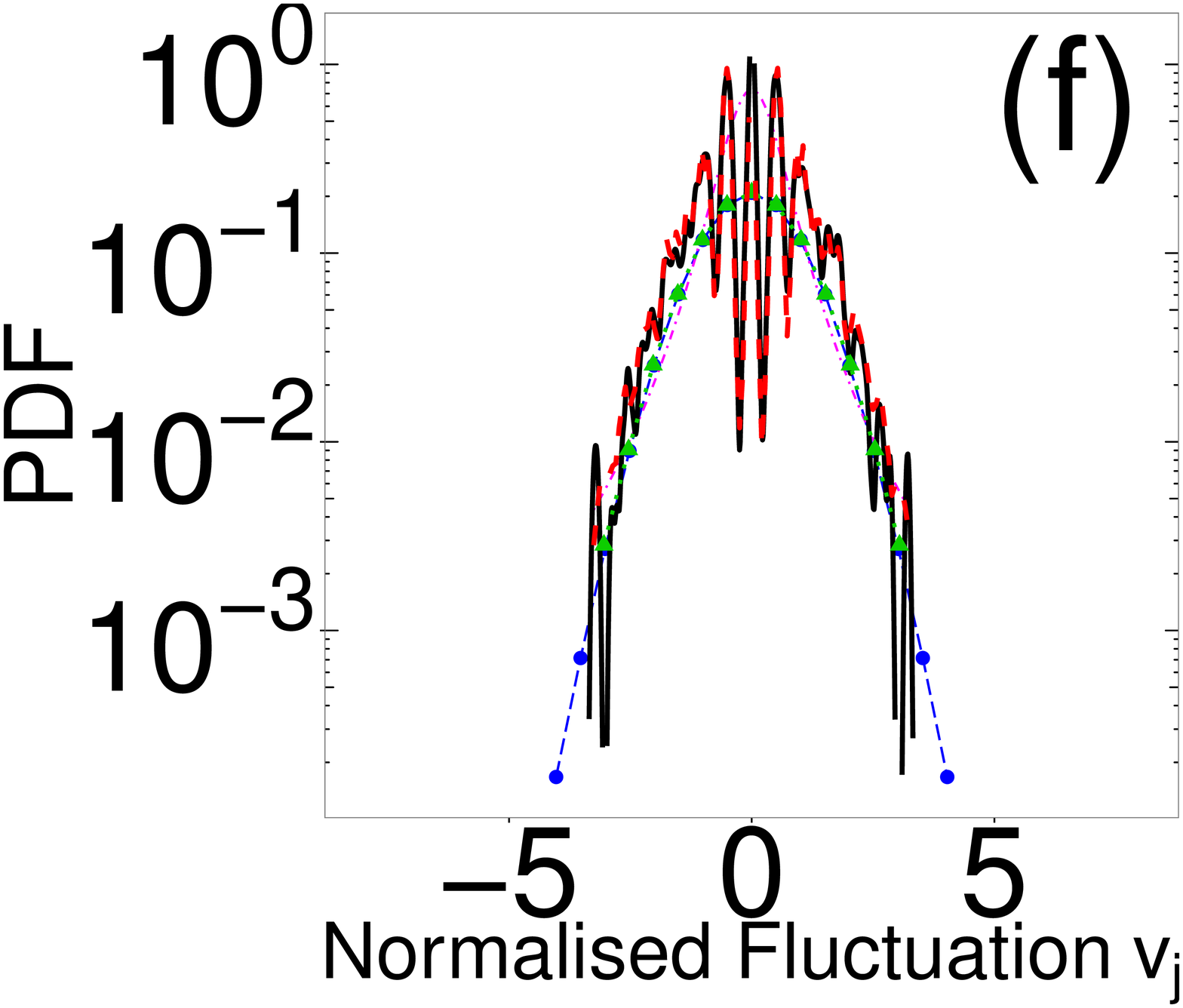}
\end{minipage}
\begin{minipage}[c]{0.25\hsize}
\includegraphics[width=4cm]{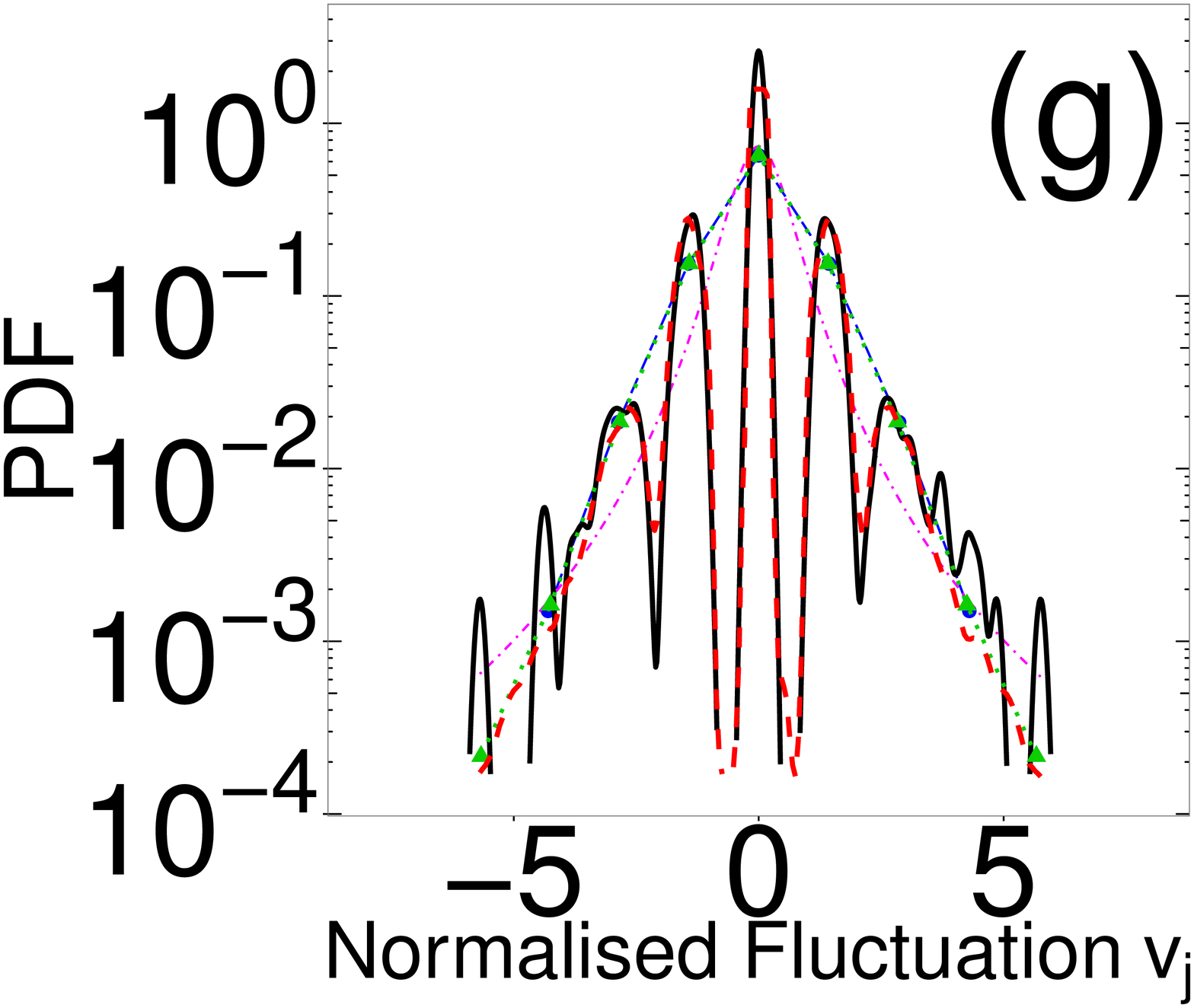}
\end{minipage}
\begin{minipage}[c]{0.25\hsize}
\includegraphics[width=4cm]{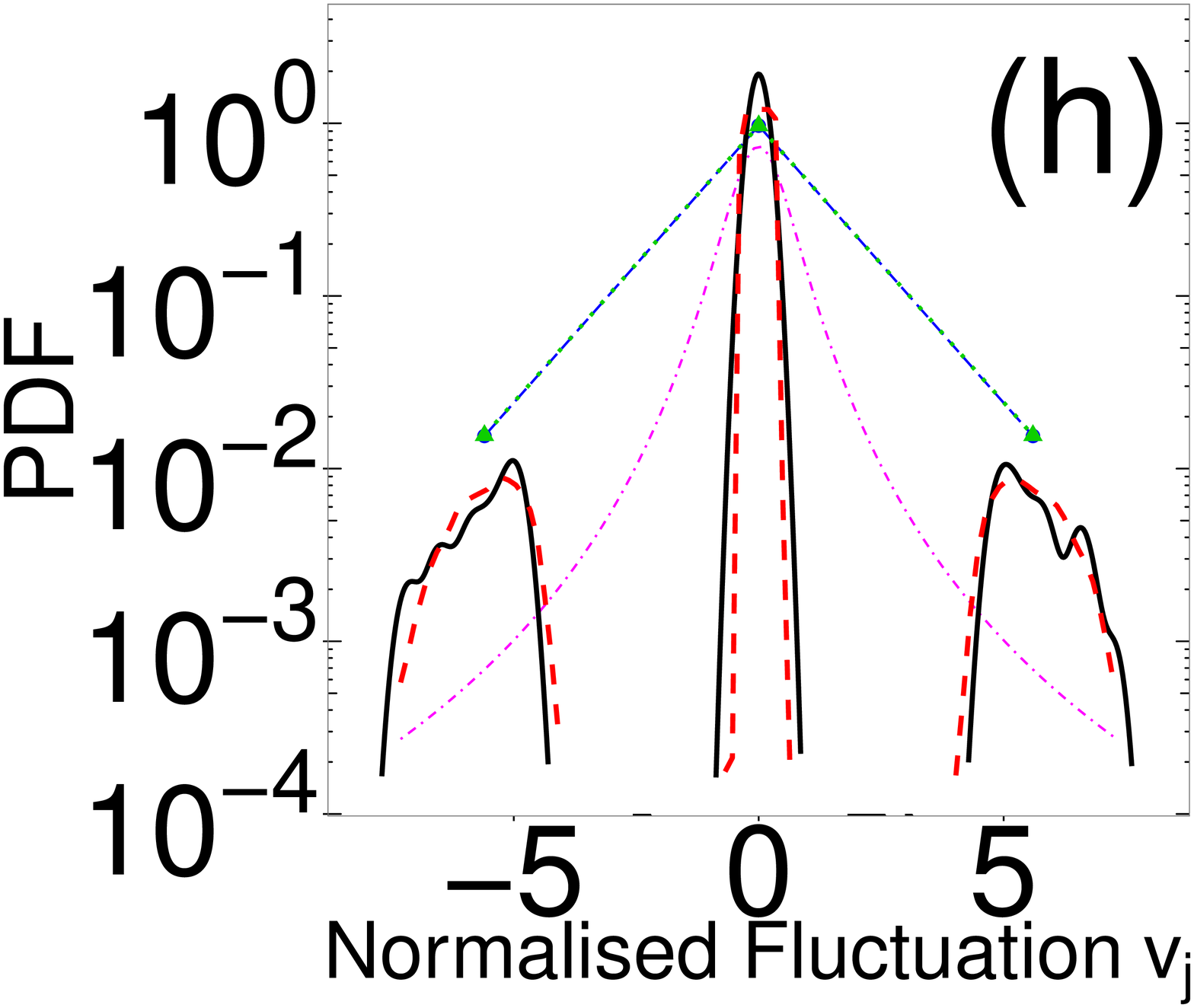}
\end{minipage}
\end{tabular}
\caption{
The probability density function of the differential of word counts scaled into the standard deviation, $v_j(t) \equiv \delta({F_j(t)/m(t)})/\sigma_{v}(c_j(t),m(t),\Delta_0(t))$, given by Eq. \ref{v_j_base}.
The red line shows the theoretical distribution given by Eq. \ref{v_pdf} for empirical parameters, the green line shows the theoretical distribution of the steady time series with the same $\check{c}_j$ as given by Eq.  \ref{v_pdf}, for $c_j(t)=\check{c}_j$ and $m(t)=1$. 
The blue line indicates the scaled differential of the Poisson distribution with the Poisson parameter $c_j$, whose standard deviation takes a value of 1; and the peach coloured 
 line indicates the differential of the t-distribution whose degree of freedom takes a value of 2.64 and has a standard deviation of 1.  
%
\textcolor{black}{The data are shown for
(a) ``ii'', which means ``good'' ($\check{c}_j=110080$),  
(b) ``sukunai'', which means ``a few'' ($\check{c}_j=7098$),
(c) ``hosoi'', which means ``slight'' ($\check{c}_j=1137$), 
(d) ``tegowai'', which means ``awkward'' ($\check{c}_j=87.00$), 
(e) ``iyarashi'', which means ``odious'' ($\check{c}_j=13.42$),  
(f) ``komuzukashii'', which means ``difficult'' ($\check{c}_j=1.97$), 
(g) ``monoui'', which means ``languid'' ($\check{c}_j=0.24$),  
    and (h) yuyusii, which means ``momentous'' ($\check{c}_j=0.016$).
}    
}
 \label{small_PDF2}
\end{figure*}

In this section, we explore some properties of fluctuations of blogs other than scalings.
First, we consider the correlation between the word appearances $F_j(t)$ and total number of blogs $m(t)$ reported in Ref. \cite{sano2010macroscopic}.
Although the steady RD model described in Ref. \cite{sano2010macroscopic,PhysRevE.87.012805} cannot depict this correlation, the extended RD model can describe it. \par
The correlation between the word appearances $F_j(t)$ and total number of blogs $m(t)$ is defined by
\begin{equation}
Cor[F_j,m] \equiv \sum_{t=1}^{T} \frac{(F_j(t)-E[F_j])\cdot (m(t)-E[m])}{ V[F_j]^{0.5} \cdot V[m]^{0.5} }.  \label{cor_def}
\end{equation}
From Eq. \ref{cor_a} in Appendix C, we can calculate the theoretical upper bound as follows:  
\textcolor{black}{
\begin{eqnarray}
&&Cor[F_j,m] \\
&\leq&   \check{c}_j \cdot \sqrt{\frac{V[m]}{\check{c}_j+\check{c}_j^2 \{(V[m]+(1+V[m]) \cdot E[\Delta_0^2]) \}}}  \nonumber \\ 
&&(T>>1), 
\label{cor1} 
\end{eqnarray}
}
where we have assumed that there is no correlation between $r_j(t)$ and $m(t)$.

From Fig. \ref{fig_mean_sd} (b), we can confirm that the observation (the black triangle) almost agrees with the theoretical curves given by Eq. \ref{cor1} (red dashed line).  \textcolor{black}{In this plot, we replace $\check{c_j}$ in Eq. \ref{cor1} with $E[F_j]$, by using Eq \ref{E_F}.} A detailed discussion of the correlation is provided in Appendix C. \par
In the above discussions, we have discussed the only summary statistics, such as means, variances, and correlation.   
Here, we investigate the probability distribution function of the fluctuation of blogs directly, where a detailed discussions including derivations is given in Appendix D.
From Fig. \ref{pdf_fig} (a)-(d), we can confirm that the distribution of word appearances $F_j(t)$ of the actual data for very small $\check{c}_j$ obeys a Poisson distribution, which is predicted by the theory.
By contrast, for very large $\check{c}_j$ we theoretically deduced that the distribution of the differential of word appearances scaled by the standard deviation $v_j(t)$ obeys the common distribution \textcolor{black}{when
we defined $v_j(t)$ as
\begin{equation}
v_j(t) \equiv \delta(F_j(t)/m(t))/\sigma_{v}(c_j(t),m(t),\Delta_0(t)) \label{v_j_base}, 
\end{equation}
\begin{eqnarray}
&&\sigma_{v}(c_j(t),m(t),\Delta_0(t))  \nonumber \\
&=&\sqrt{2 \cdot \overline{c_j(t)/m(t)}+ 2 \cdot \overline{ c_j(t)^2  \cdot \Delta_0(t)^2}}, \nonumber \\
\end{eqnarray}
and assumed that $\delta c_j(t) \approx 0$. 
}
From Fig. \ref{pdf_fig} (e), we can confirm that the probability density functions of $v_j(t)$ for real data obey the common distribution, which cannot be approximated by the normal distribution \textcolor{black}{(the thin grey dotted line in the figure)}, but can be by the Student's t-distribution scaled by the standard deviation (the peach dashed line in the figure). 
Here, the degree of freedom of the t-distribution takes a value of around 2.6. 
\textcolor{black}{We can also verify this result for the distributions quantitatively, using the Kolmogorov-Smirnov test.
The p-values of the Kolmogorov-Smirnov test are 0.00012 for the normal distribution and 0.77 for the t-distribution in case of \textcolor{black}{``ii'' (which is a Japanese colloquial word meaning ``good'')}, which is indicated by the black solid line in the figure and represents a typical example of a word with very large $\check{c}_j$. Thus, the hypothesis of the t-distribution is not rejected, but that of the normal distribution is rejected at a significance level of one percent. }  \par
\textcolor{black}{By assuming that $v_j(t)$ obeys the above-mentioned Student t-distribution for $\check{c}_j  \to \infty $, finite sized distributions for any $\check{c}_j$ can be also explained by the extended RD model. Fig. \ref{small_PDF2} demonstrates that the model can reproduce the empirical \textcolor{black}{distribution of $v_j$} for the word relating to any $\check{c}_j$ over eight orders of magnitude.  The details of these graphs are explained in Appendix G.} \par
\textcolor{black}{
Note that, even assuming that $\Delta_0(t)$ as defined by Eq. \ref{Deltam} is constant with $\Delta_0(t)=\check{\Delta}_0$, the model can reproduce all of the considered empirical observations described above. 
These results imply that the time variation of the number of blog articles is not primarily caused by the time variation of the ratio of active bloggers $M(t)/N(t)$, but rather that of the total number of bloggers $N(t)$, from the viewpoint of the random blogger model (the micro model). 
This is because we can deduce from the necessary condition that $\Delta_0(t)$, as given by Eq. \ref{Delta} is time-invariant, that $\frac{M(t)-1}{N(t)-1} \approx \frac{M(t)}{N(t)}$ is also time-invariant.  }

 \par
\section{Applications}

\begin{figure}
\includegraphics[width=9cm]{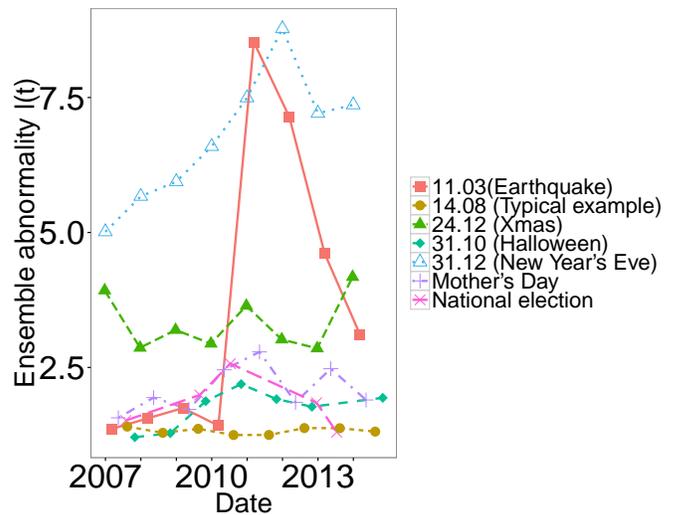}
\caption{
 Quantification of impacts of nationwide events as measured by \textcolor{black}{the ensemble abnormality $l(t)$ given by Eq \ref{median_lm}, which is quantified by the deviation of empirical ensemble scalings from the theoretical lower bound using 1771 adjectives.}
  The points indicate comparisons of the same event (or date) between years for 
the great east Japan earthquake on 11.3 (brown filled circle); the date of the lowest mean of deviation on 14.08 (grass green filled triangle); Halloween on 31.10 (emerald green filled diamond); Christmas Eve on 24.12 (aqua unfilled triangle triangle); New Year's Eve on 31.12 (blue plus); Mother's Day on 13.05.2007, 11.05.2008, 10.05.2009, 09.05.2010, 08.05.2014, 12.05.2013, and 11.05.2014 (purple cross); and the dates of National elections on 29.07.2007, 30.08.2009, 11.07.2010, 16.11.2012, and 21.07.2013 (peach coloured unfilled diamond).
From this figure, we can compare the impacts of various nationwide events by the same standard.  
}
\label{appli}
\end{figure}

\begin{figure}
\includegraphics[width=6cm, angle=270]{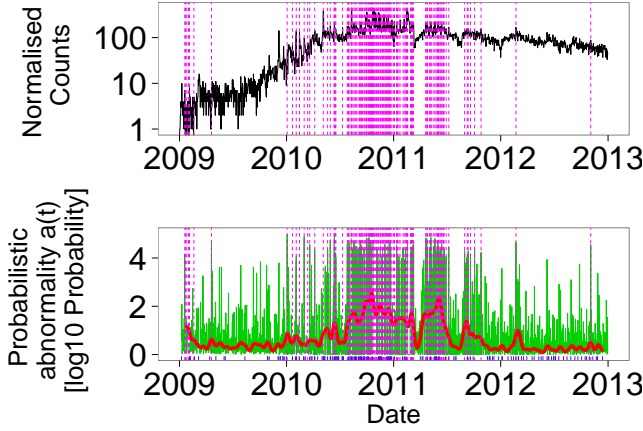}
\caption{Temporal dependency of \textcolor{black}{the probabilistic} abnormality of the word "Lekue", which mainly refers to a steamer in a microwave oven (in Japan), and was a hit product in 2010 in Japan. The top figure indicates the normalised word count of "Lekue" $\check{F}_j(t)=F_j(t)/m(t)$, and the bottom figure indicates the corresponding \textcolor{black}{probabilistic abnormality} $a_j(t)$ given by Eq. \ref{ab_index} (the green thin line). The red thick line is the abnormality smoothed by the moving weighted average $\bar{a}_j(t)=\sum_{T=-15}^{15} a_j(t) \cdot \phi_n(t-T;mean=0,sd=6)$, where $\phi_n(x)$ is the probability density function of the normal distribution. The peach dashed vertical lines indicate dates with large deviations, on which $a_j(t) \geq 4$, and the bottom blue vertical solid lines show the dates on which "Lekue" was broadcasted on television in the Tokyo area. 
From this figure, we can confirm that the \textcolor{black}{probabilistic abnormality} was high from the middle of 2010 until the middle of 2011 (the red thick line), and in the same period there were many television broadcasts regarding "Lekue", which are one of cause of abnormality.}
\label{application2}
\end{figure}

We propose some practical applications of FSs and the extended RD model. 
\subsection{Quantification of the abnormalities of dates or special nationwide events-}
First, we apply the EFS of blogs to quantify the abnormality of special nationwide events. \textcolor{black}{We quantify the abnormalities of nationwide events by using the ensemble abnormality $l(t)$ given by Eq. \ref{median_lm}, which represents the abnormal usages of 1771 basic adjectives on almost all Japanese blogs.} 
\textcolor{black}{The reason why \textcolor{black}{the ensemble abnormality $l(t)$} can be applied to measure special nationwide events that influence unusual usages of words is described in Section 4.2 (typical examples are the FIFA Word Cup and the great east Japan earthquake, described in the sixth and seventh paragraphs in that section, respectively).}
Fig. \ref{appli} shows a comparison of \textcolor{black}{the ensemble abnormality $l(t)$} of special events such as Christmas, Halloween, the national elections, and the great east Japan earthquake. From this figure, we can obtain information regarding the impacts of nationwide events.
For example, 
March 11 in 2011 (brown circle) represents a special day with a \textcolor{black}{high ensemble abnormality}, on which the great east Japan earthquake occurred, and 
the \textcolor{black}{abnormality} subsequently decreases year by year. 
In addition, Halloween (October 31), as indicated by green filled diamonds, has gradually become a more special day in Japan since 2010.
In another example, the \textcolor{black}{ensemble abnormality} of national elections (peach coloured square) rises towards a peak at the change of regime, and then decreases.  
We can also see that \textcolor{black}{the ensemble abnormality} of the first regime change in 15 years is in almost the same range as that of mother's day (the voting rate of this election, at 69.28 percent, is the highest of the last 19 years).
In contrast, \textcolor{black}{the ensemble abnormality} of the national election in 2013 is at the level of an ordinary day (the voting rate was 52.61 percent). That is, almost no abnormality is observed. \par
 We summarise the top 50 dates with largest deviations in Table \ref{abnormal_table} in Appendix E.
From this table, we can confirm that events with the highest \textcolor{black}{ensemble abnormalities} are New Year, earthquakes, terrible storms, the world cup, sudden changes in temperatures, solar eclipses, etc.
%
\textcolor{black}{Note that there is a possibility that our index incurs a bias, caused by using only adjectives.  Thus, to improve the reliability we should clarify the dependence of the index on different components of speech in the future.}
%
%
\subsection{Quantification of temporal variations of abnormalities in considered words}
\textcolor{black}{In Section 5, we calculated the distribution of word counts of an individual word.
\textcolor{black}{Thus}, we attempt to apply this result to a practical problem.} 
\textcolor{black}{In particular, we propose a measurement of the temporal variation of the abnormality of a considered word in a probabilistic manner.}
Here, we regard ``abnormal'' as meaning abrupt changes in the number of word appearances caused by a special event, such as when
television broadcasts feature news that relates to the considered word.
Thus, the measure that we will introduce in this section may be applicable for measuring the impacts of news or events. \par
We apply the theoretical probability method in the extended RD model to introduce a measure of abnormality.
In particular, we use the cumulative probability of $\delta F_j(t)$ on the condition that the parameter of the $j$-th word is same as on the prior day; namely, $c_j(t)=c_j(t-1)$.
This cumulative probability, meaning that occurrences are greater than $\delta F_j(t)$, is given by 
\begin{eqnarray}
&&A_j(t)=  \\ \nonumber 
&&P_{>\delta F}(\delta F_j(t);\bar{c}_j(t-1),\bar{c}_j(t-1)),
\end{eqnarray}
where $P_{v}(x;c,c',\Delta)$ is the cumulative distribution of $\delta F_j(t)$ with parameters $c_j(t)=c$ and $c_j(t-1)=c'$ defined by Eq. \ref{pdf_delta_f}, and $\bar{c}_j(t-1)$ is estimated by the moving average over 14 days, 
\textcolor{black}{
\begin{equation}
\bar{c}_j'(t-1)=\frac{1}{14}  \sum^{14}_{q=0}F_j(t-1-q)/m(t-1-q). \label{moving_average}
\end{equation}
}
Considering the sign and the scale, we can define the abnormality as follows:
\begin{eqnarray}
a_j(t) =
\begin{cases}
 -\log10(\frac{A_j(t)}{A_{0}})  & \text{($F_j(t)-F_j(t-1) \geq 0$)} \\
 -\log10(\frac{1- A_j(t)}{1-A_{0}})& \text{($F_j(t)-F_j(t-1) < 0$)}, 
 \label{ab_index} 
\end{cases}
\end{eqnarray}
where $A_{0}=P_{>\delta F}(0;\bar{c}_j(t-1),\bar{c}_j(t-1)) \approx 1/2$.
\textcolor{black}{Here, we refer to this value as the ``probabilistic abnormality $a_j(t)$'', in order to distinguish it from the ``ensemble abnormality $l(t)$''.} 
The theoretical minimum of the value of $a_j(t)$ is zero. \par
\textcolor{black}{Note that the characteristic features of the probabilistic abnormality $a_j(t)$ measure are as follows: \\
(i) The introduced measure describes the abnormality using quantities connected to the probability. Thus, we can compare the abnormality of considered words and dates with various values of $c_j$ by the same standard. By contrast, the commonly used simple abnormality measure, given by the growth rate of word counts on adjacent days $r_j(t)=F_j(t)/F_j(t-1)$, is not suitable.  For example, the probability of the extended RD model taking $r(t)=2$ in the case of the empirical parameters given in Appendix I is 0.33 for $c(t)=c_j(t-1)=1$, $m(t)=m(t-1)=1$, and $5 \cdot 10^{-5}$ for $c_j(t)=c_j(t-1)=100$. \\
(ii) Our measure \textcolor{black}{takes} a continuous value. Thus, we can also evaluate weak abnormalities, such as minor but continuous news. In contrast, methods that employ binary values (``abnormal'' or ``not abnormal'') based on thresholds cannot evaluate weak abnormalities that are below those thresholds.
} 
\par
Fig. \ref{application2} gives an example of this measure of abnormality \textcolor{black}{$a_j(t)$} in case of the word ``Lekue".  ``Lekue" mainly refers to a steamer in a microwave oven in Japan, which was a hit domestic product in 2010. 
The top figure indicates the normalised word count of the word $\check{F}_j(t)=F_j(t)/m(t)$, and the bottom figure indicates the corresponding \textcolor{black}{probabilistic} abnormality $a_j(t)$ (the green thin line). 
In addition, the red thick line represents the \textcolor{black}{probabilistic abnormality $a_j(t)$} smoothed by the moving weighted average \textcolor{black}{$\bar{a}_j(t)=\sum_{T=-15}^{15} a_j(t) \cdot \phi_n(t-T;mean=0,sd=6)$}, where $\phi_n(x)$ is the probability density function of the normal distribution; the peach dashed vertical lines indicate dates with large deviations, on which $a_j(t) \geq 4$; and the bottom blue vertical solid lines indicate dates on which "Lekue" was broadcasted by television in the Tokyo area. 
From this figure, we can confirm that the \textcolor{black}{probabilistic abnormality $a_j(t)$} was high from the middle of 2010 until the middle of 2011 (the red thick line), and in the same period there were many television broadcasts concerning "Lekue", which are highly correlated with the abnormality (the blue bottom vertical solid line).
\textcolor{black}{To construct the more rigorous probabilistic abnormality, we reqire a more accurate estimation of $\{c_j(t)\}$ (i.e., $\{r_j(t)\}$). The estimation is rough in this version.}
%
\begin{table*}
 \begin{tabular}{c c c |c|c}
 \hline
 $\alpha_1$ & $\alpha_2$ & $\alpha_3$ &The time series model (macro model) &The blogger model (micro model)  \\
 \hline
  0.5 & 0.5 & 0.5 & Steady Poisson process  &  Homogeneous bloggers, Invariant observable population \\
  1.0 & 0.5 & 0.5 & Unsteady Poisson process ($\Delta_n=0, \Delta_m=0$)  &  Homogeneous bloggers, Variant observable population \\
  1.0 & 1.0 & 0.5 & Unsteady Poisson process ($\Delta_n > 0, \Delta_m=0$) &  Homogeneous bloggers, Imperfectly observable population  \\
  1.0 & 1.0 & 1.0&  Random diffusion model ($\Delta_m > 0$) & Heterogeneous bloggers   \\
  \hline
 \multicolumn{5}{l}{$^{*1}$\textcolor{black}{Homogeneous corresponds to $\sigma^{(\lambda)}(t)=0$ in Eq. \ref{sigma_t}, and heterogeneous to $\sigma^{(\lambda)}(t) \neq 0$. }}   \\
  \multicolumn{5}{l}{$^{*2}$\textcolor{black}{Invariant population corresponds to $M(t)=const.$ in Eq. \ref{M_t}, and variant population to the time-variant $M(t)$. }}   \\
   \multicolumn{5}{l}{$^{*3}$\textcolor{black}{``Imperfectly observable'' means that $M(t)$ is observed with noise (see Appendix A).}}   \\
 \end{tabular}
 \caption{
 Summary of the classification of time series. The scaling exponents for $\check{c}>>1$ are $\alpha_1$ in the case of
 the raw time series $V[F_j]^{0.5}$,  $\alpha_2$ in the case of the time series scaled by the total blog $V[\tilde{F}_j]^{0.5}$, 
 and $\alpha_3$ in the case of the ensemble scaling $V_c[F]^{0.5}$.  In addition, $\Delta_m$ is given by \textcolor{black}{Eq. \ref{macro0abc}} and $\Delta_n$ is given by Eq. \ref{noise_noise}. 
 }
 \label{table1}
 \end{table*}%
\section{Conclusions and discussion}
In this paper, we have described how the properties of the fluctuations of time series representing the appearance of words can be explained by a nonstationary extension of the RD model. First, we proposed a simple nonstationary extension of the RD model (the extended RD model). Second, we showed that this extended RD model can be deduced from the model of heterogeneous bloggers who write their blogs randomly.
Third, through both a theoretical analysis of the extended RD model and an empirical data analysis, we obtained the following results: 
\begin{enumerate}
\item  The extended RD model can reproduce the following fluctuation scalings: 
The TFS of the time series of the appearances of words, the TFS of the time series of the appearances of words scaled by the total number of blogs, and the EFS at the fixed time. [Fig. \ref{fig_mean_sd} (a), Fig. \ref{fig_mean_sd_t}, Fig. \ref{EFS}.]
\item 
 The extended RD model can explain the correlation between the time series of the appearances of words and the total number of blogs [Fig. \ref{fig_mean_sd}(b)].
\item The distribution of the scaled differential of the appearances of words can be approximated by a Student's t-distribution for words with a very large frequency. This distribution is independent of the kind of word. （The former is an empirical result.） [Fig. \ref{pdf_fig}.]
\item 
By assuming that the limit distribution for a very large mean obeys the above-mentioned t-distribution, the extended RD model can explain the distribution for any finite mean. [Fig. \ref{small_PDF2}.]
\end{enumerate}
Finally, we suggested the following applications of the fluctuation scalings: (i) Quantifying the impacts of special nationwide events by measuring the deviation from the ordinary ensemble fluctuation scalings of 1771 adjectives. (ii) Quantifying the temporal variation of the abnormality of the word count of a considered word \textcolor{black}{in a probabilistic manner} [Fig. \ref{appli}, Fig. \ref{application2}.]. \textcolor{black}{Note that because our modelling framework does not consider the characteristic features of the Japanese language, it is likely that other languages can also be described using our model. Thus, investigating other languages will be an aim of our future work.}  \par
\textcolor{black}{
The extended RD model can describe non-stationary time-variances in the usages of words and total number of blogs, which the original steady RD model cannot describe. Through this extension, the model can consistently explain various properties of word appearances in blogs, including scalings and distributions.
In contrast, the original steady RD model can explain the scaling of a raw time series $V[F_j]$, but cannot explain other empirical properties. More precisely, the original model can explain scaling properties individually by changing the parameter $\Delta_u$ in Eq. \ref{r1} with each scaling, but by a using single parameter set it does not consistently explain any of the other empirical properties that we describe in this paper.
} \par
\par
The fluctuation scaling for non-steady time series is observed not just in blogs, but also other systems. For example, Ref. \cite{argollo2004separating} reported the case for the internet, electric circuits, etc. 
The authors observed changes of a scaling exponent by removing the effects of the total times series.  Then, they 
classified the results by two types: those for which the scaling exponent changes from 1.0 to 0.5, and those for which it remains unchanged.  
The extended RD model that we discussed here exhibits similar properties. 
In addition, as indicated by Table \ref{table1}, we can confirm that whether the exponent of the model changes or not depends on the model parameters. 
Moreover, we can discuss relationships to properties of bloggers (micro-structures) by using the bloggers model. 
\textcolor{black}{Because the preconditions given in Ref. \cite{argollo2004separating} are different, our result \textcolor{black}{is not directly applicable}.}
However, our study has the potential to provide suggestions for such a problem with a non-steady FS, because the extended RD model is very general. 
Therefore, further studies are required in order to clarify the relationships between the extended RD model \textcolor{black}{(i.e., the non-steady RD model)} that was proposed in this paper and the empirical observations of various actual complex systems. \par
\begin{acknowledgments}
The authors thank Dentsu Kansai Inc., Dentsu Inc., Hottolink Ultra Developers Team and Hottolink Inc. for useful discussions and providing the data.
\end{acknowledgments}

\bibliography{adj0708_pre}

\begin{thebibliography}{10}

\bibitem{preis2012quantifying}
T. Preis, H.~S. Moat, H.~E. Stanley, and S.~R. Bishop, Sci. Rep. {\bf 2},
  (2012).

\bibitem{ugander2011anatomy}
J. Ugander, B. Karrer, L. Backstrom, and C. Marlow, arXiv:1111.4503  (2011).

\bibitem{ceron2014every}
A. Ceron, L. Curini, S.~M. Iacus, and G. Porro, NEW MEDIA SOC  .

\bibitem{ginsberg2009detecting}
J. Ginsberg {\it et~al.}, Nature {\bf 457},  1012  (2009).

\bibitem{sakaki2010earthquake}
T. Sakaki, M. Okazaki, and Y. Matsuo,  in {\em Proceedings of the 19th
  international conference on World wide web}, ACM (ACM, New York, USA, 2010),
  pp.\ 851--860.

\bibitem{grajales2014social}
F.~J. Grajales~III {\it et~al.}, J. Med. Internet Res. {\bf 16},  e13  (2014).

\bibitem{yu2012survey}
S. Yu and S. Kak, arXiv:1203.1647  (2012).

\bibitem{xu2015taylor}
M. Xu, arXiv:1505.02033  (2015).

\bibitem{eisler2008fluctuation}
Z. Eisler, I. Bartos, and J. Kertesz, Adv. Phys. {\bf 57},  89  (2008).

\bibitem{onnela2010spontaneous}
J. Onnela and F. Reed-Tsochas, Proc. Natl. Acad. Sci. U. S. A. {\bf 107},
  18375  (2010).

\bibitem{sato2010fluctuation}
A.-H. Sato, M. Nishimura, and J.~A. Ho{\l}yst, Physica A {\bf 389},  2793
  (2010).

\bibitem{gerlach2014scaling}
M. Gerlach and E.~G. Altmann, New. J. Phys. {\bf 16},  113010  (2014).

\bibitem{sano2010macroscopic}
Y. Sano and M. Takayasu, JEIC {\bf 5},  221  (2010).

\bibitem{10.1371/journal.pone.0109004}
Q.~S. Hanley, S. Khatun, A. Yosef, and R.-M. Dyer, PLoS ONE {\bf 9},  e109004
  (2014).

\bibitem{taylor1961aggregation}
L.~R. Taylor, Nature {\bf 189},  732  (1961).

\bibitem{PhysRevLett.100.208701}
S. Meloni, J. G\'omez-Garde\~nes, V. Latora, and Y. Moreno, Phys. Rev. Lett.
  {\bf 100},  208701  (2008).

\bibitem{argollo2004separating}
M. Argollo~de Menezes and A.-L. Barab\'asi, Phys. Rev. Lett. {\bf 93},  068701
  (2004).

\bibitem{link1}
E.~G. Altmann and M. Gerlach, Physicists' papers on natural language from a
  complex systems viewpoint,
  http://www.pks.mpg.de/mpi-doc/sodyn/physicist-language/.

\bibitem{abrams2003linguistics}
D.~M. Abrams and S.~H. Strogatz, Nature {\bf 424},  900  (2003).

\bibitem{altmann2015statistical}
E.~G. Altmann and M. Gerlach,  in {\em Creativity and Universality in
  Language}, edited by M. Degli~Esposti, E.~G. Altmann, and F. Pachet (Springer
  International Publishing, Cham, Switzerland, 2016), pp.\ 7--26.

\bibitem{cong2014approaching}
J. Cong and H. Liu, Phys Life Rev. {\bf 11},  598  (2014).

\bibitem{sano2009}
Y. Sano, K.~K. Kaski, and M. Takayasu,  in {\em Proc. Complex '09} (Springer,
  Berlin, Germany, 2009), No.~2, pp.\ 195--198.

\bibitem{PhysRevE.87.012805}
Y. Sano {\it et~al.}, Phys. Rev. E {\bf 87},  012805  (2013).

\bibitem{giometto2015sample}
A. Giometto {\it et~al.}, Proc. Natl. Acad. Sci. U. S. A. {\bf 112},  7755
  (2015).

\bibitem{cohen2013taylor}
J.~E. Cohen, Theor. Popul. Bioly. {\bf 88},  94  (2013).

\bibitem{fronczak2010origins}
A. Fronczak and P. Fronczak, Phys. Rev. E {\bf 81},  066112  (2010).

\bibitem{kendal2011taylor}
W.~S. Kendal and B. J{\o}rgensen, Phys. Rev. E {\bf 83},  066115  (2011).

\bibitem{kendal2004taylor}
W.~S. Kendal, ECOL. COMPLEX. {\bf 1},  193  (2004).

\bibitem{cohen2015random}
J.~E. Cohen and M. Xu, Proc. Natl. Acad. Sci. U. S. A. {\bf 112},  7749
  (2015).

\end{thebibliography}
\appendix
\renewcommand{\theequation}{A.\arabic{equation}}
\renewcommand{\thefigure}{A-\arabic{figure}}
\setcounter{figure}{0}
\setcounter{equation}{0}
\section{Derivation of the random diffusion model from the blogger model} 
\subsection{Model}
Here, we demonstrate that the extended RD model can be deduced from a simple model of the behaviour of bloggers.   
We consider a system that consists of $N(t)$ bloggers ($N(t)>>1$).
In this system, the bloggers perform the following behaviour from time $t=1$ to $t=T$: 
\begin{enumerate}
\item The $i$-th blogger writes his or her blog randomly with the probability $p^{(i)}(t)$ ($0 \leq p^{(i)}(t) \leq 1$) $(i=1,2,3,\cdots,N(t))$. Here, we define a random variable $q_i(t)$ that takes a value of $1$ when the $i$-th blogger writes his or her blog and $0$ when he or she does not write at time $t$.
\item The $i$-th blogger who writes a blog in step 1 writes the $j$-th word  $f^{(i)}_j(t)$ times, where $f^{(i)}_j(t)$ is sampled from a Poisson distribution with the Poisson parameter $\lambda^{(i)}_j(t)$. $(j=1,2,3,\cdots,W(t))$, 
\end{enumerate}
where we denote the number of words at time $t$ by $W(t)$.
We consider two macro random variables deduced from the micro model.
The first is the number of bloggers that have written at time $t$ (the active number of bloggers) $M(t)$, and is defined by 
\begin{equation}
M(t)=\sum^{N(t)}_{i=1}q_i(t). \label{M_t}
\end{equation}
The second is the total number of appearances of the $j$-th word in all blogs $F_j(t)$, which is defined by 
\begin{equation}
F_j(t)=\sum^{N(t)}_{i=1}q_i(t) \cdot f^{(i)}_j(t).
\end{equation}
\par
In this paper, for simplicity, we introduce the following assumptions concerning $p^{(i)}(t)$ and $\lambda_j^{(i)}(t)$:
\begin{enumerate}
\item $\mathbf{p}(t)=(p^{(1)}(t),p^{(2)}(t),\cdots,p^{(N(t))}(t))$ and $\mathbf{\lambda}_j(t)=(\lambda^{(1)}_j(t),\lambda^{(2)}_j(t) \cdots \lambda^{(N(t))}_j(t))$ are independent. That is, $\mathbf{\lambda}_j(t)$ and $\mathbf{p}(t)$ are sampled from random variables whose joint probability density function is defined by 
\begin{eqnarray}
&&P_{\{\mathbf{\lambda_j}(t),\mathbf{p}(t)\}}(\lambda^{(1)}_j(t),\lambda^{(2)}_j(t),\cdots,\lambda^{(N)}_j(t),  \nonumber \\
&&p^{(1)}(t),p^{(2)}(t),\cdots,p^{(N)}(t))  \\
&=&P_{\mathbf{\lambda_j}(t)}(\lambda^{(1)}_j(t),\lambda^{(2)}_j(t),\cdots,\lambda^{(N)}_j(t)) \nonumber \\
&\cdot& P_{\mathbf{p}(t)}(p^{(1)}(t),p^{(2)}(t),\cdots,p^{(N)}(t)) \nonumber \\
&=&P_{\mathbf{\lambda_j}(t)}(\mathbf{\lambda_j}(t)) \cdot P_{\mathbf{p}}(\mathbf{p}(t)).  
\end{eqnarray}
\item 
The mean and the variance of the mean of the appearances of the $j$-th word over all bloggers,  $\lambda_j(t) \equiv \sum^{N(t)}_{i=1}\lambda_j^{(i)}(t)/N(t)$, are assumed to be $\mu_j^{(\lambda)}(t)$ and $[\mu_j^{(\lambda)} \cdot \sigma^{(\lambda)}(t)]^2$, respectively.
That is, the mean of $\lambda_j(t)$ is assumed to be
\begin{eqnarray}
&&\int \int \sum_{k=1}^{N(t)}\frac{\lambda_j^{(k)}}{N(t)} \cdot P_{(\mathbf{\lambda_j}(t),\mathbf{p(t)})}(\mathbf{\lambda_j}(t),\mathbf{p(t)})) \cdot d \mathbf{\lambda_j} d\mathbf{p}  \nonumber  \\
&=&\left<\sum_{k=1}^{N(t)} \lambda_j^{(k)}/N(t)\right>_{(\mathbf{\lambda_j},\mathbf{q})} \\ 
&=&\mu_j^{(\lambda)}(t),
\end{eqnarray}
and the variance of $\lambda_j(t)$ is assumed to be
\begin{eqnarray}
&& \int \int \sum_{k=1}^{N(t)} \frac{(\lambda_j^{(k)}(t)-\mu_j^{(\lambda)}(t))^2}{N(t)} \nonumber \\ 
&\cdot&  P_{(\mathbf{\lambda_j}(t),\mathbf{p(t)})}(\mathbf{\lambda_j}(t),\mathbf{p(t)})) \cdot d \mathbf{\lambda_j} d\mathbf{p} \nonumber \\
&=&\left<\sum_{k=1}^{N(t)} (\lambda_j^{(k)}(t)-\mu_j^{(\lambda)}(t))^2/N(t) \right>_{(\mathbf{\lambda_j},\mathbf{q})}. \nonumber \\
&=&\{\mu_j^{(\lambda)} \cdot \sigma^{(\lambda)}(t)\}^2. \label{sigma_t}
\end{eqnarray}
\item The mean and the variance of the mean for the probability that a blogger writes his or her blog over all bloggers, $p(t) \equiv \sum^{N(t)}_{i=1}p^{(i)}(t)/N(t)$, are assumed to be $\mu^{(p)}(t)$ and $\sigma^{(p)}(t)^2$, respectively.
That is, the mean of $p(t)$ is assumed to be
\begin{eqnarray}
&&\int \int \sum_{k=1}^{N(t)}\frac{p^{(k)}}{N(t)} \cdot P_{(\mathbf{\lambda_j}(t),\mathbf{p(t)})}(\mathbf{\lambda_j}(t),\mathbf{p(t)})) \cdot d \mathbf{\lambda_j} d\mathbf{p} \nonumber \\
&=&\left<\sum_{k=1}^{N(t)}p^{(k)}/N(t)\right>_{(\mathbf{\lambda_j},\mathbf{q})}=\mu_j^{(p)}(t) ,
\end{eqnarray}
and the variance of $p(t)$ is assumed to be
\begin{eqnarray}
&& \int \int \sum_{k=1}^{N(t)} \frac{(p^{(k)}(t)-\mu_j^{(p)}(t))^2}{N(t)} \nonumber \\ 
&\cdot&  P_{(\mathbf{\lambda_j}(t),\mathbf{p(t)})}(\mathbf{\lambda_j}(t),\mathbf{p(t)})) \cdot d \mathbf{\lambda_j} d\mathbf{p} \nonumber \\
&=&\left<\sum_{k=1}^{N(t)} (p^{(k)}(t)-\mu_j^{(\lambda)}(t))^2/N(t) \right>_{(\mathbf{\lambda_j},\mathbf{q})} \nonumber \\
&=& \sigma^{(p)}(t)^2.
\end{eqnarray}
\end{enumerate}
\par
\subsection{Distribution of $F_j(t)$}
Here, we calculate the distribution of the number of appearances of the $j$-th word, $F_j(t)$, at time $t$.
For this, we investigate two cases: (i) $M(t)$ is observable, and (ii) $M(t)$ is not observable.
\subsubsection{The case that $M(t)$ is observable}

First, we calculate the distribution of $F_j(t)$ in the case that $M(t)$ is observable.
The condition that $M(t)$ is observable corresponds to the case that $q_i(t)$ satisfies the following restraint condition:
\begin{equation}
M(t)=\sum^{N(t)}_{i=1}q_i(t). \label{blogger_fix}
\end{equation}
$F_j(t)$ is defined as
\begin{equation}
F_j(t)=\sum^{M}_{i=1}q_i(t) \cdot f^{(i)}_j(t).
\end{equation}
Because of the additive property of the Poisson distribution of $f^{(i)}_j(t)$, 
$F_j(t)$ obeys a Poisson distribution with the Poisson parameter
\begin{equation}
\Pi_j(t)=\sum^{M}_{i=1}q_i(t) \cdot \lambda^{(i)}_j(t).
\end{equation}
Considering the restraint condition on $q_i(t)$ given by Eq. \ref{blogger_fix}, the distribution of $\Pi_j(t)$ can be denoted by 
the value $\Gamma^{(k)}_j(t)$ with the probability $P^{(k)}(t)$ as follows:
%
\begin{equation}
\Gamma^{(k)}_j(t)=\sum_{i=1}^{M(t)}\lambda^{(l^{(k)}_i(t))}_j(t), 
\end{equation}
and the probability is given as 
\begin{eqnarray}
P^{(k)}(t)&=&\frac{Q^{(l^{(k)})}(t)}{Q(t)}   
\end{eqnarray}
\begin{eqnarray}
Q^{(l^{(k)})}(t) &=&\prod_{\{i|1<=i<=N;i \in l^{(k)}(t)\}}p^{(i)}(t) \nonumber \\
&\cdot& \prod_{\{i|1<=i<=N;i \notin l^{(k)}(t)\}}(1-p^{(i)}(t)). 
\end{eqnarray}
\begin{equation}
Q(t)=\sum_{i=1}^{L(t)}Q_i(t),
\end{equation}
where $k$ is the label $k=1,2, \cdots,L(t)$ and $l^{(k)}(t)=\{l^{(k)}_1(t),l^{(k)}_2(t),\cdots,l^{(k)}_{M(t)}(t)\}$ are the set of indexes of a blogger such that $M(t)$ bloggers are chosen from the total $N(t)$. Here, different labels $k$ corresponds sets $l^{(k)}$ that are always different. In addition,   
$L(t)$ represents the number of all combinations, $L(t) \equiv {}_{N(t)}C_{M(t)}=\frac{N!}{(N-M)!M!}$, and $Q(t)$ is a normalization factor.\par
First, we define the two variables $P(t)$ and $\Gamma_j(t)$, for convenience of the calculation.
$P(t)$ is defined by
\begin{eqnarray}
P(t) \equiv \sum^{L}_{l=1}\frac{P^{(l)}(t)}{L}=\frac{1}{L}=\frac{1}{{}_{N(t)}C_{M(t)}}. 
\end{eqnarray}
We also define $\bar{P}^{(l)}(t) \equiv P^{(l)}(t)-P(t)$. \par
Similarly, $\Gamma_j(t)$ is defined by $\Gamma_j(t) \equiv \sum^{L(t)}_{i=1} \Gamma^{(l)}_j(t)/L$,
where 
\begin{eqnarray}
 \sum^{L(t)}_{i=1} \Gamma^{(l)}_j(t) ={}_{N(t)-1}C_{M(t)-1} \sum^{N(t)}_{i=1} \lambda^{(i)}_j(t).
\end{eqnarray}
Here, the number of $\lambda^{(i)}_j(t)$ in $\sum^{L(t)}_{i=1} \Gamma^{(l)}_j(t)$ is calculated by the number of combination given by choosing $(M(t)-1)$ from the total of $(N(t)-1)$ bloggers, where $-1$ indicates that we chose the $i$-th blogger.  Thus, we obtain the coefficients  ${}_{N(t)-1}C_{M(t)-1}$.
Therefore, we can obtain that
\begin{eqnarray}
 \Gamma_j(t) \equiv \sum^{L(t)}_{i=1} \Gamma^{(l)}_j(t)/L =M(t) \sum^{N(t)}_{i=1} \frac{\lambda^{(i)}_j(t)}{N(t)}=M(t) \mu^{(\lambda)}_j(t).  \nonumber \\ 
  \label{gamma_j}
\end{eqnarray}
\par
Using these values, we can obtain the mean $<\Pi_j(t)>_{\{q_i(t)\}}$ of $\Pi_j(t)$, as
\begin{eqnarray}
&&<\Pi_j(t)>_{\{q_i(t)\}}=<\Gamma_j(t)> \\
&=&\sum^{L(t)}_{l=1}\Gamma^{(l)}_j(t) \cdot P^{(l)}(t) \nonumber \\
&\approx& P(t) \cdot \sum^{L(t)}_{i=1} \Gamma_j(t)+ <\sum^{L(t)}_{i=1} \bar{\Gamma}^{(l)}_j(t) \cdot \bar{P}^{(l)}(t)>+O(\sqrt{L(t)})  \nonumber  \\
&=& M(t) \cdot \mu_j^{(\lambda)}(t) \quad (L(t)>>1).
\end{eqnarray}
Next, we use $\Gamma_j(t)$ to calculate the variance as
\begin{eqnarray}
&&<(\Pi_j(t)-<\Pi_j(t)>_{\{q_i(t)\}})^2>_{\{q_i(t)\}} \\
&=& \sum^{L(t)}_{l=1}(\Gamma^{(l)}_j(t)-<\Gamma_j(t)>)^2 \cdot P^{(l)}(t) \\
&\approx&  \sum^{L(t)}_{i=1} \Gamma^{(l)}_j(t)^2 \cdot P^{(l)}(t) -M(t)^2 \cdot \mu^{(\lambda)}_j(t)^2.  \nonumber \\
\label{variance_0}
\end{eqnarray}
Here, we used the fact that $\sum^{L}_{i=1}P^{(l)}(t)=1$ and Eq. \ref{gamma_j}.
Then, we can decompose the first term as follows:  
\begin{eqnarray}
&&\sum^{L(t)}_{l=1} \Gamma^{(l)}_j(t)^2 \cdot P^{(l)}(t) \\
&=&  \sum^{L(t)}_{i=1} \{\sum_{k=1}^{M(t)}[\bar{\lambda}^{(l^{(i)}_k(t))}_j(t)+\lambda_j(t)]\}^2  \cdot P^{(i)}(t) \\  
&\approx& \sum^{L(t)}_{i=1} P(t) \cdot  \left\{ \sum_{k=1}^{M(t)}\bar{\lambda}^{(l^{(i)}_k(t))}_j(t)^2 \right. \nonumber \\
&+& 2\cdot \sum_{m=1}^{M(t)} \sum_{n=1}^{n<m}\bar{\lambda}^{(l^{(i)}_m(t))}_j(t) \cdot \bar{\lambda}^{(l^{(i)}_n(t))}_j(t)   \nonumber \\ 
&+&  M(t)^2 \lambda_j(t)^2  \Biggr\},  
\end{eqnarray}
where $\bar{\lambda}^{(l^{(i)}_k(t))}_j(t) \equiv \lambda^{(l^{(i)}_k(t))}_j(t) -\lambda_j(t)$. \par
Taking the sum over $i$, we obtain
\begin{eqnarray}
&&<(\Pi_j(t)-<\Pi_j(t)>_{\{q_i(t)\}})^2>_{\{q_i(t)\}} \nonumber \\
&\approx&  P(t)  \cdot \left\{{}_{M(t)-1}C_{N(t)-1} \cdot \sum_{k=1}^{N(t)} \bar{\lambda}^{(l^{(i)}_k(t))}_j(t)^2 \right. \nonumber \\
&+& 2\cdot {}_{M(t)-2}C_{N(t)-2} \cdot \sum_{m=1}^{N(t)} \sum_{n=1}^{n<m}\bar{\lambda}^{(l^{(i)}_m(t))}_j(t) \cdot \bar{\lambda}^{(l^{(i)}_n(t))}_j(t)   \nonumber \\ 
&+&  {}_{M(t)}C_{N(t)} \cdot M(t)^2 \lambda_j(t)^2  \Biggr\} \\
&=&     M(t) \cdot \sum_{k=1}^{N(t)} \frac{\bar{\lambda}^{(l^{(i)}_k(t))}_j(t)^2}{N(t)}+ M(t)^2 \lambda_j(t)^2  \nonumber \\
&+& 2\cdot M(t) \cdot \frac{M(t)-1}{N(t)-1}  \cdot \sum_{m=1}^{N(t)} \sum_{n=1}^{n<m} \frac{\bar{\lambda}^{(l^{(i)}_m(t))}_j(t) \cdot \bar{\lambda}^{(l^{(i)}_n(t))}_j(t)}{N(t)}   \nonumber  \\
\end{eqnarray}
Thus, we can write the first term of Eq. \ref{variance_0} as
\begin{eqnarray}
&&<(\Pi_j(t)-<\Pi_j(t)>_{\{q_i(t)\}})^2>_{\{q_i(t)\}}  \nonumber \\
&\approx&  M(t) \cdot \sum_{k=1}^{N(t)} \frac{\bar{\lambda}^{(l^{(i)}_k(t))}_j(t)^2}{(N(t))^2} +  M(t)^2 \lambda_j(t)^2  \nonumber \\
&+& M(t) \cdot \frac{M(t)-1}{N(t)-1}  \cdot \sum_{k=1}^{N(t)} \frac{\bar{\lambda}^{(l^{(i)}_k(t))}_j(t) \cdot (\lambda_j(t)-\bar{\lambda}^{(l^{(i)}_k(t))}_j(t)) }{N(t)} \nonumber \\   
&\approx&  M(t) (1-\frac{M(t)-1}{N(t)-1}) \cdot \mu_j^{(\lambda)}(t)^2 \cdot \sigma^{(\lambda)}_j(t)^2 \nonumber \\ 
&+& M(t)^2 \mu^{(\lambda)}_j(t)^2, 
\end{eqnarray}
where $\lambda_j=\sum^{N}_{i=1}\lambda^{(i)}_j/N(t)$,$\bar{\lambda}^{(i)}_j(t) \equiv \lambda^{(i)}_j(t)-\lambda_j(t)$．
Hence, the variance can be written as 
\begin{eqnarray}
&&<(\Pi_j(t)-<\Pi_j(t)>_{q_i(t)})^2>_{q_i(t)}  \nonumber \\ 
&\approx& M(t) \cdot (1-\frac{M(t)-1}{N(t)-1})  \mu_j^{(\lambda)}(t)^2 \cdot \sigma^{(\lambda)}(t)^2. 
\end{eqnarray}
 \par 
Consequently, from these results we can obtain the extended RD model (the macro model) for $pN>>1$, as 
\begin{equation}
F_j(t) \sim Poi(c_j(t) \cdot \Lambda_j(t))
\end{equation}
where the formulas connecting the micro parameters and the macro parameters are approximately given by
\begin{equation}
c_j(t) =  \mu^{(\lambda)}_j(t) \cdot \sum_{t=1}^{T}M(t)
\end{equation}
\begin{equation}
m(t) = \frac{M(t)}{\sum_{t=1}^{T}M(t)} 
\end{equation}
\begin{equation}
\Delta_m(t)=m(t)^{\beta_m(t)} \cdot \Delta_0(t) 
\end{equation}
\begin{equation}
\Delta_0(t) = \sqrt{(1-\frac{M(t)-1}{N(t)-1})} \cdot \frac{1}{\sum_{t=1}^{T}M(t)} \cdot \sigma^{(\lambda)}(t)
\end{equation}
\begin{equation}
\beta_m(t) = 0.5, 
\end{equation}
\textcolor{black}{where the mean of $\Lambda_j(t)$ is $m(t)$, and the standard deviation is $\Delta_m(t)$. }
Note that we use the constraint condition $\sum^{T}_{t=1}m(t)/T=1$. \par
\subsubsection{The case that $M(t)$ is unobservable}
Next, we calculate the distribution $F_j(t)$ on the condition that $M(t)$ cannot be observed. \par
$F_j(t)$ is defined by
\begin{equation}
F_j(t)=\sum^{N}_{i=1}q_i(t) \cdot f^{(i)}_j(t).
\end{equation}
%
Using the properties of Poisson distribution obeyed by $f^{(i)}_j(t)$, we can deduce that $F_j(t)$ obeys a Poisson distribution with the Poisson parameter
\begin{equation}
\Pi_j(t)=\sum^{N}_{i=1}q_i(t) \cdot \lambda^{(i)}_j(t).
\end{equation}
The mean of $\Pi_j(t)$ is calculated by 
\begin{eqnarray}
&&<\Pi_j(t)>_{\{q_j\}} \nonumber \\
&=& \sum^{N}_{i=1}(\bar{p}_i(t)+p(t)) \cdot (\bar{\lambda}^{(i)}_j(t)+\lambda_j(t)) \nonumber \\
&\approx& N \cdot p(t) \cdot \lambda_j(t) +  \sum^{N}_{i=1} <\bar{p}_i(t)> \cdot <\bar{\lambda}^{(i)}_j(t)> \nonumber \\
&\approx& N \cdot \mu^{(p)}(t) \cdot \mu^{(\lambda)}_j(t),
\end{eqnarray}
where $p(t) \equiv \sum^{N(t)}_{i=1}p_i(t)/N(t)$ and $\bar{p}_i(t) \equiv p_i(t)-p(t)$.\\
The variance of $\Pi_j(t)$ is calculated by  
\begin{eqnarray}
&&\left<(\Pi_j(t)-<\Pi_j(t)>_{\{q_i\}}\right)^2>_{\{q_i\}} \\
&=&\left<\Pi_j(t)^2 \right>_{\{q_i\}}- <\Pi_j(t)>_{\{q_i\}}^2\\
&\approx& \left(\sum^{N(t)}_{i=1} \mu^{(p)}(t) \cdot \bar{\lambda}^{(i)}_j(t) \right)^2 \nonumber \\
&+& \left<\left(\sum^{N(t)}_{i=1}  \bar{q}_i(t)^2 \cdot( \lambda_j(t) +\bar{\lambda}_j^{(i)}(t))^2\right) \right>_{\{q_i\}} \nonumber \\
&\approx& \left(\mu^{(p)}(t) \cdot \sum^{N(t)}_{i=1} \bar{\lambda}^{(i)}_j(t) \right)^2 \nonumber \\
&+& \sum^{N(t)}_{i=1} \left< \bar{q}_i(t)^2 \right>_{\{q_i\}}/N \cdot \sum^{N(t)}_{i=1}( \mu_j^{(\lambda)}(t) +\bar{\lambda}_j^{(i)}(t))^2, \nonumber 
\end{eqnarray}
where $\bar{q}_i(t) \equiv q_i(t)-\sum^{N(t)}_{i=1}q_i(t)/N(t)$, and we use the fact that $\sum^{N(t)}_{i=1}q_i(t)/N(t) \approx \mu^{(p)}(t)$ ($N>>1$). \par 
Using the facts that $\sum^{N}_{i=1}\bar{\lambda}^{(i)}_j(t) \approx 0$ and $\left< \bar{q}_i(t)^2 \right>=p_i(t)\cdot(1-p_i(t))$; that is, considering the variance of the Bernoulli distribution with the probability $p_i(t)$, 
we get that
\begin{eqnarray}
&&\left<\Pi_j(t)^2-<\Pi_j(t)>_{\{q_i\}}\right>_{\{q_i\}} \nonumber  \\
&\approx& \sum^{N(t)}_{i=1}(p_i(t)-p_i(t)^2)/N \cdot \sum^{N(t)}_{i=1}(\lambda_j(t) +\bar{\lambda}_j^{(i)}(t))^2   \nonumber \\
&\approx& N(t)  \cdot (\mu^{(p)}(t)-\mu^{(p)}(t)^2-\sigma^{(p)}(t)^2) \nonumber \\
&\cdot& \mu_j^{(\lambda)}(t)^2 \cdot (1+\sigma_j^{(\lambda)}(t)^2).   \nonumber
\end{eqnarray}
Thus, we obtain the variance of $\Pi_j(t)$ as
\begin{eqnarray}
&&\left<\Pi_j(t)^2-<\Pi_j(t)>_{\{q_i\}}\right>_{\{q_i\}} \nonumber \\
&=&N(t) \cdot  \mu_j^{(\lambda)}(t)^2 \cdot(1+\sigma_j^{(\lambda)}(t)^2) \nonumber \\
& \cdot& (\mu^{(p)}(t)-\mu^{(p)}(t)^2-\sigma^{(p)}(t)^2 ).  
\end{eqnarray}
\par
%
Consequently, from these results we can write the extended RD model (the macro model) for $pN>>1$ as
\begin{equation}
F_j(t) \sim Poi(c_j(t) \cdot \Lambda_j(t)),
\end{equation}
where the formulas connecting the micro parameters with the macro parameters are approximately given by
\begin{equation}
c_j(t) = \mu_j^{\lambda}(t) \cdot \sum^{T}_{t=1}(\mu^{(p)}(t)N(t)) 
\end{equation}
\begin{equation}
m(t) = \frac{N(t) \cdot \mu^{(p)}(t)}{\sum^{T}_{t=1}(\mu^{(p)}(t)N(t))} 
\end{equation}
\begin{equation}
\Delta_m(t)=m(t)^{\beta_m(t)} \cdot \Delta_0(t) 
\end{equation}
\begin{eqnarray}
\Delta_0(t)&=&\frac{1}{\sqrt{\mu^{(p)}(t) \cdot \sum^{T}_{t=1}(\mu^{(p)}(t)N(t)) }  } \cdot \{(1+\sigma_j^{(\lambda)}(t)^2) \nonumber \\
& \cdot& (\mu^{(p)}(t)-\mu^{(p)}(t)^2-\sigma^{(p)}(t)^2 )\}^{0.5}  
\end{eqnarray}
\begin{eqnarray}
\beta_{m}=0.5,
\end{eqnarray}
\textcolor{black}{where the mean of $\Lambda_j(t)$ is $m(t)$, and the standard deviation is $\Delta_m(t)$.} 
\par
\renewcommand{\theequation}{B.\arabic{equation}}
\renewcommand{\thefigure}{B-\arabic{figure}}
\setcounter{figure}{0}
\setcounter{equation}{0}
\section{Derivation of fluctuation scaling of the extended random diffusion model} 
In this section, we calculate the FSs of the extended RD model. 
%
Note that we employ the following assumptions in this section.
First, we assume that $F_j(t)$ obeys the following extended RD model:
\begin{equation}
F_j(t) \sim Poi(c_j(t) \cdot \Lambda_j(t)), 
\end{equation}
\textcolor{black}{where $c_j(t) \geq 0$ is the scale factor, and $\Lambda_j(t) \geq 0$ is the random variable with the mean $m(t) \geq 0$ and standard deviation $\Delta_m(t)=m(t)^{\beta_m} \cdot \Delta_0(t) \geq 0$.}
Second, $F_j(t)$ is decomposed into two parts: the scale part $c_j(t) \cdot m(t)$ and the random part $\epsilon_j(t)$. That is,
\begin{eqnarray}
F_j(t)=c_j(t) \cdot m(t)+\epsilon_j(t),
\end{eqnarray}
where  $c_j(t)$ is also decomposed into two parts: the scale factor  $\check{c}_j$ and the time variation factor $r_j(t)$. That is,
\begin{equation}
c_j(t)=\check{c}_j \cdot r_j(t),
\end{equation}
where $m(t)$ and $r_j(t)$ are independent and satisfy $E[r_j]=\sum^{T}_{t=1}r_j(t)/T=1$ and $E[m]=\sum^{T}_{t=1}m(t)/T=1$. 
From these assumptions and the formula for products of independent random variables, we obtain the formula  
\textcolor{black}{
\begin{eqnarray}
&&V[m \cdot r]= \nonumber \\ 
&&V[m] \cdot V[r_j]+E[m]^2 \cdot V[r_j]+E[r_j]^2 \cdot V[m]+O(\frac{1}{\sqrt{T}})  \nonumber \\ 
&\approx& V[r_j] \cdot V[m]+V[r_j]+V[m], \nonumber
\end{eqnarray}
\begin{eqnarray}
&&V[A+B]=V[A]+V[B]+O(\frac{1}{\sqrt{T}}) \approx V[A]+V[B], \nonumber
\end{eqnarray}
and 
\begin{eqnarray}
V_c[A+B] &=&V_c[A]+V_c[B]+O(\frac{1}{\sqrt{N_c}}) \approx V_c[A]+V_c[B], \nonumber   
\end{eqnarray}
}
 where $N_c$ is defined in Appendix E.  In addition, $E[\epsilon_j]$, $V[\epsilon_j]$, $E_c[\epsilon_j]$, and $V[\epsilon_j]$ are also given in Appendix E. 
\par
\subsection{\textcolor{black}{Temporal mean and the temporal variance of raw time series, $E[F_j]$ and $V[F_j]$} }
We calculate the temporal mean and the temporal variance of the raw time series of word appearances $F_j(t)$. 
$E[F_j(t)]$ is written as 
\begin{eqnarray}
E[F_j(t)]&=&E[\check{c}_j \cdot r_j \cdot m] + E[\epsilon_j] \\
&\approx&E[\check{c}_j] \cdot E[r_j] \cdot E[m] + E[\epsilon_j] \\
&\approx& \check{c_j}.
\end{eqnarray}
\par
%
Similarly, $V[F_j(t)]$ is calculated as
\begin{eqnarray}
&&V[F_j] \nonumber \\
&=&  V[\check{c}_j \cdot r_j \cdot m+\epsilon_j] \\
&\approx& V[\check{c}_j \cdot r_j]V[m]+E[\check{c}_j \cdot r_j]^2 \cdot V[m]+E[m]^2 \cdot V[\check{c}_j \cdot r_j]  \nonumber \\
&+&V[\epsilon_j] \\
&\approx& \check{c}_j^2 \cdot \{V[r_j]V[m]+E[r_j]^2 \cdot V[m]+ E[m]^2 \cdot V[r_j] \} \nonumber  \\
&+&E[\check{c}_j \cdot r_j \cdot m + m^{2 \beta_{m}} \cdot \Delta_0^2 \cdot \check{c}_j^2 \cdot r_j^2   ] \\
&\approx&\check{c}_j+ \nonumber   \\
&& \check{c}_j^2 \{ V[r_j]+(V[r_j]+1)\cdot (V[m]+E[m^{2 \beta_{m}}] \cdot E[\Delta_0^2]) \}. \nonumber \\
\label{V_F_ex}
\end{eqnarray} 
Here, we can substitute in $\beta_{m}=0$ or $\beta_{m}=0.5$ to obtain the simple equation
\begin{eqnarray}
&&V[F_j] \nonumber \\
&\approx&\check{c}_j+ \nonumber   \\
&& \check{c}_j^2 \{ V[r_j]+(V[r_j]+1)\cdot (V[m]+E[\Delta_0^2]) \}. \nonumber \\
\label{V_F_ex2}
\end{eqnarray}
Furthermore, in the case that $\beta_{m}=1.0$ we obtain 
\begin{eqnarray}
&&V[F_j] \approx \nonumber \\
&& \check{c}_j^2 \{ V[r_j]+(V[r_j]+1)\cdot (V[m]+(1+V[m]) \cdot E[\Delta_0^2]) \}. \nonumber \\
\end{eqnarray}
\subsection{\textcolor{black}{Temporal means and temporal variances of time series normalised by their system sizes: $E[\tilde{F}_j]$ and $V[\tilde{F}_j]$, $E[\hat{F}_j]$ and $V[\hat{F}_j]$}} 
\textcolor{black}{\subsubsection{  The case that the system size is observed without observation errors, $\tilde{F}_j(t)$: }}  
\textcolor{black}{We calculate the temporal mean and the temporal variance of the time series $\tilde{F}_j(t) \equiv F_j(t)/m(t)$ of word appearances normalised by the scaled total number of blogs (i.e., the system size) $m(t)$, in the case that $m(t)$ does not have observation errors. }
$E[\tilde{F}_j]$ is written as   
\begin{eqnarray}
E[\tilde{F}_j]&=&E[\check{c}_j \cdot r_j \cdot m)/m+\epsilon_j/m]  \\
&\approx& \check{c}_j \cdot E[r_j] \cdot E[\epsilon_j/m] \\
&\approx& \check{c_j},
\end{eqnarray}
where we use the fact that $\sum^{T}_{t=1}\frac{\epsilon_j(t)}{T} \approx 0$  $(T>>1)$. \par 
Similarly, $V[\tilde{F}_j]$ is derived as  
\begin{eqnarray}
V[\tilde{F}_j]&=&  V[(c_j \cdot r_j \cdot m)/m+\epsilon_j/m] \\
&\approx&  V[c_j \cdot r_j]+V[\epsilon_j/m] \\
&\approx& E[\frac{1}{m}]\check{c}_j+ \nonumber \\
&\check{c}_j^2& \cdot\{ V[r_j]+ E[ m^{2 \beta_{m}-2}]E[r_j^2] \cdot E[\Delta_0^2]) \\
&\approx& E[\frac{1}{m}]\check{c}_j+ \nonumber \\
&\check{c}_j^2& \{V[r_j]+ (V[r_j]+1)\cdot E[ m^{2 \beta_{m}-2}] \cdot E[\Delta_0^2] )\}. \nonumber  \\
\label{V_tilde_F_ex}
\end{eqnarray}
\textcolor{black}{\subsubsection{The case that the system size is observed with observation errors, $\hat{F}_j(t)$: }}
We now consider the case where the number of bloggers $m(t)$ includes observational errors.
We denote the scaled total number of blogs with noise by $n(t)$. Then, 
the normalised time series of word appearances is defined by  
\begin{equation}
\hat{F}_j(t) \equiv F_j(t)/n(t).
\end{equation}
Here, for simplicity  
we assume that the relationship between $m(t)$ and $n(t)$ is given by 
\begin{equation}
m(t)-n(t)=n(t)^{\beta_{n}} \cdot u(t), \label{noise_noise}
\end{equation}
where $u(t)$ is the random noise that satisfies the following conditions: (i) $\sum^{T}_{t=1}u(t)/T = O(1/\sqrt{T}) \approx 0$ and $\sum^{T}_{t=1}u(t)^2/T = \check{\Delta}_n^2+ O(1/\sqrt{T}) \approx \check{\Delta}_n^2$ for $T>>1$, (ii) $ |e(t)| <<  1 $, and (iii) $\beta_{n}$ is a constant, which determines the properties of the noise.
 \par
Then, $E[\check{F}_j]$ is approximated as  
\begin{eqnarray}
E[\hat{F}_j(t)]&=&E[\check{c}_j \cdot r_j \cdot m/n+  \epsilon_j/n] \\
&\approx& \check{c}_j \cdot E[r_j] \cdot 1 \\
&\approx& \check{c}_j.
\end{eqnarray}
Similarly, $V[\check{F}_j]$ is estimated as  
\begin{eqnarray}
V[\hat{F}_j(t)]&=&V[\check{c}_j \cdot r_j \cdot m/n+\epsilon_j/n] \\
&\approx& V[c_j \cdot r_j \cdot m/n]+V[\epsilon_j/n] \\
&\approx& \check{c}_j \cdot E[\frac{m}{n^2}]  \nonumber \\
&+& \check{c}_j^2 \cdot \{V[r_j]\cdot E[\frac{m}{n}]^2 \nonumber \\
&+&(V[r_j]+1)(V[\frac{m}{n}]+E[\frac{m^{ 2 \beta_{m}}}{n^2}] \cdot E[\Delta_0^2]) \}. \nonumber \\
\label{V_hat_F_ex}
\end{eqnarray}
\par
{\bf The formula expressed by $V[n]$} \\
As in the case that $m(t)$ is unobservable, we can express Eq. \ref{V_hat_F_ex} in terms of $n(t)$ only (without using $m(t)$).
In this section, we consider only simple cases, for which $\beta_m=0,0.5,1.0$. \par
Using the relationship in Eq. \ref{noise_noise}, we can obtain the following results:
\begin {itemize}
\item The temporal mean of $E[\frac{m}{n}]$ 
\begin{equation}
E[\frac{m}{n}] \approx E[\frac{n+n^{\beta_{n}} \cdot u}{n}] \approx E[\frac{n}{n}] =1
\end{equation}
\item  The temporal variance of $V[\frac{m}{n}]$ 
\begin{eqnarray}
V[\frac{m}{n}] &\approx& V[\frac{n+ n^{\beta_{n}} \cdot u}{n}] \approx V[\frac{n}{n}]+V[n^{\beta_{n}-1} u] \nonumber \\ 
&\approx& E[n^{2\beta_{n}-2} \cdot \Delta_n^2] \approx E[n^{2\beta_{n}-2}] \cdot E[\Delta_n^2] \nonumber \\
\end{eqnarray}
\item  The temporal mean of $E[\frac{m^{ 2 \beta_{m}}}{n^2}]$  \par
For $\beta_{m}=0$, we have
\begin{eqnarray}
&&E[\frac{m^{ 2 \beta_{m}}}{n^2}] \approx E[\frac{(n+n^{\beta_{n}} \cdot u)^{0} }{n^2}] \\
&=& E[n^{-2}].  \label{beta00}
\end{eqnarray}
For $\beta_{m}=0.5$, we have
\begin{eqnarray}
&&E[\frac{m^{ 2 \beta_{m}}}{n^2}] \approx E[\frac{(n+n^{\beta_{n}} \cdot u)^{1} }{n^2}] \\
&\approx&  E[1/n]+E[n^{\beta_{n}-2}] \cdot E[ u^{1}] \\
& \approx& E[1/n].   \label{beta05}
\end{eqnarray}
For $\beta_{m}=1.0$, we have
\begin{eqnarray}
&&E[\frac{m^{ 2 \beta_{m}}}{n^2}] \approx E[\frac{(n+n^{\beta_{n}} \cdot u)^{2} }{n^2}] \\
&\approx&  1+E[n^{2\beta_{n}-2}] \cdot E[ u^{2}] \\
& \approx& 1+ E[n^{2\beta_{n}-2}] \cdot E[\Delta_n^2].  \label{beta10}
\end{eqnarray}
\end{itemize}
Thus, substituting these expressions into Eq. \ref{V_hat_F_ex}, we obtain the variance of $\hat{F}_j(t)$ as follows.
For $\beta_{m}=0$,
\begin{eqnarray}
&&V[\hat{F}_j(t)] \nonumber \\
&\approx& \check{c}_j \cdot E[\frac{1}{n}]+  \nonumber \\
&\check{c}_j^2&  \{V[r_j]+(V[r_j]+1) \cdot (E[n^{2\beta_{n}-2}]E[\Delta_n^2] \nonumber \\
&+& E[1/n^2]E[\Delta_0^2]))  \}.  \nonumber \\ 
\end{eqnarray}
For $\beta_{m}=0.5$,
\begin{eqnarray}
&&V[\hat{F}_j(t)] \nonumber \\
&\approx& \check{c}_j \cdot E[\frac{1}{n}]+  \nonumber \\
&\check{c}_j^2&  \{V[r_j]+(V[r_j]+1) \cdot (E[n^{2\beta_{n}-2}]E[\Delta_n^2] \nonumber \\
&+& E[1/n]E[\Delta_0^2]))  \}.  \nonumber \\ 
\end{eqnarray}
For $\beta_{m}=1.0$,
\begin{eqnarray}
&&V[\hat{F}_j(t)] \nonumber \\
&\approx& \check{c}_j \cdot E[\frac{1}{n}]+  \nonumber \\
&\check{c}_j^2&  \{V[r_j]+(V[r_j]+1) \cdot (E[n^{2\beta_{n}-2}]E[\Delta_n^2] \nonumber \\
&+& E[\Delta_0^2](1+E[n^{2\beta_{n}-2}]E[\Delta_n^2])))  \}.  \nonumber \\ 
\end{eqnarray}
Considering the fact that $E[n^{2\beta_{n}-2}]E[\Delta_n^2]<<1$, we substitute $E[n^{2\beta_{n}-2}]E[\Delta_n^2]$ for $1$. Then,
\begin{eqnarray}
V[\hat{F}_j(t)] &\approx& \check{c}_j \cdot E[\frac{1}{n}]+  \nonumber \\
&\check{c}_j^2&  \{ V[r_j]+(V[r_j]+1)  \nonumber \\
&\cdot& (E[n^{2\beta_{n}-2}] \cdot E[\Delta_n^2]+ E[\Delta_0^2])))  \}.  \nonumber \\
\end{eqnarray}
\par
\subsection{The temporal variances of the differential of a normalised time series: $V[\delta \hat{F}_j]$ and $V[\delta \tilde{F}_j]$}
\subsubsection{The case that the system size is observed with observation errors, $\delta \hat{F}_j(t)$: }  
Next, we calculate the temporal variance of the differential of the time series of normalised word appearances
 $\delta \hat{F}_j(t)$, which is defined by 
 \begin{equation}  
\delta \hat{F}_j(t)=\frac{F(t)}{n(t)}-\frac{F(t-1)}{n(t-1)}.
\end{equation}
Then, we can estimate the variance of $\delta \hat{F}_j(t)$ as 
\begin{eqnarray}
&V[\delta \hat{F}_j(t)]& \nonumber \\ 
&=&V[\delta(c_j \cdot r_j \cdot m/n)-\delta{\epsilon/n}] \\
&\approx& V[c_j \cdot \delta( \cdot r_j \cdot m/n)]+ 2 \cdot V[ \epsilon/n] \\
&\approx& 2 \cdot \check{c}_j \cdot E[\frac{m}{n^2}]  \nonumber \\
&+&\check{c}_j^2 \cdot \{V[\delta (r_j \cdot m/n)]  \nonumber \\
&+&2 \cdot (V[r_j]+1) \cdot E[\frac{m^{2 \beta_{m}}}{n^2}] \cdot  E[\Delta_0^2] \}.  \nonumber \label{delta_V_hat_F_ex}   \\
\end{eqnarray}
%
\par
{\bf The formula expressed by $V[n]$} \\
Now, we express $V[\delta \hat{F}_j(t)]$ without using $m(t)$.
First, we calculate $V[\delta(r_j \cdot m/n)]$ without using $m(t)$:
\begin{eqnarray}
&&V[\delta(r_j \cdot m/n)]  \nonumber \\
&\approx& V[ \delta( r_j (n+n^{\beta_{n}}\cdot e)/n))] \\
&\approx& V[\delta r]+2 \cdot V[r_j \cdot (e) \cdot n^{\beta_{n}-1}] \\
&\approx& V[\delta r]+2 \cdot (V[r_j]+1)  \cdot E[n^{2\beta_{n}-2}] \cdot E[\Delta_n^2]. \nonumber \\
\label{m_2_n_2}
\end{eqnarray}
Then, we substitute this result and Eq. \ref{m_2_n_2} into Eq. \ref{delta_V_hat_F_ex}.
For $\beta_{m}=0$, we can use Eq. \ref{beta00} to obtain:
\begin{eqnarray}
&V[\delta \hat{F}_j(t)]& \nonumber \\ 
&\approx& 2 \cdot \check{c}_j \cdot E[\frac{1}{n}]  \nonumber \\
&+&\check{c}_j^2 \cdot \{V[\delta r_j]  \nonumber \\
&+& 2 \cdot (V[r_j]+1) \cdot (E[1/n^2] \cdot E[\Delta_0^2] \nonumber \\
&+&E[n^{2\beta_{n}-2}] \cdot ( E[\Delta_n^2] ))\}.  \nonumber \label{delta_V_hat_F_ex00}    \\
\end{eqnarray}
For $\beta_{m}=0.5$, we can use Eq. \ref{beta05} to obtain
\begin{eqnarray}
&V[\delta \hat{F}_j(t)]& \nonumber \\ 
&\approx& 2 \cdot \check{c}_j \cdot E[\frac{1}{n}]  \nonumber \\
&+&\check{c}_j^2 \cdot \{V[\delta r_j]  \nonumber \\
&+& 2 \cdot (V[r_j]+1) \cdot (E[1/n] \cdot E[\Delta_0^2] \nonumber \\
&+&E[n^{2\beta_{n}-2}] \cdot ( E[\Delta_n^2] ))\}.  \nonumber \label{delta_V_hat_F_ex05}    \\
\end{eqnarray}
For $\beta_{m}=1.0$, we can use Eq. \ref{beta10} to obtain
\begin{eqnarray}
&V[\delta \hat{F}_j(t)]& \nonumber \\ 
&\approx& 2 \cdot \check{c}_j \cdot E[\frac{1}{n}]  \nonumber \\
&+&\check{c}_j^2 \cdot \{V[\delta r_j]  \nonumber \\
&+& 2 \cdot (V[r_j]+1) \cdot (E[\Delta_0^2] \nonumber \\
&+&E[n^{2\beta_{n}-2}] \cdot E[\Delta_n^2] \cdot (1+ E[\Delta_0^2]))\}.  \nonumber \label{delta_V_hat_F_ex10}   \\
\end{eqnarray}
 \begin{eqnarray}
&V[\delta \hat{F}_j]& \nonumber \\ 
&\approx& 2 \cdot \check{c}_j \cdot E[\frac{1}{n}]  \nonumber \\
&+&\check{c}_j^2 \cdot \{V[\delta r_j]  \nonumber \\
&+& 2 \cdot (V[r_j]+1) \cdot ( E[\Delta_0^2] +E[n^{2\beta_{n}-2}] \cdot E[\Delta_n^2] )\}.  \nonumber \label{delta_V_hat_F_ex2}.   \\
\end{eqnarray}
\par
Here, we can use the facts that $V[r_i]>0$ and $V[\delta r_j]>0$ for $\beta_{m}=0.5$ or $\beta_{n}=0.5$ to obtain the inequality 
\begin{eqnarray}
&V[\delta \hat{F}_j]& \nonumber \\ 
&\geq& 2 \cdot \check{c}_j \cdot E[\frac{1}{n}]  \nonumber \\
&+&2\cdot \check{c}_j^2 \cdot E[1/n] \{(E[\Delta_0^2] +E[\Delta_n^2] )\}. \nonumber \\ 
\label{delta_V_hat_F_ex_geq2} 
\end{eqnarray}
\subsubsection{The case that the system size is observed without observation errors, $\delta \tilde{F}_j(t)$: } 
Next, we calculate the differential of the time series of word appearances scaled by the total number of blogs in the case that the 
total number of blogs can be observed, $\delta \tilde{F}_j(t)$.
Here, $\delta \tilde{F}_j(t)$ is defined by
\begin{equation}
\delta \tilde{F}_j(t)=\frac{F(t)}{m(t)}-\frac{F(t-1)}{m(t-1)}. \label{def_dtF2}
\end{equation}
Formally, $\delta \tilde{ F}_j(t)$ corresponds to $\delta \hat{F}_j(t)$ under the condition that $e(t)=0$.
Thus, replacing $n(t)$ with $m(t)$ in Eq. \ref{delta_V_hat_F_ex}, we obtain 
we can obtain the variance of $\tilde{F}_j(t)$, 
\begin{eqnarray}
&V[\delta \tilde{F}_j(t)]& \nonumber \\ 
&\approx& 2 \cdot \check{c}_j \cdot E[\frac{1}{m}]  \nonumber \\
&+&\check{c}_j^2 \cdot \{V[\delta r_j]  \nonumber \\
&+& 2 \cdot (V[r_j]+1) \cdot  E[m^{2 \beta_{m}-2}]E[\Delta_0^2] \}. \nonumber \\  \label{delta_V_mf_a}
\end{eqnarray}

\subsection{Ensemble mean and the ensemble variance of raw time series, $E_c[F(t)]$ and $V_c[F(t)]$ }
We calculate the ensemble mean and the ensemble variance of the raw time series of word appearances $\{F_j(t)\}$. 
The ensemble mean and the ensemble variance of $\{F_j(t)\}$ are defined by 
\begin{equation}
E_c[F(t)]=\frac{\sum_{\{i:\check{c}_i=c \}}F_i(t)}{\sum_{\{i:\check{c}_i=c \}}1}
\end{equation}
\begin{equation}
V_c[F(t)]=E_c[\{F(t)-E_c[F(t)]\}^2].
\end{equation}
Here, we assume that $E_c[r(t)] \approx 1$. \par
Then, $E_c[F(t)]$ can be written as   
\begin{eqnarray}
E_c[F(t)]&=&E_c[c \cdot r(t) \cdot m(t)+\epsilon(t)] \\
&\approx& c \cdot  m(t) \cdot E_c[r(t)] \\
&\approx& c \cdot m(t).
\end{eqnarray}
In addition, $V_c[F(t)]$ is estimated by  
\begin{eqnarray}
&&V_c[F(t)] \nonumber \\
&=&V_c[c \cdot r(t) \cdot m(t)+\epsilon(t)] \\
&\approx& V_c[c \cdot r(t) \cdot m(t)]+V_c[\epsilon(t)] \\
&\approx& c \cdot m(t) \nonumber \\
&+& c^2\{ m(t)^2 \cdot V_c[r(t)]+m(t)^{2 \beta_{m}} \Delta_0(t)^2 \cdot (1+V_c[r(t)])\}. \nonumber \\
\end{eqnarray}
\textcolor{black}{\subsection{Ensemble variance of the differential of raw time series, $V_c[\delta F(t)]$}}
We calculate the ensemble variance of the differential of the raw time series of word appearances $\{\delta F_j(t)\}$. 
The ensemble mean and the ensemble variance of  $\{\delta F_j(t)\}$ are defined by 
\begin{equation}
V_c[\delta F(t)]=E_c[\{\delta F(t)-E_c[\delta F(t)]\}^2].
\end{equation}
$V_c[\delta F(t)]$ is written as   
\begin{eqnarray}
&&V_c[\delta F(t)] \nonumber \\
&=&V_c[\delta( c \cdot r(t) \cdot m(t))+\delta \epsilon(t)] \\
&\approx&  2 \cdot c \cdot \{\overline{m(t)} \} \nonumber \\
&+& c^2 \cdot \{V_c[\delta (r(t) \cdot m(t))]+ 2 \cdot (\overline{m(t)^{2 \beta_{m}}  \cdot \Delta_0(t)^2} \nonumber \\
&+&\overline{m(t)^{2 \beta_{m}} \cdot \Delta_0(t)^2 \cdot V_c[r(t)]} ) \}. \label{DESF_t} 
\end{eqnarray}
\par
\textcolor{black}{
Under the conditions that $\Delta_0(t) \approx {\Delta}_0(t-1)$ and \textcolor{black}{$V_c[r(t)] \approx V_c[r(t-1)]$}, we can obtain the simple expression 
\begin{eqnarray}
&&V_c[\delta F(t)] \nonumber \\
&\approx& 2 \cdot c \cdot \{\overline{m(t)} \} \nonumber \\
&+& c^2 \cdot \{V_c[\delta (r(t) \cdot m(t))]+ 2 \cdot (\overline{m(t)^{2 \beta_{m}}}\Delta_0(t)^2 \nonumber \\
&+&\overline{m(t)^{2 \beta_{m}}} \cdot \Delta_0(t)^2 \cdot V_c[r(t)] \}, 
\end{eqnarray}
and }under the condition that $\Delta_0(t)=\check{\Delta}_0$ and \textcolor{black}{$V_c[r(t)] \approx V_c[r(t-1)]$}, we can obtain the simple expression 
\begin{eqnarray}
&&V_c[\delta F(t)] \nonumber \\
&\approx& 2 \cdot c \cdot \{\overline{m(t)} \} \nonumber \\
&+& c^2 \cdot \{V_c[\delta (r(t) \cdot m(t))]+ 2 \cdot (\overline{m(t)^{2 \beta_{m}}}\check{\Delta}_0^2 \nonumber \\
&+&\overline{m(t)^{2 \beta_{m}}} \cdot \check{\Delta}_0^2 \cdot V_c[r(t)] \}.
\end{eqnarray}
\textcolor{black}{\subsection{Box ensemble mean and ensemble variance of raw time series, $E^{\zeta}_c[F(t)]$ and $V^{\zeta}_c[F(t)]$ }}
We calculate the box ensemble mean and the ensemble variance of the raw time series of word appearances $\{F_j(t)\}$. 
The box ensemble mean and the box ensemble variance of $\{F_j(t)\}$ are defined by 
\begin{equation}
E_c^{\zeta}[F(t)]=\frac{\sum_{\{i:(c - \zeta) \leq \check{c}_i < (c + \zeta) \}}F_i(t)}{\sum_{\{i:(c - \zeta) \leq \check{c}_i < (c + \zeta) \}}1}
\end{equation}
\begin{equation}
V_c^{\zeta}[F(t)]=E_c^{\zeta}[\{F(t)-E_c^{\zeta}[F(t)]\}^2].
\end{equation}
Here, we introduce the following notation to calculate the statistics.
First, we assume that the box size is proportional to $c$. That is, $\zeta \equiv c \cdot \zeta_0$, where $\zeta_0$ is a given constant.
Second, we define $\check{\zeta}_j$ such that $\check{c}_j=c \cdot \check{\zeta}_j$, and 
we assume for simplicity that $E^{\zeta}_c[\zeta_i] \approx 0$.
\par
Using the introduced notation, we can obtain that
\begin{equation}
c_j(t)=c \cdot (1+\check{\zeta}_j) \cdot r_j(t)=c \cdot r_j'(t),
\end{equation}
where $r'_j(t) \equiv (1+\check{\zeta}_j) \cdot r_j(t)$．
From these assumptions, we can calculate the following statistics:
\textcolor{black}{
\begin{eqnarray}
E^{\zeta}_c[r'(t)]&=&E^{\zeta}_c[(1+\check{\zeta}) \cdot r(t)] \\
& \approx& E^{\zeta}_c[(1+\check{\zeta})] \cdot E[r(t)] \approx 1 
\end{eqnarray} 
 and 
\begin{eqnarray} 
V^{\zeta}_c[r'(t)] &\approx& V_c[r(t)] \cdot V_c[\zeta]+V_c[r]+V_c^{\zeta}[\zeta] \\
&\approx& V_c[r]+V^{\zeta}_c[\zeta](1+V_c[r]).
\end{eqnarray}
}
\par
Thus, $E^{\zeta}_c[F(t)]$ can be written as
\begin{eqnarray}
E^{\zeta}_c[F(t)]&=&E_c[c \cdot r'(t) \cdot m(t)+\epsilon(t)]  \\
&\approx& c \cdot E_c[r'(t)] \cdot m(t)  \\
&\approx& c \cdot m(t).
\end{eqnarray}
Similarly, $V^{\zeta}_c[F(t)]$ can be written as
\begin{eqnarray}
&&V^{\zeta}_c[F(t)] \\
&=&V_c[c \cdot r'(t) \cdot m(t)+\epsilon(t)]  \\
&\approx& c \cdot m(t) \nonumber \\
&+& c^2  \{ m(t)^2 \cdot  V_c[r'(t)]+m(t)^{2\beta_{m}}  \Delta_0(t)^2 \cdot (1+V_c[r'(t))]\} \nonumber \\
&\approx& c \cdot m(t) \nonumber \\
&+& c^2 \{ m(t)^2 V_c[r(t)]+m(t)^{2\beta_{m}} \Delta_0(t)^2 \cdot (1+V_c[r(t))] \nonumber \\
&+&V_c^{\zeta}[\zeta]( (1+V_c[r(t)])]) \cdot (m^2+m(t)^{2\beta_{m}}\Delta_0(t)^2) \}. \label{EFS_zeta_a} 
\end{eqnarray}
\textcolor{black}{\subsection{Ensemble variance of the differential of raw time series, $V^{\zeta}_c[\delta F(t)]$ }
}
We calculate the ensemble variance of the differential of the raw time series of word appearances $\{F_j(t)\}$.
The variance of the differential of the time series of word appearances is defined as 
\begin{equation}
V^{\zeta}_c[\delta F(t)]=E^{\zeta}_c[\{\delta F(t)-E^{\zeta}_c[\delta F(t)]\}^2].
\end{equation}
\par
Note that because $r_i(t)$ and $r_i(t-1)$ are correlated, we cannot employ the replacement of $r$ by $r'$ in this case.  \par 
For simplicity, assuming that $\zeta_j$ obeys the uniform distribution whose domain is $c-c\zeta_0 \leq c \cdot \check{\zeta}_j < c+c\zeta_0$, we can calculate $V^{\zeta}_c[\delta F(t)]$ as the mixture distribution in regard to $\check{\zeta}_j$. 
In order to calculate this mixture distribution, we employ the following general formula for the variance of the mixture distribution. 
If a general random variable $X$ obeys the mixture distribution 
\begin{equation}
P_{X}(x)=\sum_{i=1}^{W}\frac{1}{W}P_{X^{(i)}}(x;\mu^{(i)}),
\end{equation}
where $X_i$ is the random variable with mean $\mu^{(i)}$ and probability density function $P_{X^{(i)}}(x;\mu^{(i)})$, 
then the mean of $X$ is given by $E_{X}[X]=\int_{-\infty}^{-\infty}xP_{X}(x)dx=\mu$. Furthermore, the variance $V_{X}[X]=E_{X}[(X-\mu)^2]=\int_{-\infty}^{-\infty}(x-\mu)^2P_{X}(x)dx$ of $X$ is given by
%
\begin{eqnarray}
&&V_{X}[X] \\
&=&E_{X}[(X-\mu)^2] \\
&=& \sum_{i=1}^{W}E_{X^{(i)}}[(X^{(i)}-\mu^{(i)}+\mu^{(i)}-\mu)^2]  \nonumber \\
&=&  \sum_{i=1}^{W} V_{X^{(i)}}[X^{(i)}]+V_{X^{(i)}}[\mu^{i}]. \label{miture}
\end{eqnarray}
Here, calculating $X^{(i)}$ we have
\begin{eqnarray}
X^{(i)}&=& \delta F_i(t)=\delta( c \cdot \check{\zeta}_i \cdot r_i(t) \cdot m(t))+\delta \epsilon_i(t) \nonumber  \\
\mu^{(i)}&=&c \cdot \check{\zeta}_i \cdot \delta(r_i(t) \cdot m(t)), \nonumber \\
\end{eqnarray}
and by applying the formula in Eq. \ref{miture} we can calculate
$E_{X^{(i)}}[(\mu^{(i)}-\mu)^2]$,
\begin{eqnarray}
&&E_{X^{(i)}}[(\mu^{(i)}-\mu)^2] \approx c^2 \cdot\{ V_{c}^{\zeta}[\zeta_] \cdot V_{c}^{\zeta}[\delta (r(t) m(t))] \nonumber \\
&+&E^{\zeta}_{c}[\zeta]^2 \cdot V_{c}^{\zeta}[\delta (r(t) m(t))]+V_{c}^{\zeta}[\zeta] \cdot E_{c}^{\zeta}[\delta (r(t) m(t))]^2 \}. \nonumber \\
\end{eqnarray}
Replacing $c$ with $c\cdot\zeta_i$ in Eq. \ref{DESF_t}, 
we obtain 
\begin{eqnarray}
&&E_{X^{(i)}}[(X^{(i)}-\mu^{(i)})^2] \nonumber \\
&=& E_c^{\zeta}[c \check{\zeta} \cdot {(m(t-1)+m(t))}  \nonumber \\
&+&c^2 \check{\zeta}^2 \cdot \{V_c^{\zeta}[\delta (r(t) \cdot m(t))] \nonumber \\
&+& 2 (\overline{\Delta_0(t)^2}+ \overline{\Delta_0(t)^2 \cdot V_c^{\zeta}[r_i(t)]}])] \\
&=& 2 \cdot c \cdot E_c^{\zeta}[\check{\zeta}] \cdot {  (\overline{m(t)})}  \nonumber \\
&+& c^2 \cdot(E_c^{\zeta}[\check{\zeta}]^2+V_c^{\zeta}[\check{\zeta}]) \cdot \{V_c^{\zeta}[\delta (r(t) \cdot m(t))] \nonumber \\
&+& 2 \cdot (\overline{m(t)^{2\beta_{m}} \cdot \Delta_0(t)^2} \nonumber \\
&+&\overline{m(t)^{2\beta_{m}} \cdot \Delta_0(t)^2 \cdot V_c^{\zeta}[r(t)]} ) \}.  
\end{eqnarray}
Using the assumption that $E_c^{\zeta}[\check{\zeta}] \approx 1$, we get that
\begin{eqnarray}
&&E_{X^{(i)}}[(X^{(i)}-\mu^{(i)})^2] \nonumber \\
&\approx& 2 \cdot c \cdot (\overline{m(t)})  \nonumber \\
&+&c^2 \cdot(1+V_c^{\zeta}[\check{\zeta}]) \cdot \{V_c^{\zeta}[\delta (r(t) \cdot m(t))] \nonumber \\
&+& 2 \cdot (\overline{m(t)^{2\beta_{m}} \cdot \Delta_0(t)^2} \nonumber \\
&+&\overline{m(t)^{2\beta_{m}} \cdot \Delta_0(t)^2 \cdot V_c^{\zeta}[r(t)]} ) \}.
\end{eqnarray}
Thus, we have that
\begin{eqnarray}
&&V_X[X]\\
&\approx&V^{\zeta}_c[\delta F(t)] \nonumber \\
&\approx& 2 \cdot c \cdot (\overline{m(t)})  \nonumber \\
&+&c^2 \cdot \{V^{\zeta}_c[\check{\zeta}] \cdot V^{\zeta}_c[\delta (r(t) m(t))] \nonumber \\
&+&E^{\zeta}_c[\check{\zeta}]^2 \cdot V^{\zeta}_c[\delta (r(t) m(t))]+V^{\zeta}_c[\check{\zeta}] \cdot E^{\zeta}_c[\delta (r_i(t) m(t))]^2 \nonumber \\
&+&(1+V^{\zeta}_c[\check{\zeta}]) \cdot (V^{\zeta}_c[\check{\delta} (r(t) \cdot m(t))] \nonumber \\
&+& 2 \cdot (\overline{m(t)^{2\beta_{m}} \cdot \Delta_0(t)^2} \nonumber \\
&+&\overline{m(t)^{2\beta_{m}} \cdot \Delta_0(t)^2 \cdot V^{\zeta}_c[r(t)]} )\}.
\end{eqnarray}
Here, in the case that $\Delta_0(t)=\check{\Delta}_0$ and \textcolor{black}{$V_c^{\zeta}[r(t)] \approx V_c^{\zeta}[r(t-1)]$,} we can obtain the following expression:
\begin{eqnarray}
&&V^{\zeta}_c[\delta F(t)] 
\\ &\approx& 2 \cdot c \cdot (\overline{m(t)})  \nonumber \\
&+&c^2 \cdot \{V^{\zeta}_c[\check{\zeta}] \cdot V^{\zeta}_c[\delta (r(t) m(t))] \nonumber \\
&+&E^{\zeta}_c[\check{\zeta}]^2 \cdot V^{\zeta}_c[\delta (r(t) m(t))]+V^{\zeta}_c[\check{\zeta}] \cdot E^{\zeta}_c[\delta (r(t) m(t))]^2 \nonumber \\
&+&(1+V^{\zeta}_c[\check{\zeta}]) \cdot (V^{\zeta}_c[\delta (r(t) \cdot m(t))] \nonumber \\
&+& 2 \cdot \check{\Delta}_0^2 \cdot (\overline{m(t)^{2\beta_{m}}} \nonumber \\
&+&\overline{m(t)^{2\beta_{m}}}  \cdot V^{\zeta}_c[r(t)] \}. \label{EFS_zeta_delta_F}
\end{eqnarray}
\renewcommand{\theequation}{C.\arabic{equation}}
\renewcommand{\thefigure}{C-\arabic{figure}}
\setcounter{figure}{0}
\setcounter{equation}{0}
\section{Correlation between word counts and the total number of blogs, $Cor[F_j,m]$ } 
We calculate the correlation between the time series of word appearances $F_j(t)$ and the total number of blogs $m(t)$, which is defined by
\begin{eqnarray}
Cor[F_j,m]&=&\frac{E[(F_j(t)-E[F_j]) \cdot(m(t)-E[m])]}{(V[F_j] \cdot V[m])^{1/2}}. \nonumber \\ 
\label{cor_def} 
\end{eqnarray}
The numerator of Eq. \ref{cor_def} is calculated as   
\begin{eqnarray}
&&E[(F_j(t)-E[F_j]) \cdot(m(t)-E[m])] \\
&=& E[(c_j \cdot r_j(t) \cdot m(t)+\epsilon_j(t)-c_j) \cdot(m(t)-1)] \\
&\approx& E[c_j\{(r_j \cdot m(t)^2 )-(r_j \cdot m(t) )-(m(t)-1) \}] \\
&\approx& \check{c}_j \{ E[r_j]\cdot E[m^2]+Cor[r_j,m^2]\sqrt{V[r_j]V[m^2]} \nonumber \\
&-&E[r_j]E[m]-Cor[r_j,m]\sqrt{V[r_j]V[m]}-E[m-1]\}  \nonumber \\
&\approx& \check{c}_j \{  V[m]+Cor[r_j,m^2]\sqrt{V[r_j]V[m^2]} \nonumber \\
&-&Cor[r_j,m]\sqrt{V[r_j]V[m]}\}.  \nonumber \\
\label{cor_nomi}
\end{eqnarray}
Hence, by substituting Eqs. \ref{cor_nomi} and \ref{V_F_ex} into Eq. \ref{cor_def}, we obtain the correlation
\begin{eqnarray}
&&Cor[F_j,m]  \nonumber  \\ 
&&\approx \nonumber  \\
&& \check{c}_j \cdot \frac{1}{\sqrt{ V[m] \{ \check{c}_j+\check{c}_j^2 \{V[r_j]+(V[r_j]+1)}} \nonumber \\
&&\cdot \frac{1}{\sqrt{(V[m]+E[m^2] \cdot E[\Delta_0^2]) \}\}}} \nonumber \\
&& \cdot \left\{V[m]+Cor[r_j,m^2]\sqrt{V[r_j]V[m^2]} \right. \nonumber \\ 
&-&\left. Cor[r_j,m]\sqrt{V[r_j]V[m]}\right\}. \label{cor_cor}
\end{eqnarray}
Under the condition that $r_j(t)$ and $m(t)$ are independent, we have that $Cor[r_j,m] \sim 0$, $Cor[r_j,m^2] \sim 0$ for $T>>1$.  
Under these conditions, we obtain that
\begin{eqnarray}
&&Cor[F_j,m] \approx \nonumber \\ 
&&  \check{c}_j \cdot \sqrt{V[m]} \cdot 1/[\check{c}_j+ \nonumber \\
&&\check{c}_j^2 \{V[r_j]+(V[r_j]+1) \cdot(V[m]+E[m^{2 \beta_{m}}] \cdot E[\Delta_0^2]) \}]^{0.5}. \nonumber \\
\end{eqnarray}
In addition, by using the fact that $V[r_j] \geq 0$ we obtain the following simpler expression \textcolor{black}{for the theoretical upper bound}:  
\textcolor{black}{
\begin{eqnarray}
&&Cor[F_j,m] \nonumber \\
&& \leq   c \cdot \sqrt{\frac{V[m]}{\check{c}_j+\check{c}_j^2 \{(V[m]+E[m^{2 \beta_{m}}] \cdot E[\Delta_0^2]) \}}}. \quad (T>>1) \nonumber \\\label{cor_a}
\end{eqnarray}
}
\par
However, in our observation the assumption that $T$ is very large is not sufficiently satisfied ($T \approx 2000$). That is, the approximations $Cor[r_j,m_j] \approx 0$ and $Cor[r_j,m_j^2] \approx 0$ are not always accurate, because $r_j(t)$ is sometimes contingently similar to $m(t)$.  Thus, the upper bound of $Cor[F_j,m_j]$ given by Eq. \ref{cor_cor} is not always highly accurate.
Thus, we calculate the case with 
$r_j(t)=m(t)$, for which  $Cor[r_j,m_j]$ is the largest.
Under this condition,  we obtain the correction for the upper bound as    
\textcolor{black}{
\begin{eqnarray}
&&Cor[F_j,m] \nonumber \\
& \leq & \frac{\check{c}_j \cdot \{E(m^3)-E[m^2]\}}{\sqrt{V[m]  [ \check{c}_j (1+V[m])+\check{c}_j^2 \{ V[m^2]+E[\Delta^2_2] (1+V[m^2]) \}]}}. \nonumber \\\label{cor_up}
\end{eqnarray}
}
In order to obtain this result, we calculate the numerator in Eq. \ref{cor_def} as follows:
\begin{eqnarray}
&&E[(F_j(t)-E[F_j]) \cdot (m(t)-E[m])] \\
&\approx&E[ \check{c}_j \{(m_j^3)-(m_j^2)-(m-1) \}] \\
&\approx& \check{c}_j \{E [m_j^3]-E[m_j^2] \}.
\end{eqnarray}
Furthermore, the denominator in Eq. \ref{cor_def} is calculated as follows:
\begin{eqnarray}
&&V[F_j] \nonumber \\
&\approx&V[c \cdot m \cdot m+\epsilon_j] \\
&\approx&c^2 \cdot V[m^2]+V[\epsilon_j] \\
&\approx&c(V[m]+1) \nonumber \\
&+&c^2\{V[m^{2}]+E[\Delta_0^2](1+V[m^{\beta_{m}+1}])\}.  
\end{eqnarray}
\renewcommand{\theequation}{D.\arabic{equation}}
\renewcommand{\thefigure}{D-\arabic{figure}}
\setcounter{figure}{0}
\setcounter{equation}{0}
\section{Probability density function}
We calculate the temporal probability density function of the time series of word appearances, $F_j(t)$. 
The temporal probability density function of the time series of word appearances $F_j(t)$ is given by 
\begin{eqnarray}
&&P_{F_j(t)}(F_j(t))= \nonumber \\
&&\int^{\infty}_{0} \frac{(c_j(t) \cdot m')^{F_j(t)} \cdot \exp(-c_j(t) \cdot m') }{F_j(t)!} \phi_{m(t)}(m') dm', \nonumber \\
\label{def_rd}
\end{eqnarray} 
where $\phi_{m(t)}(x)$ is the probability density function with mean $m(t)$ and standard deviation $\Delta_m(t)$． 
%
\subsection{Case of very large $c_j(t)$}
First, we consider the density function for very large $c_j(t)$. 
When the Poisson parameter is very large, the Poisson distribution can be approximated by the normal distribution. Hence, by approximating the Poisson distribution by the normal distribution with mean $c_j(t) \cdot m'$ and standard deviation $\sqrt{c_j(t) \cdot m'}$, we have
\begin{eqnarray}
&&P_{F_j(t)}(F_j(t)) \approx \nonumber \\
&&\int^{\infty}_{0} \frac{1}{\sqrt{2 \pi c_j(t) m' }} \nonumber \\
&\cdot & \exp{\left( \frac{-(F_j(t)-c_j(t) m')^2}{\sqrt{2 c_j(t) m'}}\right)} \phi_{m(t)}(m') dm'  \\
&=&\int^{\infty}_{0} \frac{ \frac{1}{c_j(t)} }{ \sqrt{\frac{2 \pi m'}{c_j(t)} }} \exp{ \left(\frac{-( m'-\frac{F_j(t)}{c_j(t)} )^2 }{ \sqrt{2 \frac{m'}{c_j(t)} }}\right)} \phi_{m(t)}(m') dm'. \nonumber \label{eq_pdf_large0} \\
\end{eqnarray}
The part of $1/\sqrt{\cdot} \exp(\cdot)$ in Eq. \ref{eq_pdf_large0} is regarded as the probability density function of the normal distribution with mean $F_j(t)/c_j(t)$ and standard deviation $1/c_j(t)$. The probability density function of the normal distribution with a very small standard deviation can be approximated by the delta function. Thus, we get   
\begin{eqnarray}
&&P_{F_j(t)}(F_j(t))  \nonumber \\
&\approx& \frac{1}{c_j(t)}  \int^{\infty}_{0} \delta(m'-\frac{F_j(t)}{c_j(t)}) \phi_{m(t)}(m') dm' \nonumber \\
&=& \frac{1}{c_j(t)} \phi_{m(t)}(\frac{F_j(t)}{c_j(t)}). \nonumber \\ \label{eq_pdf_large}
\end{eqnarray}
\par
This result indicates that $F_j(t)/c_j(t)$ obeys the distribution with the probability density function $\phi_{m(t)}(x)$. 
Moreover, by using the assumption that $\Delta_m(t)=m(t)^{\beta_m} \cdot \Delta_0$, $\phi_m(t)$ can be written as
\begin{equation}
\phi_{m(t)}(x)=1/m(t)^{\beta_{m}} \cdot \phi_0(x-m(t)/m(t)^{\beta_{m}}(t)), \label{phi_m}.
\end{equation}
where $\int^{\infty}_{-\infty}x \phi_{0}(x)dx=0$ and $\int^{\infty}_{-\infty}x^2 \phi_{0}(x)dx=\Delta_0^2$.
Thus, $F_j/(c_j(t) \cdot m(t)^{\beta_m})$ obeys the distribution with the probability density function $\phi_0$ that does not depend on the time and words. \par
\par
Here, because $c_j(t)$ cannot be precisely observed, we consider the effect of its observation errors.  
We assume that the random variable $X$ has the probability density function
\begin{equation}
P_{X}(x)=1/C \cdot \phi(x/C).
\end{equation}
Then, the random variable $X_0=X/C$ obeys
\begin{equation}
P_{X_0}(x)= \phi(x).
\end{equation}
However, when we only observe $C$ with an observation error $q$ $(q<<1)$, where $C'=C \cdot (1+q)$, the probability density function of $X$ normalised by $C'$, $X_0'=X/C'=C \cdot X_0/(C(1+q))=X_0/(1+q)$, is written as 
\begin{equation}
P_{X_0'}(x)= (1+q) \cdot \phi_0((1+q) \cdot x).
\end{equation}
By using the Taylor expansion in terms of $q$, we can obtain 
\begin{eqnarray}
&&P_{X_0'}(x)=P_{X_0}(x) 
+q \cdot  (x \cdot P_{X_0}'(x)+P_{X_0}(x))  \nonumber \\
&+& \frac{1}{2} \cdot q^2  \cdot (x^2 \cdot P_{X_0}''(x)+2 \cdot x \cdot P_{X_0}'(x)) \nonumber \\
&+&\frac{1}{6} \cdot  q^3 \cdot ( x^3 \cdot  P_{X_0}'''(x)+3 \cdot x^2 \cdot P_{X_0}''(x)) \nonumber  \\
&+&O(q^{4}).
\end{eqnarray}
From this equation, we can see that deviation from $P_{X_0}(x)$ is proportional to 
$x$, such that $x \cdot P_{X_0}'(x)$, $x^2 \cdot P_{X_0}''(x)$, etc.
Therefore, when $P_{X_0}(x)$ is concentrated near the origin, the effects of the observation error $q$ become smaller. 
Thus, in order to the confirm the properties of the probability distribution for very large $c_j(t)$, we introduce $w(t)$ given by
\begin{equation}
w_j(t)=\frac{F_j(t)/m(t)-F_j(t-1)/m(t-1)}{c_j(t) \cdot m(t)^{{\beta_m}-1}+c_j(t-1) \cdot m(t-1)^{{\beta_m}-1}}.
\end{equation}
Here, under the condition that $c_j(t) \approx c_j(t-1)$ and $c_j(t)>>1$, $w_j(t)$ approximately obeys the distribution with the probability density function $\phi_2(x)=\int^{\infty}_{0}\phi_0(x+p) \phi_0(p) dp$, which does not depend on the time and words.
Because this value is more closely concentrated near to zero than the direct observation, we can reduce the effect of the observation error $q(t)$.
\subsection{Case of small $c_j(t)$}
Here, we calculate the temporal distribution of counts of word appearances $F_j(t)$ for very small $\check{c}_j$.
Under the condition that $\check{c}_j$ is very small (i.e., $\check{c}_j<<1$ ) and the sample is finite, we consider the case that $F_j(t)$ takes values of only $0$ or $1$.
Thus, the probability that $F_j(t)$ takes a value of 0 is given by 
\begin{eqnarray}
&&P_{F_j(t)}(0)= \nonumber \\
&&\int^{\infty}_{0} \frac{(c_j(t) \cdot m')^{0} \cdot \exp(-c_j(t) \cdot m') }{F_j(t)!} \phi_{m(t)}(m') dm' \nonumber \\
&=&\int^{\infty}_{0} 1 \cdot \exp(-c_j(t) \cdot m')  \phi_{m(t)}(m') dm' \nonumber \\
&\approx&\int^{\infty}_{0}  (1-(c_j(t) \cdot m'))  \phi_{m(t)}(m') dm' \nonumber \\
&=&1-c_j(t) m(t), 
\end{eqnarray} 
and the probability that $F_j(t)$ takes a value of 1 is given by 
\begin{eqnarray}
&&P_{F_j(t)}(1)= \nonumber \\
&=&\int^{\infty}_{0} \frac{(c_j(t) \cdot m')^{1} \cdot \exp(-c_j(t) \cdot m') }{F_j(t)!} \phi_{m(t)}(m') dm' \nonumber \\
&\approx&\int^{\infty}_{0} (c_j(t) \cdot m')^{1} \cdot (1-c_j(t) \cdot m')  \phi_{m(t)}(m') dm' \nonumber \\
&\approx& c_j(t) m(t).
\end{eqnarray} 
This probability distribution corresponds to the Poisson distribution with Poisson parameter $c_j(t) m(t)$.
\par 
Next, we consider the temporal probability distribution of  $\{F_j(t)\}$. That is, the mixture distribution of $F_j(1),F_j(2),\cdots,F_j(T)$.
From the definition, the temporal probability distribution of $\{F_j(t)\}$ is given by 
\begin{equation}
P(F)_{\{F_j(t)\}} = \frac{1}{T} \cdot \sum^{T}_{t=1}P(F)_{F_j(t)}.
\end{equation}
Owing to the assumption that $\check{c_j}<<1$, we have that $P_{F_q(t)}(2)<<0$.
Therefore, we can obtain the probability  
\begin{equation}
P_{\{F_j(t)\}}(0)=\frac{1}{T} \cdot \sum^{T}_{t=1}P(0)_{F_j(t)}=\sum^{T}_{t=1}(1-c_j(t))/T=1-\check{c}_j, 
\end{equation}
\begin{equation}
P_{\{F_j(t)\}}(1)=\frac{1}{T} \cdot \sum^{T}_{t=1}P(0)_{F_j(t)}=\sum^{T}_{t=1}c_j(t)/T=\check{c}_j. \label{eq_pdf_small}.
\end{equation}
This probability distribution corresponds to the Poisson distribution with Poisson parameter $\check{c}_j$. \par
However, from the actual observations depicted in Fig. \ref{pdf_fig}, we can confirm that the distribution also obeys a Poisson distribution under the condition that $c_j(t) \approx 5$,  which is not satisfied by the above condition $c_j(t)<<1$. Now, we consider the reason for these observations. \par
We consider a random variable $U$ that obeys the random diffusion model with   
the scale parameter $C \geq 0$, and a random variable $M \geq 0$ whose probability distribution is $\phi_{M}(x)$ with the mean $<M>=\int_{0}^{\infty}x\phi_{M}(x)dx=1$. \par
We can calculate the distribution of $U$ as follows:
\textcolor{black}{
\begin{eqnarray}
&&P_{U}(F)= \nonumber \\
&&\int^{\infty}_{0} \frac{(C \cdot m')^{F} \cdot \exp(-C \cdot m') }{F!} \phi_{M}(m') dm' \nonumber \\
&=&\frac{C^{F}}{F!}\int^{\infty}_{0}m'^{F}  \sum^{\infty}_{q=0}\frac{(-C  m')^q}{q!} \phi_{M}(m') dm' \nonumber \\
&=& \frac{C^{F}}{F!} \sum^{\infty}_{q=0} \frac{(-C )^q}{q!} <M^{F+q}> \nonumber \\
&=&  \frac{C^{F}}{F!} \exp(C)  \nonumber \\
&\cdot& \sum^{\infty}_{q=0}(-1)^q \cdot P_{Poi}(q;C) <M^{F+q}>, \nonumber \\
\label{approx_pdf}
\end{eqnarray}
}
where $< \cdot>$ represents the mean of the distribution. 
We can confirm that this equation is in agreement with the Poisson distribution under the condition that $<M^{F+q}>=1$ (for any q).
In addition, we can also obtain the condition that the distribution of $P(F)$ is approximated by a Poisson distribution $<M^{F+r}>=1$ ($0 \leq r \leq q^{(*)}$), 
because the probability density function $P_{Poi}(q;C)$ takes values close to $0$ except for near to $q=C$, and we can neglect the terms with $q^{(*)}>>C$ in the summation of Eq. \ref{approx_pdf}.\par
In the case of $F_j(t)$, which is shown in Figs. \ref{pdf_fig} (a)-(d), $M$ almost corresponds to $m(t)$ and $C$ corresponds to $\check{c}_j$.
Thus, we can obtain that $E[m]=1$, $E[m^2]=1.05$, $E[m^3]=1.16$,
$E[m^3]=1.33$, $E[m^4]=1.57$, $\cdots$, and $E[m^{10}]=4.86$. 
From this series of moments, we can confirm that the distribution of $F_j(t)$ is approximated by the Poisson distribution on the condition that $\check{c}_j \lessapprox 1$. 
 \par
However, from the results shown in Fig. \ref{pdf_fig} (a), we find that approximation of the Poisson distribution nearly holds true for $\check{c}_j \approx 5 \geq 1$.  The reason for this is that the coefficient $(-1)^j$ in Eq. \ref{approx_pdf} reduces the effect of $<M^q> =1$.
For example, under the conditions that $C=5$ and $F=5$, we can use the parameters of actual observations to numerically calculate $P_{U}(5)/P_{Poi}(5) \approx 0.878$.  
This result indicates that the distribution $F_j$ is in approximately a 90 percent agreement with the Poisson distribution under these conditions.
 \par
%
\renewcommand{\theequation}{E.\arabic{equation}}
\renewcommand{\thefigure}{E-\arabic{figure}}
\setcounter{figure}{0}
\setcounter{equation}{0}
\section{Basic fluctuation scaling (the mean and the variance of the random variable $F_j(t)$).}
We calculate the mean and variance of the distribution of $F_j(t)$.
Here, the mean $<F_j(t)>$ is defined as 
\begin{eqnarray}
&&<F_j(t)>=\int^{\infty}_{0}xP_{F_j(t)}(x)dx, \\
\end{eqnarray}
and the variance $<F_j(t)>$ is defined as 
\begin{eqnarray}
&&\left<(F_j(t)-<F_j(t)>)^2\right> \nonumber \\
&=&\int^{\infty}_{0}(x-<F_j(t)>)^2P_{F_j(t)}(x)dx. \nonumber \\
\end{eqnarray} 
\par
From the definition given by \ref{def_rd}, we obtain the mean as
\textcolor{black}{
\begin{eqnarray}
&&<F_j(t)>=\int^{\infty}_{-\infty}x P_{F_j(t)}(x)dx= 
 c_j(t) \cdot m(t).
\end{eqnarray}
}
\textcolor{black}{
Next, we calculate the variance.   
Here, the second moment of $F_j(t)$ is obtained as follows:
\begin{eqnarray}
\left<F_j(t)^2\right>&=&\int^{\infty}_{0} x^2 \int^{\infty}_{0} P_{poi}(x;m' \cdot c_j(t)) \phi_{m(t)}(m') dm' dx \nonumber \\
&=&c_j(t) \cdot m(t)+c_j(t)^2 <m(t)^2>. 
\end{eqnarray}
}
\par
\textcolor{black}{Thus, the variance of $F_j(t)$ is written as 
\begin{eqnarray}
&&\left<(F_j(t)-<F_j(t)>)^2\right> \nonumber \\ 
&=& <F_j(t)^2>-<F_j(t)>^2  \\
&=&c_j(t) \cdot m(t)+c_j(t)^2 \cdot (<m(t)^2> - <m(t)>^2) \nonumber \\
&=& c_j(t) \cdot m(t)+c_j(t)^2 \cdot \Delta_m(t)^2, \nonumber 
\end{eqnarray}
where $\Delta_m(t)^2$ is the variance of $m(t)$, $\Delta_m(t)^2 \equiv <m(t)-<m(t)>^2>$.} \par 
Under the condition that $\Delta_m(t)=m(t)^{\beta_{m}} \cdot \Delta_0(t)$, we obtain the variance as
\begin{eqnarray}
&&\left<(F_j(t)-<F_j(t)>)^2\right>  \nonumber \\
&\approx& c_j(t) \cdot m(t)+(c_j(t) \cdot m(t)^{\beta_{m}})^2 \cdot \Delta_0(t)^2. \nonumber \\
 \label{delta_0_base_sd} 
\end{eqnarray}
%
\par
In the case that $F_j(t)$ can be expressed as   
\begin{equation}
F_j(t)=c_j(t) m(t)+\epsilon_j(t), 
\end{equation}
we can calculate  
\textcolor{black}{
 the temporal first moment of $\epsilon_j(t)$ as follows: 
\begin{equation}
E[\epsilon_j] \equiv \sum^{T}_{t=1}\epsilon_j(t)/T = O(\frac{1}{\sqrt{T}}) \approx 0, 
\label{epsilon_et_then}
\end{equation} 
where $E[\epsilon_j] \rightarrow 0$ for $T \rightarrow \infty$．  
Moreover, the second moment is given by  
\begin{eqnarray}
&&E[\epsilon_j^2] \equiv \sum^{T}_{t=1}\epsilon_j(t)^2/T \\
&=& E[<\epsilon_j^2>]+O(\frac{1}{\sqrt{T}}) \\
&=& E[<(F_j-<F_j>)^2>]+O(\frac{1}{\sqrt{T}})  \\
&=& E[c_j \cdot m+c_j^2 \cdot m^{2 \beta_{m}} \cdot \Delta_0^2]+O(\frac{1}{\sqrt{T}})  \\
&\approx& \check{c}_j \cdot E[r_j] + \check{c}_j^2 \cdot E[r_j^2 \cdot m^{2 \beta_{m}} \cdot \Delta_0^2],  
\label{epsilon_vt}
\end{eqnarray}
where $E[\epsilon_j^2] \rightarrow  \check{c}_j \cdot E[r_j] + \check{c}_j^2 \cdot E[r_j^2 \cdot m^{2 \beta_{m}} \cdot \Delta_0^2]$ in the limit of $T \rightarrow \infty$. 
Here, the temporal variance of $\epsilon_j$, $V[\epsilon_j]$ can also be obtained as follows:
\begin{eqnarray}
V[\epsilon_j]&=&E[\epsilon_j^2]-E[\epsilon_j]^2 \\
&=&\check{c}_j \cdot E[r_j] + \check{c}_j^2 \cdot E[r_j^2 \cdot m^{2 \beta_{m}} \cdot \Delta_0^2]+O(\frac{1}{\sqrt{T}}) \nonumber \\
&\approx& \check{c}_j \cdot E[r_j] + \check{c}_j^2 \cdot E[r_j^2 \cdot m^{2 \beta_{m}} \cdot \Delta_0^2].
\end{eqnarray}
}
 \par
\textcolor{black}{Next, we calculate the corresponding ensemble first and second moments of $\epsilon_j(t)$.
The first moment $E_c[\epsilon(t)]$ is calculated as
\begin{equation}
E_c[\epsilon(t)] \equiv \frac{\sum_{j \in \{j:\check{c}_j=c\}}\epsilon_j(t)}{\sum_{j \in \{j:\check{c}_j=c\}}1}=O(\frac{1}{\sqrt{N_c}}) \approx 0, 
\label{epsilon_ee}
\end{equation}
and the second moment $E_c[\epsilon(t)^2]$ is given by 
\begin{eqnarray}
&&E_c[\epsilon(t)^2] \equiv \frac{\sum_{j \in \{j:\check{c}_j=c\}}\epsilon_j(t)^2}{\sum_{j \in \{j:\check{c}_j=c\}}1} \\
&=& E_c[<\epsilon(t)^2>]+O(\frac{1}{\sqrt{N_c}}) \\
&=& E_c[<(F(t)-<F(t)>)^2>]+O(\frac{1}{\sqrt{N_c}}) \\
&=& E_c[ c r(t) \cdot m(t)+c^2 \cdot (r(t) m(t)^{\beta_{m}})^2  \Delta_0(t)^2]+O(\frac{1}{\sqrt{N_c}}) \nonumber  \\
&\approx& c \cdot m(t) \cdot E_c[r(t)]+c^2 \cdot  m(t)^{2 \beta_{m}}  \cdot \Delta_0(t)^2 \cdot E_c[r(t)^2] \nonumber \\ 
\label{epsilon_ve}
\end{eqnarray}
where $c_j(t)=c\cdot r_j(t)$, $\Delta_m(t)=m(t)^{\beta_{m}} \cdot \Delta_0(t)$, and $N_c$ is the number of the word with mean $c$ (i.e., the number of samples).} \par
\textcolor{black}{In addition, the ensemble variance of $\epsilon_j(t)$, $V_c[\epsilon(t)]$, can be obtained as follows: 
\begin{eqnarray}
&&V_c[\epsilon(t)]=E_c[\epsilon(t)^2]-E_c[\epsilon(t)]^ \\
&=& c m(t) E_c[r(t)]+c^2 \cdot m(t)^{2 \beta_{m}} \Delta_0(t)^2 E_c[r(t)^2] +O(\frac{1}{\sqrt{N_c}}) \nonumber \\
&\approx& c \cdot m(t) \cdot E_c[r(t)]+c^2 \cdot  m(t)^{2 \beta_{m}}  \cdot \Delta_0(t)^2 \cdot E_c[r(t)^2].  \nonumber \\
\label{epsilon_ve}
\end{eqnarray}
}
\renewcommand{\theequation}{F.\arabic{equation}}
\renewcommand{\thefigure}{F-\arabic{figure}}
\setcounter{figure}{0}
\setcounter{equation}{0}
\section{Top 50 dates of high abnormality in uses of adjectives}
Table \ref{abnormal_table} presents the top 50 dates on which high abnormalities in the uses of adjectives were observed from 1.11.2006 to 31.12.2012.
Here, abnormalities in uses of adjectives are measured by deviation from the theoretical lower bound of the ensemble scaling given by Eq. \ref{median_lm}.
From the table, we confirm that usage of adjectives is affected by special events, such as Christmas and New Year, significant earthquakes, the world cup, and meteorological phenomena such as typhoons and high temperature differences.
\begin{table*}[ht]
\centering
\begin{tabular}{clclc|c}
  \hline
  Ranking & Date & Deviation $l$ & Event \\ 
  \hline
1 & 2011-03-13 & 14.62 & Great east Japan earthquake  \\ 
  2 & 2011-03-12 & 10.86 & Great east Japan earthquake   \\ 
  3 & 2011-12-31 & 8.77 &  Year-end and New Year season \\ 
  4 & 2011-03-11 & 8.53 & Great east Japan earthquake  \\ 
  5 & 2012-01-01 & 7.88 & Year-end and New Year season  \\ 
  6 & 2010-12-31 & 7.49 & Year-end and New Year season  \\ 
  7 & 2011-01-02 & 7.22 & Year-end and New Year season  \\ 
  8 & 2012-12-31 & 7.21 & Year-end and New Year season  \\ 
  9 & 2011-01-01 & 7.16 & Year-end and New Year season  \\ 
  10 & 2012-03-11 & 7.14 & Great east Japan earthquake (one year memory)  \\ 
  11 & 2009-12-31 & 6.59 &Year-end and New Year season   \\ 
  12 & 2012-03-12 & 6.31 &Great east Japan earthquake (one year memory)   \\ 
  13 & 2008-12-31 & 5.94 &Year-end and New Year season   \\ 
  14 & 2012-01-02 & 5.78 &Year-end and New Year season    \\ 
  15 & 2007-12-31 & 5.67 & Year-end and New Year season   \\ 
  16 & 2011-03-14 & 5.49 &  Great east Japan earthquake   \\ 
  17 & 2012-04-03 & 5.40 & \textcolor{black}{Terrible storm (by the explosive low-pressure system)} \\ 
  18 & 2010-06-26 & 5.31 & FIFA World Cup \\ 
  19 & 2011-09-21 & 5.30 & Typhoon  \\ 
  20 & 2011-09-22 & 5.04 & Typhoon  \\ 
  21 & 2006-12-31 & 5.01 & Year-end and New Year season    \\ 
  22 & 2010-01-01 & 4.99 & Year-end and New Year season    \\ 
  23 & 2009-10-09 & 4.98 & Typhoon  \\ 
  24 & 2010-01-02 & 4.94 & Year-end and New Year season   \\ 
  25 & 2008-01-02 & 4.76 & Year-end and New Year season   \\ 
  26 & 2009-01-01 & 4.68 & Year-end and New Year season   \\ 
  27 & 2010-04-22 & 4.66 & \textcolor{black}{Large temperature difference from the previous day} \\ 
  28 & 2010-06-25 & 4.61 & FIFA World Cup  \\ 
  29 & 2010-06-30 & 4.59 &  FIFA World Cup \\ 
  30 & 2008-01-01 & 4.57 & Year-end and New Year season  \\ 
  31 & 2010-07-01 & 4.57 &  FIFA World Cup  \\ 
  32 & 2010-12-03 & 4.32 &  Explosive low-pressure system  \\ 
  33 & 2010-07-13 & 4.19 & National election?  \\ 
  34 & 2010-04-07 & 4.14 & \textcolor{black}{Large temperature difference from the previous day} \\ 
  35 & 2009-01-02 & 4.10 & Year-end and New Year season   \\ 
  36 & 2007-01-02 & 3.94 & Year-end and New Year season   \\ 
  37 & 2006-12-24 & 3.93 & Christmas  \\ 
  38 & 2009-08-12 & 3.92 &  2009 Shizuoka earthquake (Magnitude of 6.4) \\ 
  39 & 2010-03-21 & 3.91 &   Typhoon? \\ 
  40 & 2009-03-20 & 3.86 &  \textcolor{black}{Large temperature difference from the previous day} \\ 
  41 & 2010-09-23 & 3.84 &  \textcolor{black}{Large temperature difference from the previous day} \\ 
  42 & 2011-02-14 & 3.83 & St. Valentine's Day  \\ 
  43 & 2010-05-07 & 3.83 &  \textcolor{black}{Large temperature difference from the previous day} \\ 
  44 & 2012-05-22 & 3.82 &  Solar eclipse \\ 
  45 & 2007-01-01 & 3.75 &  Year-end and New Year season   \\ 
  46 & 2012-04-09 & 3.72 &  A new school term and large temperature difference \\ 
  47 & 2008-08-30 & 3.69 &  Unknown \\ 
  48 & 2008-09-27 & 3.66 &  Difference in temperature  \\ 
  49 & 2011-07-18 & 3.64 &  FIFA Women's World Cup, Difference in humidity \\ 
  50 & 2010-12-24 & 3.64 & Christmas   \\ 
   \hline
\end{tabular}
\caption{
The top 50 dates on which high abnormalities in the uses of adjectives measured by the deviation $l$ given by Eq. \ref{median_lm} in the period 1.11.2006 to 31.12.2012.}
\label{abnormal_table}
\end{table*}

\renewcommand{\theequation}{G.\arabic{equation}}
\renewcommand{\thefigure}{G-\arabic{figure}}
\setcounter{figure}{0}
\setcounter{equation}{0}
%
\section{Comparison of the probability density function between theory and observation for various $\check{c}_j$}
Fig. \ref{small_PDF2} presents a comparison between the theoretical distribution and corresponding empirical observations for various $\check{c}_j$ from $\check{c}_j=0.016$ to $\check{c}_j=11080$.
For observational convenience, we introduce a normalised differential value
\begin{equation}
v_j(t)=\delta({F_j(t)/m(t)})/\sigma_{v}(c_j(t),m(t),\Delta_0(t)), \label{v_j}
\end{equation}
and the probability density function of the $j$-th word $\{v_j(t)\}$ is obtained by producing a histogram from the data $v_j(1),v_j(2) v_j(3),\cdots,v_j(T)$.
Here, $v_j(t)$ is defined to fulfil the condition that the temporal standard deviation takes a value of $1$.
\textcolor{black}{
The normalised factor in Eq. \ref{v_j} was calculated by using the conditions
\begin{eqnarray}
\sigma_{\check{F_j}}(t)^2&\equiv&<F_j(t)/m(t)-<F_j(t)/m(t)>>^2 \nonumber \\
&=& (c_j(t)/m(t)+c_j(t)^2 \cdot m(t)^{2 \beta_{m}-2} \cdot \Delta_0(t)^2), \nonumber 
\end{eqnarray}
and 
\begin{eqnarray}
\sigma_{v_j}(t)^2& \equiv &<(\delta \{F_j(t)/m(t)\}-<\delta \{F_j(t)/m(t)\}>)^2> \nonumber \\
&\approx& \sigma_{\check{F_j}}(t)^2+\sigma_{\check{F_j}}(t-1)^2,
\end{eqnarray}
which are derived from Eq. \ref{delta_0_base_sd} with the assumption that $Cor(F_j(t)/m(t),F_j(t-1)/m(t-1)) \approx 0$.
}
In addition, $c_j(t)$ in Eq. \ref{v_j} is estimated 
\textcolor{black}{by the moving median 
\begin{eqnarray}
&&c_j(t)= \nonumber \\
&&Median\{F_j(t-7)/m(t-7),F_j(t-6)/m(t-6),\nonumber  \\ 
&&\cdots,F_j(t+6)/m(t+6),F_j(t+7)/m(t+7)\}. \label{moving_median}
\end{eqnarray}
Here, we employ the median in order to decrease the effects of outliers. 
} 
\textcolor{black}{
Note that in order to avoid the median taking a value of zero under the condition that $\check{c_j} \leq 4$, we exceptionally employ the simple mean
$c_j(t)=\sum^{T}_{t=1}F_j(t)/m(t)$.
}
%
%
The theoretical probability density distribution corresponding to the above mentioned observable empirical distribution is 
the mixture distribution of $\{v_j(t)\}$, and is given by
\begin{equation}
P_{\{v_j\}}(x)=\sum^{T}_{t=1}P_{v_j(t)}(x)/T, \label{v_pdf}
\end{equation}
where 
\begin{eqnarray}
P_{v_j(t)}(x)&=& \sigma(c_j(t),m(t),\Delta_0(t)) \nonumber \\
&\cdot& P_{d_j(t)}(x \cdot \sigma(c_j(t),m(t),\Delta_0(t))) \label{pdf_v_j}
\end{eqnarray}
\begin{equation}
P_{d_j(t)}(x)=\int^{\infty}_{-\infty} P_{u_j(t)}(x+q) \cdot P_{u_j(t)}(q)dq \label{d_j}
\end{equation}
\begin{equation}
P_{u_j(t)}(x)= m(t) \cdot P_{F_j(t)}(x \cdot m(t)). \label{u_j} 
\end{equation}
Here, we apply the formulas for the probability density distribution of the differential and for translations of random variables to $P_{F_j(t)}(x)$.
In addition, we assume that $\phi_{m(t)}(x)$ in $P_{F_j(t)}$, as given in Eq. \ref{def_rd}, is $\phi_{m(t)}(x)=1/m(t) \cdot \phi_0(x/m(t))$, and that $\phi_0(x)$ is a scaled t-distribution with degree of freedom 2.64 and standard deviation $\Delta_0(t)=0.021$. \par
Note that the reason for using the differential $v_j(t)$ is to reduce the effects of observational errors of $c_j(t)$ (see Appendix D)．
In addition, we choose the words used in Fig. \ref{small_PDF2} from those on the bottom curves in Fig. \ref{fig_mean_sd_t} in order to satisfy the condition that $v_j(t)$ obeys the unique probability density function $c_j(t) \approx c_j(t-1)$ for very large $c_j$ (see Appendix D). \par
From Fig. \ref{small_PDF2}, we can confirm that the theoretical curves given by Eq. \ref{v_pdf} (red thick dashed line) are in good agreement with the corresponding empirical observations (black solid line) over a  range of eight digits from $\check{c}_j=0.016$ to $11080$.
Moreover, we can confirm the transition from the (scaled) t-distribution (peach dash-dotted line) to the (scaled) Poisson distribution (blue thin dashed line).
In addition, we can see that the theoretical distribution in the case of a steady time series (i.e., $c_j(t)=\check{c}_j$ and $m(t)=1$) agrees with the empirical distributions 
 in the domain of $\check{c}_j$ in which we neglect discreteness.
\par
\renewcommand{\theequation}{H.\arabic{equation}}
\renewcommand{\thefigure}{H-\arabic{figure}}
\setcounter{figure}{0}
\setcounter{equation}{0}
\section{Estimation of scaled total number of blogs $m(t)$ from the data}
  Here, we estimate the scaled total number of blogs $m(t)$ by using the moving median, as follows: 
\textcolor{black}{
\begin{enumerate}
\item[Step 1.] We create a set $S$ consisting of indexes of words such that $c_j$ takes a value larger than the threshold $\check{c}_j(t) \geq 100$.
\item[Step 2.] We estimate $m(t)$ as the median of $\{F_j(t)/c_j:j \in S \}$ with respect to $j$.
\item[Step 3.] For $t=1,2,\cdots,T$, we calculate $m(t)$ using step 2. 
\end{enumerate}
}
Here, we use the only words with $\check{c}_j(t) \geq 100$ in step 1 because we neglect discreteness. In step 2, we apply the median because of its robustness to outliers.
\par

\renewcommand{\theequation}{I.\arabic{equation}}
\renewcommand{\thefigure}{I-\arabic{figure}}
\setcounter{figure}{0}
\setcounter{equation}{0}
\section{The cumulative distribution of $\delta F_j(t)$}
Here, we present the calculation of the cumulative distribution of $\delta F_j(t)=F_j(t)-F_j(t-1)$.
Using the probability density function $F_j(t)$ given by Eq. \ref{def_rd} and the formula for the probability density function of the differential of two random variables, the probability density function of  $\delta F_j(t)=F_j(t)-F_j(t-1)$ is given by
\begin{equation}
P_{\delta F_j(t)}(x)=\int_{0}^{\infty}P_{F_j(t)}(x+y)P_{F_j(t-1)}(y) dy.
\end{equation}
Then, by taking the integral we can write the cumulative distribution $P_{>\delta F_j(t)}(x)$ as
\begin{equation}
P_{> \delta F_j(t)}(x; c_j(t-1),c_j(t))=\int^{\infty}_{x}P_{\delta F_j(t)}(y)dy. \label{pdf_delta_f}
\end{equation}
Note that this distribution depends on the parameters $c_j(t-1)$ and $c_j(t)$ and the distributions $\phi_{m(t-1)}(x)$ and $\phi_{m(t)}(x)$ (see Eq. \ref{def_rd}). 
In the same manner as in Appendix G, we assume that $\phi_{m(t)}(x)=1/m(t) \cdot \phi_0(x/m(t))$, where $\phi_0(x)$ is a scaled t-distribution with degree of freedom 2.64 and standard deviation $\Delta_0(t)=0.021$, when we apply this equation to the actual blog data.
%
\end{document}